\documentclass[review,3p,times]{elsarticle}

\usepackage{amssymb}
\usepackage{amsmath}

\usepackage{url}

\usepackage{color}
\usepackage{pdflscape}
\usepackage{multirow}
\usepackage[export]{adjustbox}

\begin{document}

\begin{frontmatter}



\title{Battery Second Life : A Review of Experimental Studies}


\author[inst1,inst2,inst3]{Marwan Hassini}
\author[inst1,inst3]{Eduardo Redondo-Iglesias}
\author[inst2,inst3]{Pascal Venet}

\affiliation[inst1]{organization={Université Gustave Eiffel, ENTPE, EMob-Lab, Lyon, France}}
\affiliation[inst2]{organization={Universite Claude Bernard Lyon 1, Ampère, UMR5005, INSA Lyon, Ecole Centrale de Lyon, CNRS,
Villeurbanne, F-69100, France}}
\affiliation[inst3]{organization={EMob-Lab/Ampère Joint Research Team for Energy Management and Storage for Transport), 69500 Bron, France}}

\begin{abstract}
As electric vehicle volume increase, the interest for reusing their retired batteries grow. Giving a second life to batteries reduce their environmental impact and provide access to more affordable energy storage systems. This lifespan extension is of interest in a wide range of fields as the reuse offer a solution for end-of-life batteries management and the development of renewable energies. Although several review articles have described opportunities and barriers related to reuse, to our knowledge none have synthesized insights from studies that conducted experiments on retired batteries. This article aims to address this gap by providing a detailed analysis of more than 150 articles. This article's main findings are the proposition of a new vocabulary for batteries end-of-life, an evaluation of state of health and dispersion in retired batteries as well as the synthesis of aging and safety studies on these batteries. In a nutshell, this review show that the performance of retired batteries can be high, their dispersion should be managed and their degradation can be slow and safe. Limitations of current battery research  ideas for future experiments and ideas for future studies are also presented.
\end{abstract}


\begin{highlights}
\item The widespread definition of battery end-of-life is not consistent with experimental findings
\item Retired batteries can be reused for several thousands of cycles
\item Important limitations of current battery research are highlighted
\end{highlights}


\end{frontmatter}



\section{Introduction}
\label{sec:introduction}

The versatility of lithium-ion batteries has made them ubiquitous in the energy storage market. Historically, consumer electronics, such as mobile phones and laptops, dominated the market because these systems require only a limited amount of energy and power \cite{sun2017}. In recent decades, electric vehicles and renewable energy have been widely adopted as solutions for reducing greenhouse gas emissions \cite{ipcc2022}. The demand for batteries in the electric vehicle and renewable energy industries has significantly increased the required energy and power, leading to an increase in the number and size of batteries. This evolution raises concerns about the production of new batteries, which requires the extraction and processing of large quantities of raw materials \cite{baars2021,bridge2022}. To ensure the environmental benefits of transitioning to electrified mobility and energy production, the environmental impact of batteries must be reduced. Reusing retired batteries is desirable for achieving sustainability because it reduces the need to extract virgin raw materials for new batteries. It also leads to more efficient use of available resources \cite{faessler2021stationary,wraalsen2021circular}.

Several review articles have discussed the advantages and barriers limiting the batteries reuse. These articles provide an overview of the challenges and remaining research questions related to the second life of batteries \cite{martinez2018,haram2021,cready2003technical,junior2023towards,hossa19,al2022techno,kampker2023identification,zhao2021review}. While these review articles cover a significant amount of literature on various aspects of second-life batteries, including opportunities, applications, and barriers, they mostly present an overview of the challenges. They often present the technical capabilities of batteries and the need for experimental studies as paramount for the development of battery reuse. However, in the literature there is a lack of detailed descriptions of the results obtained in the growing number of articles presenting experimental results on retired batteries. To the best of the authors' knowledge, no review article has synthesized their findings. This is an important gap in the literature.

One consequence of this gap is that many authors must make assumptions about battery performance, degradation and safety. In research works related to the design and management of energy systems, or the economic and environmental costs, it is common to assume that batteries cannot be used beyond a certain level of performance degradation, and that replacement is required \cite{heymans2014economic,seger2023storage,chaud2025techno,pilali2026techno}. It is also common to neglect cell-to-cell performance variation in battery packs, since there are no measures available to quantify this phenomenon. These assumptions have significant implications for a large number of studies within a wide range of research communities. Reviewing experimental studies on retired and reused batteries could help to make these assumptions more realistic.

A second consequence of this literature review gap is that the existing experimental studies remain one-off studies that lack visibility and comparison to other works. The diversity of retired battery technologies and the technical challenges they face require an incremental research approach, as no single study can cover such a broad spectrum. Synthesis works are necessary to organize research results and provide a clear overview of the current state of knowledge, possible controversies, and information gaps.

This article aims to address this gap by synthesizing and critically evaluating the most significant findings from experimental studies on retired and reused batteries. The authors hope that this work will contribute to gauging the progress of research, identifying new research opportunities, and increasing the visibility of existing works, resulting in the possibility of producing a realistic and incremental research. The article covers the most prominent topics related to retired batteries and their reuse. Table~\ref{tab:research_questions} shows the research questions investigated in the different sections of the article.

\begin{table}[h!]
    \centering
    \begin{tabular}{ ll } 
    \textbf{Open research questions } & \textbf{Section}\\                   \hline

        \textit{How to define second life batteries ?} &II-III \\
        \hline
         \textit{How performant are retired batteries ?}& III \\
         \hline
         \textit{How to assess rapidly the performance of a retired battery ?} & III \\
        \hline
         \textit{How reused batteries age ?}  & IV\\
        \hline
         \textit{ How retired batteries should be reused and managed to ensure a satisfying lifespan ?} &IV\\
         \hline
         \textit{How battery aging affects their safety ?}  &IV \\
         \hline
   

    \hline       
    \end{tabular}
    \caption{Research questions related to second life batteries}
    \label{tab:research_questions}
\end{table}

\newpage

\section{Method}

\subsection{Literature review}

\begin{figure}[h!]
    \centering
    \includegraphics[width=0.75\columnwidth]{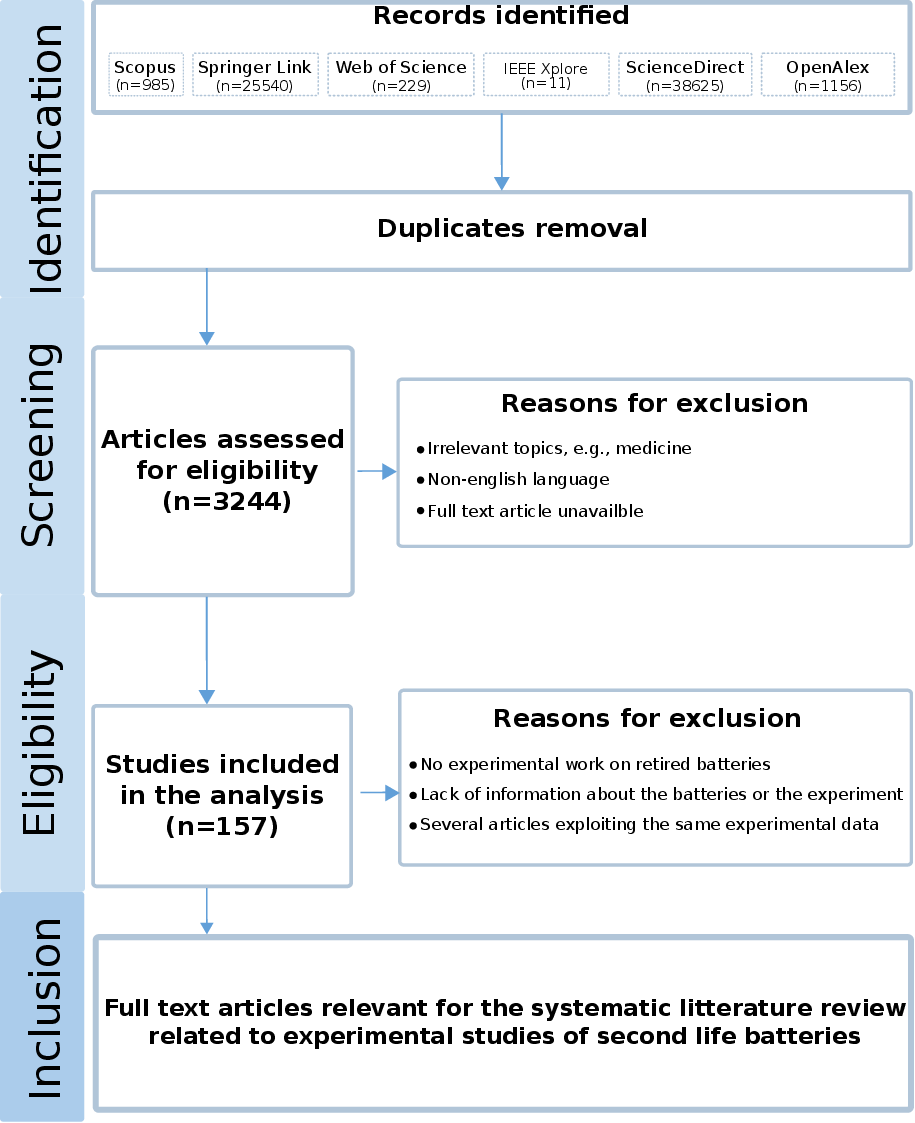}
    \caption{The systematic review results overview shown in the PRISMA flow diagram.}
    \label{fig:review_methodology_carac}
\end{figure}

The literature review method followed the Preferred Reporting Items for Systematic Reviews and Meta-Analyses (PRISMA) principles in order to provide a reliable and comprehensive assessment of existing works \cite{mishra2023prisma}. Figure~\ref{fig:review_methodology_carac} gives an overview of the method used.

Records were identified by browsing six of the most popular scientific literature databases. The last browsing was done on the 4 of june 2026. The table~\ref{tab:keywords_database} shows the keywords and boolean operators used for each database browsed.

\begin{table*}[h!]
    \centering
\resizebox{\columnwidth}{!}{
    \begin{tabular}{ cc } 
    \textbf{Database}&\textbf{Keywords and Boolean Operators}\\
    \hline
        \textbf{OpenAlex} & \shortstack{(second-life  OR repurposed OR retired) AND batter* AND  (state of health OR soh*) AND (experiment* OR test) \\ second-life  OR repurposed OR retired) AND batter* AND (aging OR aging) AND (experiment* OR test)}\\
        \hline
         \textbf{Scopus} &  
        \shortstack{ TITLE-ABS-KEY ( second-life OR repurposed OR retired ) AND batter* AND ALL ( state of health OR soh ) AND ALL ( experiment* OR test  )  \\   TITLE-ABS-KEY ( second-life OR repurposed OR retired ) AND batter* AND ALL ( aging OR aging ) AND ALL ( experiment* OR test ) }  \\
                 \hline

         \textbf{SpringerLink} & \shortstack{( second-life OR repurposed OR retired)  batter*   AND (state of health OR soh*) AND (experiment* or test)\\ ( second-life OR repurposed OR retired)  batter*  AND aging OR aging AND (experiment* or test)}\\
                  \hline

         \textbf{Web of Science} & \shortstack{ALL=((second-life  OR repurposed  OR retired)  AND batter* AND (state of health OR soh) AND (experiment* OR test)) \\   ALL=((second-life  OR repurposed  OR retired)  AND batter* AND (aging OR aging) AND (experiment* OR test))}\\
                  \hline

         \textbf{IEEE Xplore}&\shortstack{(second-life OR retired OR repurposed ) AND batter* AND (state of health OR soh) AND (experiment* or test)\\  (second-life OR retired OR repurposed) AND batter* AND (aging OR aging) AND (experiment* OR test)}\\
         \hline
         \textbf{ScienceDirect} & \shortstack{(second-life OR repurposed OR retired) AND batter* AND (state of health OR soh) AND (experiment or test)\\ (second-life OR repurposed OR retired) AND batter* AND (aging OR aging) AND (experiment* or test)} \\
          \hline
    \end{tabular}
}
    \caption{Search database and keywords}
    \label{tab:keywords_database}
\end{table*}
\newpage
To avoid an incomplete search, the wildcard * was added to match all possibilities, e.g, “batter*”, which included all possible words such as “battery”, “batterie” or “batteries”. The articles collected are then evaluated according to their relevance to the subject of the study. Articles outside the scope, duplicates and non-peer-reviewed articles were excluded from the analysis. For each paper selected, the references and the papers citing it were also checked to add any relevant publications. Finally, the content and experimental data of each paper are analyzed and presented in the following sections of this article.

\subsection{General review of existing studies about second-life batteries}

 In total, 157 articles were investigated \cite{abdel17,aeppli2025aging, ahmei22, amuta2023state,amuta2025novel,attidekou2020methodologies,attidekou2026rapid,azizi22,bauma18,beatty2025long,braco22,braco2021experimental,brand2012aging,camboim2024state,cao2024second,castro2022air,chang2022experimental,chang2026novel,chen2017degradation,chen2022efficient,chung2021charge,cui2024taking,dang2024optimal,de2025diagnostics,drallmeier2022challenges,du2021state,du2026efficient,ezpeleta2025rapid,fan2022fast,fan2025generalized,gao2022annual,gao2024comprehensive,gao2024evaluation,garg2020development,gu2023quick,guenther2024second,haram2023second,hassini2023capacity,he2025state,hosen2026data,huang2022state,huang2024flexible,huang2025rapid,jiang2017recognition,jiang2024state,kehl2021electrical,kemeny2026impact,lacap2021development,lai2019rapid,li2017lifetime,li2020fast,li2022screening,li2024electrochemical,li2025lifetime,li2025state,li2026rapid,liao2017performance,lin2023state,lin2023retired,liu2022characterization,luo2021rapid,lv2022investigation,ma2015investigation,marini2026design,michelini2024impact,montes2025multi,montoya2024quantitative,najjar2026fast,ni2023rapid,peng2026fast,perez2025novel,quinard2019fast,ramirez2023statistical,ran2020data,rasheed2023active,raza2026ev,salinas2019second,sarker2025performance,schaltz2021incremental,scholz2023experimental,schuster2016correlation,she2021battery,sordi2026degradation,sun2022quick,tao2024generative,tong2013off,tran2022used,vasquez2025,vignesh2024need,wang2020,wan2026enhanced,wang2023rapid,wang2023state,wang2026adaptive,wen2023fast,white2020repurposed,xiao2021discharge,xiong2020online,xu2019study,xu2020novel,xu2026regenerate,yang2020layered,yang2023fast,yuwono2026characterization,yuwono2026performance,zhang2014novel,zhang2020performance,zhang2020quick,zhang2021state,zhang2022reliable,zhang2023state,zhang2024_2,zhang2025two,zhang2026developing,zhou2020efficient,zhou2023health,zou2023fast,zhu2025bayesian,zhu2025rapid,ezeigwe2026degradation,kebede2023optimal,ohrelius2025polarization,wang2019study,wheeler2025impact,kuntz2023,jesus2026impact,liu2023,zhang2024_1,waldmann2017,borner2017,wang2025research,carter2019,friesen2016,gerosa2025,kabra2025,zhao2026evolution,abada2018,leonhardt2026effects,liu2020,cruz2025evaluating,burke2013life,faraji2023accelerated,jung2023rapid,liu2023fast,locorotondo2020electrical,lucu2020dataa,lyu2022synchronous,ma2020multilayer,martinez2018technical,mohsin2022new,mowri2021modification,oeser2018effectiveness,pang2024effects,wanner2024diagnosis,wu2023balancing,moy2024second,wang2025deep,timke2024impact}.  To the best of author knowledge, it is one the largest dataset collected for a literature review in the battery field. Figure~\ref{fig:characteristics_studies} shows the characteristics of the studies reviewed.

\begin{table}[h!]
\begin{tabular}{cc}
    \begin{tabular}{c}
     \textbf{Nutshell Summary Statistics}\\
     \hline
        157 studies from 2012 to 2026.  \\
        $>$15,000 retired batteries tested \\
        mostly unknown first life duration (90\%) \\
        1st use : 61\% EV,  21\% unknown\\

        cylindrical dominance (43\%) \\
        pos.~electrode: 37\% LFP, 26\% NMC, 10\% LMO-LNO \\
      7\% with associated open dataset \\
           \hline
    \end{tabular}
&
    \includegraphics[scale=0.5,valign=m]{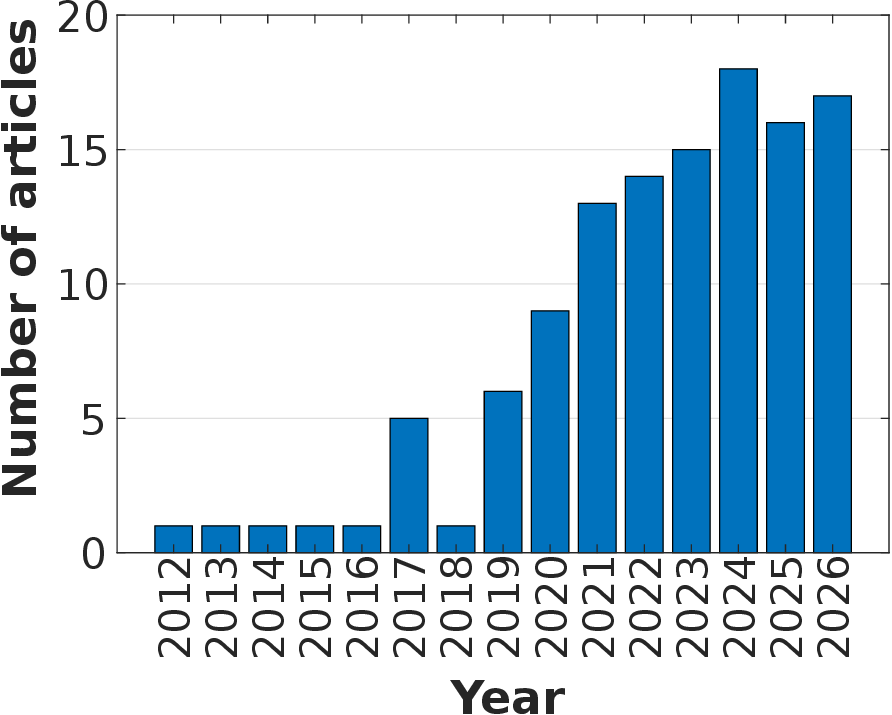}
    \label{fig:number_of_publication_year}
\end{tabular}
\label{OSM_tree}

    \centering
    \includegraphics[width=\columnwidth]{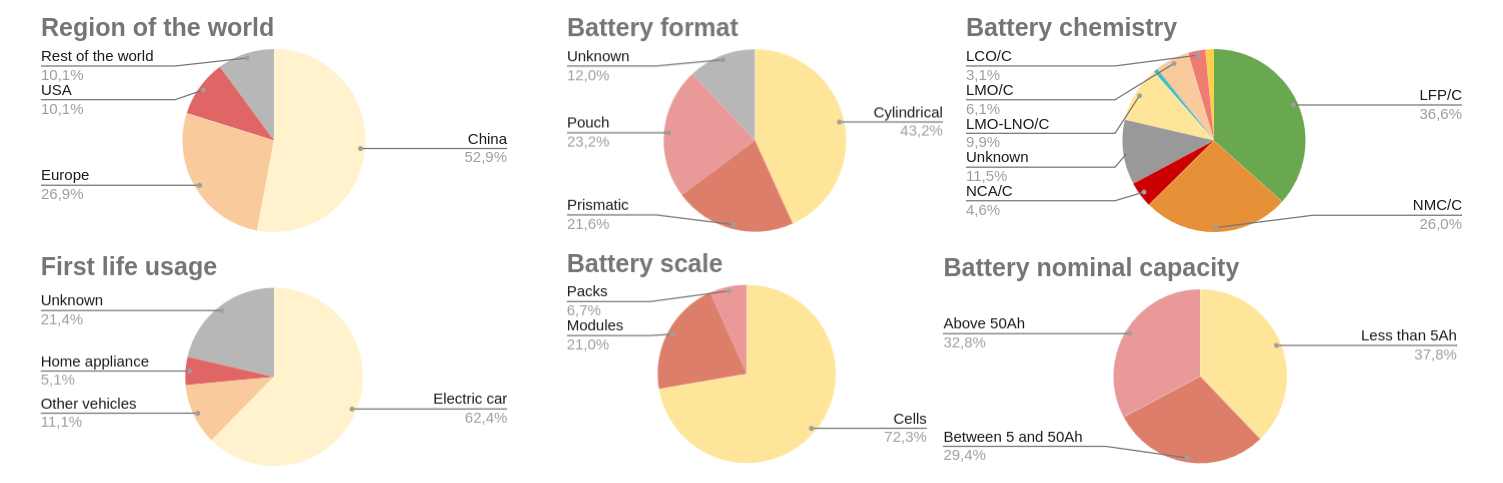}
    \caption{Main characteristics of the studies reviewed in this article. Data collection was stopped in june 2026.}
    \label{fig:characteristics_studies}

\end{table}
\newpage

This synthesis show that this review is timely as the first experimental studies on retired batteries were published but it is only since 2021 that this topic gained interest. In the last five years (2021-2026), 102 articles have been published. Chinese research teams are giving important interest to this topic. This synthesis also illustrates the fact that battery technology is constantly evolving. Since the first study in 2012, the market shares of battery technologies evolved from cobalt-rich technologies such as Lithium Cobalt Oxide (LCO) or Lithium Manganese Oxyde (LMO) to low or free cobalt ones such as Lithium Fer Phosphate (LFP). LFP technology is largely favoured by Chinese teams, which is why it represents almost half of the technologies studied. This choice may be explained by easier access to this technology, as LFP batteries are favoured by Chinese car and bus manufacturers. In terms of cell format, cylindrical is the most common format in retired batteries. This can be explained by their popularity in the automotive sector and the tendency of researchers to choose them as they are easier to test \cite{hassini2023lithium}. This may also explain why the majority of studies have tested cells with limited capacities instead of battery packs or modules. Cell-level studies are necessary to gain an understanding of the performance and durability of different elements. However, cell assembly studies are needed to understand the evolution of cell-to-cell variation. Retired batteries were mostly coming from electric vehicles (61\%) and in the large majority (90\%) batteries first life history is unknown. 

\subsection{Study limitations} \label{sec:definitions}

Before diving into the analysis of the studies, the authors would like to highlight the unavoidable limitations of this study. First, this review may have missed some relevant studies. This can be attributed to the lack of a common vocabulary in the battery field. Since the early 2010' \cite{wolfs2010economic,burke2009performance}, second-life batteries have received increasing attention in various application domains, such as automobiles, energy production, and manufacturing \cite{cusenza2019reuse,bai2024techno,zhu2025towards}. The lack of a common vocabulary on a global scale has made the term "second life battery" polysemous and ambiguous, designating a wide variety of battery types. 

\begin{figure}[h!]
    \centering
    \includegraphics[width=\linewidth]{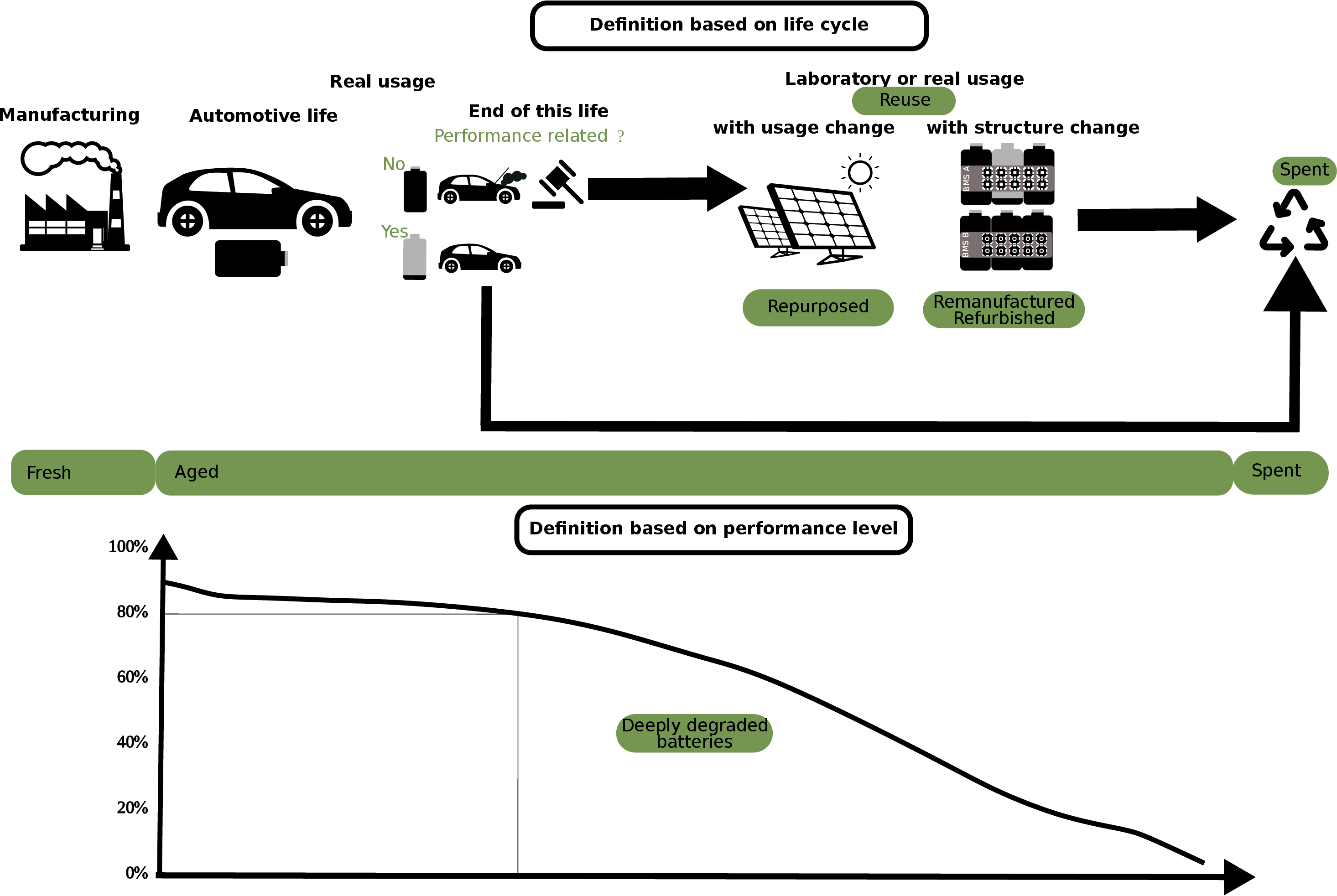}
    \caption{Vocabulary used to define a battery according to its position on life cycle and performance}
    \label{fig:definitions}
\end{figure}

This article proposes a clarified vocabulary to distinguish the different meanings of the term and create a consensus on its use. This is necessary for conducting cumulative research, as studies of the same type may be referenced under different terms, which makes difficult the comparison between different studies. Fig.~\ref{fig:definitions} presents a proposition of the vocabulary to favor according to the battery.

A "fresh" battery is one that is up to a few months old and has just come out of the manufacturer. Any battery that has been used for a longer period of time could be designated as an "aged" battery. A "retired" battery is one that is no longer suitable for its original purpose for any reason. A "highly degraded" battery is one that has a state of health below a certain performance threshold. A "reused" battery is a retired battery that has been deemed suitable for a new life, whether it be the second, third, etc., with or without a change in structure or usage. A "remanufactured" battery is one whose structure or elements have been changed for reuse. A "repurposed" battery is one that has experienced a change in usage (e.g., from a vehicle in its first life to stationary in its second life) for reuse. A "spent" battery is one that has been used to the point that it no longer has any power or energy. The term "second-life battery" should be used with caution due to its popularity. It is used in very different contexts and can have different meanings depending on the author. With this limitation in mind, it can be used to designate "repurposed" or "remanufactured" batteries. However, it should not be used to describe a battery whose performance falls below a certain threshold. The studies that used this definition and for which the first life was in a laboratory were considered in the separated section~\ref{sec:fresh_batteries}.

\subsection{How to compare results ?}

In addition to the lack of information on methodological analysis, the battery field suffers from a lack of sharing of raw data or at least detailed results, eg. capacity evolution over time / cycles . Only 7\% of the articles reviewed in this study shared the raw data they generated. The data used in this study were obtained using WebPlotDigitizer, software that extracts numerical data from images \cite{rohatgi2021webplotdigitizer}. Drevon et al. demonstrated that the software is a reliable and valid tool for extracting data \cite{drevon2017intercoder}. In this article, the accuracy was validated based on previous works by the authors \cite{hassini2023capacity}. The collected data were in exact agreement with experimental measurements. Using this software to collect data drastically reduced the time, energy, and money necessary to produce the same data through experiments. One limitation of this method is its inapplicability to studies publishing very large, homogeneous datasets. Some data points may have been lost due to overlap. For such studies, the authors recommend publishing experimental results in tables or in a separate file alongside the article.

\subsection{A representative sample ?}

There is also no certainty that the collected data are representative of the batteries that will retire in the coming years. An important bias is related to the dismantling of retired batteries. In this review, most studies were conducted at cell scale (72\%). However it is well established that dismantling batteries and especially batteries from electric vehicles is a complex task~\cite{harper2019recycling,hertel2024towards}. Several authors have shown that the cost of disassembling activity may limit significantly the economic interest of reused batteries \cite{rallo2020economic,lander2023breaking}.

The primary objective of this work is to collect the experimental data from the literature on retired batteries. Experimental data from over an hundred studies are gathered and investigated to highlight their main findings, consensus and controversies. This review should be seen as a snapshot of current knowledge regarding retired batteries. Results presented in this review will need updates as retired batteries types evolves and volumes grow but also as research techniques and solid knowledge are developed.

\section{Choosing retired batteries} \label{sec:choosing_batteries}

From both a research and industry standpoint, choosing the right batteries for reuse is a significant challenge. The growing number of studies presenting experimental measurements on retired batteries provides an opportunity to address this issue. This review addresses three common practices. First, studies on retired batteries often assume that a battery has reached the end of its first life after experiencing a 20-30\% capacity loss. This section will compare this theoretical threshold to experimental data. Second, many studies omit an assessment of the level and influence of cell dispersion on retired battery performance. This section will evaluate whether this assumption is reasonable. Finally, this section will review the most promising characterization techniques that could be used on retired batteries to maximize diagnostic accuracy while avoiding destruction and resource intensity
\cite{dubarry2020perspective}. In this section, authors aim to contribute to answering the following research questions : 
\begin{itemize}
    \item Are the experimental data from literature consistent with the definition of a 20-30\% loss of capacity threshold for retired batteries ? 
   \item How important is the performance spread among retired batteries ?
   \item How to assess rapidly the performance of a retired battery ?
\end{itemize}


\subsection{State of health evaluation of retired batteries}
A popular assumption is to consider that battery end of first life is reached after a capacity loss of 20-30\% \cite{lee2021lithium,huang2017online}. This theoretical threshold was defined in automotive standards in the 1990' \cite{USABC1996}. At the time, they were in line with the reduced range of vehicles available on the market. Since then, and despite the continuous increase in vehicle autonomy, this threshold has remained very popular. Although it may seem reasonable to use it as a first approach, it is important to compare this theoretical definition with experimental results. Different research communities are using this threshold in their work for estimating the rest of useful life of batteries \cite{li2021, zhou2020,deng2021}, capacitors \cite{el2019lifetime,ayob2022soc,rochefort2025forecasting} or fuel cells \cite{jouin2016estimating,hua2022review}. Aging campaigns are generally stopped before this fixed threshold \cite{severson2019,wildfeuer2023}. This decision leads to an important gap in the understanding of battery degradation beyond this threshold. Aging models are also affected as they are less reliable for highly degraded batteries. The same threshold is also used whatever the use and acceleration factor studied. This choice undoubtedly affects our understanding of the aging mechanisms that can affect deeply degraded batteries.  The use of this fixed threshold does not only affect aging studies and battery research community. Several research communities working on battery-powered systems also use this assumption. Electric and hybrid vehicles, as well as photovoltaic installations, are a few examples of uses for which the threshold influence has been demonstrated \cite{salim2019, anselma2021, gomez2023}. Scientists studying the environmental and economic impact of systems using batteries are also affected as the use of a fixed threshold impact the quality of their work. For this type of study, the threshold is commonly used to define the point at which the storage system must be replaced \cite{marques2019,le2020}. It is clear that for such study a stochastic approach would be more convenient. It is also true for works focusing on the management of retired batteries stocks \cite{xu2020, chai2021,cheng2022,chauhan2024energy, fallah2024,ma2024}.\\

Despite the fact that this threshold is by far the most common, some authors have tried to raise a debate about it. Martinez-Laserna et al. suggest that its validity may be debated, as batteries with different levels of degradation are likely to be available for potential second life use \cite{martinez2018technical}. Several simulation studies also showed that end-of-life could be reached at different health states depending on battery size, characteristics and uses\cite{saxen15,arrin21,casal19a,etxandi2024extending}. These initiatives have not yet convinced the research community to leave this theoretical threshold. This may be explained by the fact that these studies were based on observations and simulations work only. Supporting the statements with experimental results could give them greater leverage. The wealth of experimental studies on retired batteries conducted over the past decade is an interesting source  of information that could be used in this purpose. It may also enable to define a probability distribution function of the performance of retired batteries that researchers could  use for stochastic studies related to retired batteries \cite{piombo2025novel,zhang2025prediction,zhao2025lstm,hu2025mathematical,qaadan2025weibull}. \\

Experimental data from 104 articles presenting results of characterization test on more than 10,000 retired cells were collected. These number of articles and measurements help to give a first estimate of the performance of retired batteries and limit the influence of any article on the general trend. To the best of author knowledge, it is one the largest dataset collected for a review in the battery field. Producing these data through experiment would have been impossible for a single research group as years of continuous experiment would be necessary. Fig.~\ref{fig:sohq_review_carac} show the data gathered from literature.

\begin{table}[h!]
        \label{tab:parameters_gev}

\begin{tabular}{cc}

    \includegraphics[scale=0.25,valign=m]{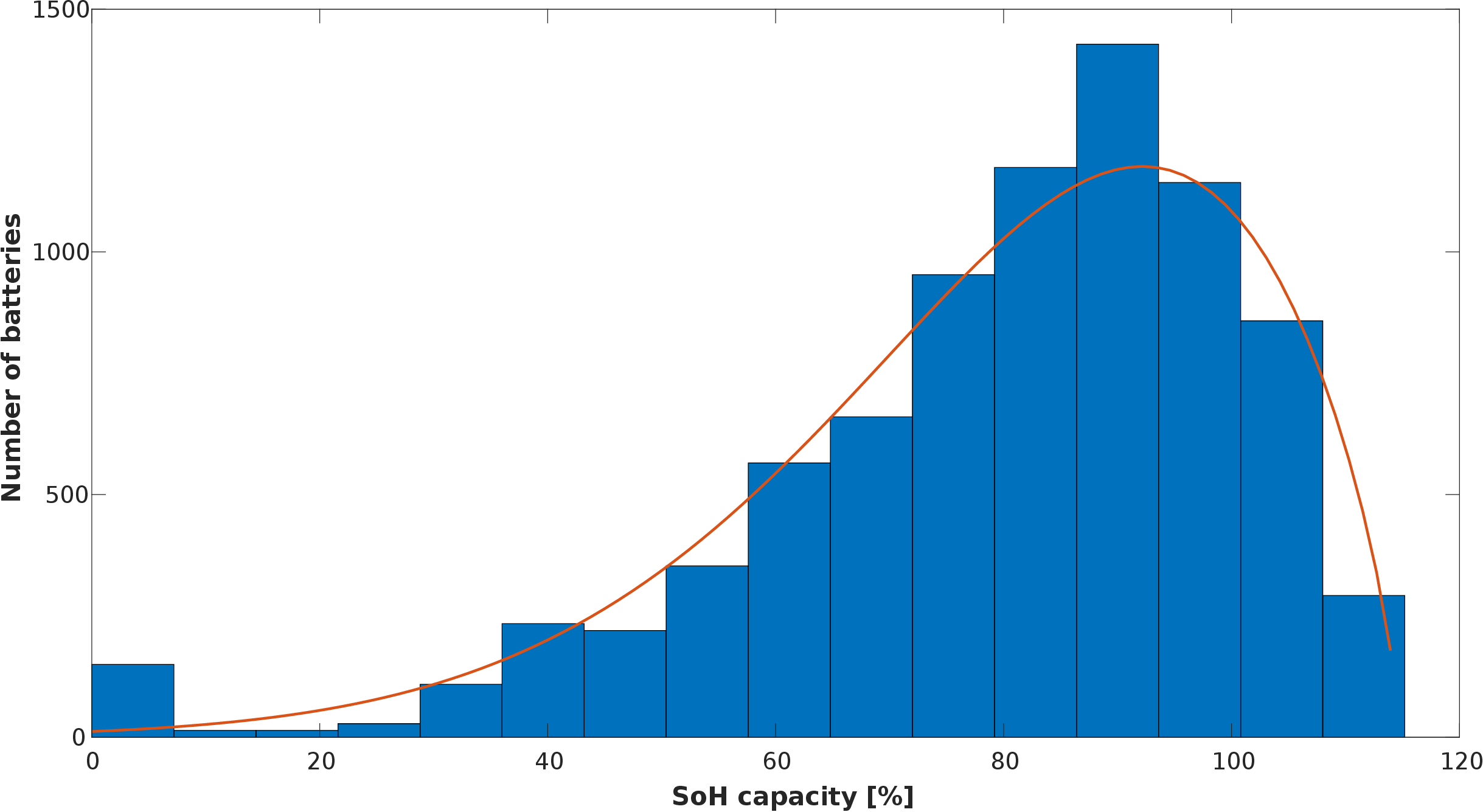}
    \label{fig:density_probability_review_carac}
&

\begin{tabular}{cc}
    \textbf{Parameter}& \textbf{Value}\\
\hline
    mean & 79.9\%\\  \hline
    median  & 84.5\% \\  \hline
    standard deviation  & 21.7 \\  \hline
        $k$ & -0.60  [-0.61, -0.59]\\
                \hline
    $\sigma$ &   23.3 [22.9, 23.7]\\
            \hline
       $\mu$&  75.8  [75.3, 76.3]\\
        \hline     

    \end{tabular}

\end{tabular}
    \caption{Distribution of the performance of the retired batteries. The Weibull distribution that fits the data is described in appendix~\ref{eq:weibull} with the parameters $k$, $\sigma$  and $\mu$ shown in the table.}
\label{fig:sohq_review_carac}
\end{table}

The figure shows that retired batteries exhibit a wide range of performance levels. The distribution ranges from 0\% to over 110\%. The distribution is negatively skewed, with a median of 84.5\% and a mean of 79.9\%. The distribution of the performance of the retired batteries can be described by a Weibull distribution as described in \ref{eq:weibull} with the parameters $k$, $\sigma$  and $\mu$ presented in the table. This result aligns well with previous studies on second life batteries \cite{schuster2015lithium,casals2017electric,bauma18,zhao2022research}. A significant amount of studies have tested retired batteries with a state of health above 100\% \cite{fan2022fast,hosen2026data,li2021,gao2024evaluation,hassini2023capacity,li2025state,schuster2016correlation,lacap2021development,kehl2021electrical,chen2022efficient,guenther2024second,ma2021study,lv2022investigation,de2025diagnostics}. This observation can be explained by the origin of the retired batteries. As figure~\ref{fig:definitions} shows, battery retirement is not always due to a performance that no longer meets users' and applications' needs. Battery and systems that still meet user requirements can retire because of the completion of a project or contract, the publication of restrictive legislation, or an accident that did not directly affect the battery. Retirement can also occur when battery performance no longer allows for the economic viability of the system. Battery degradation is just one of the possible reasons that can lead to a system retirement but it attracts the most research interest as such retirement raise questions related to safety concerns, or technical faults and failures.  These questions will be addressed in the rest of this article.

\subsection{Evaluation of performance dispersion}
This section presents the dispersion of performance inside retired batteries. Heterogeneity between elements can be explained by different factors. During the production of batteries, imperfections in the process generate heterogeneity between cells. In \cite{paul13}, 20,000 cylindrical LFP cells were characterized at the beginning of their lifecycle. Their performance followed a normal distribution, with standard deviations of 1.3\% for capacity and 5.8\% for resistance. In \cite{rumpf17}, Rumpf et al. tested 10,000 cylindrical LFP cells, the performance of which followed a normal distribution with standard deviations of 0.28\% and 0.72\%, respectively, for capacity and resistance. Heterogeneity between cells is accentuated by assembly and use in an electric vehicle. The position of the cells in the assembly may significantly influences how they age. In a battery pack, there can be significant temperature variations. Jiang et al. showed that battery module parameter dispersion increases with battery aging and that state of charge imbalance is the main factor contributing to capacity fading in the battery pack \cite{jiang2017recognition}. While this topic has gained limited interest, it may be a major concern. As the performance of an assembly is directly affected by cell-to-cell variations, solutions to mitigate their effects are needed \cite{hassini2023capacity,etxandi2024extending,di2023novel,jeanneret2025ageing}.

\begin{figure}[h!]
    \centering
    \includegraphics[width=\columnwidth]{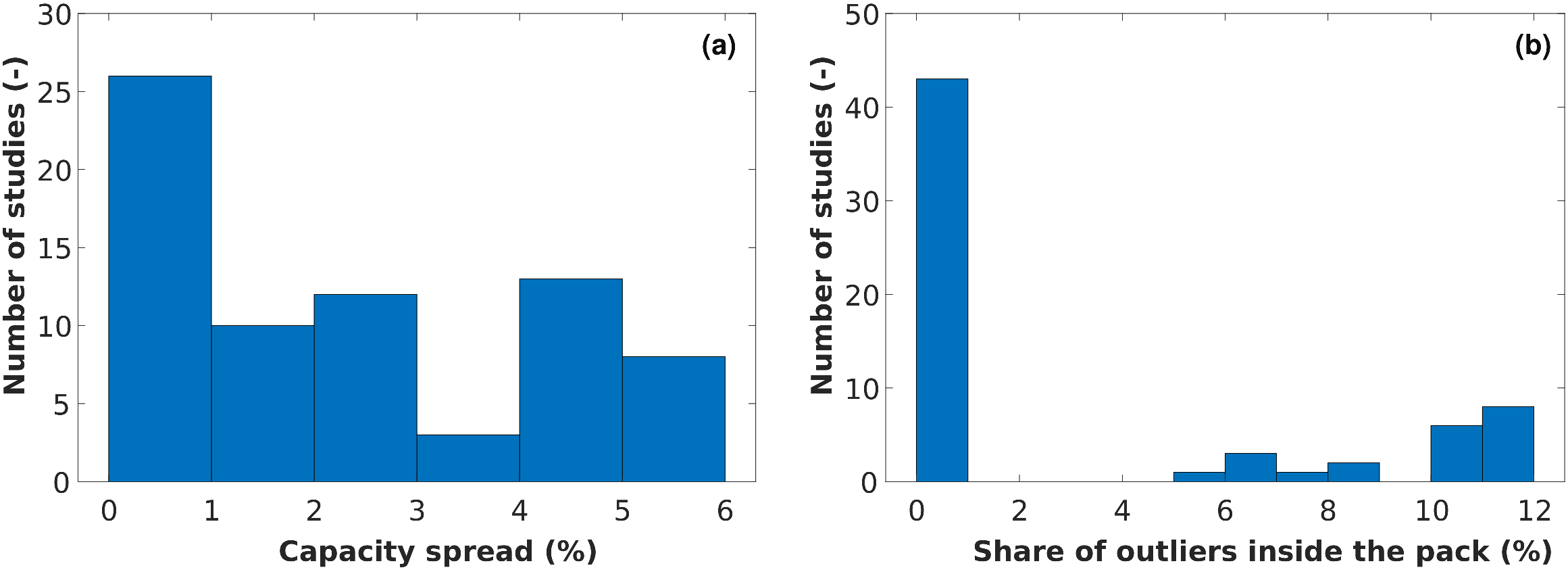}
    \caption{Capacity dispersion (a) and share of outliers (b) in the studies}
    \label{fig:dispersions}
\end{figure}

Performance dispersion was evaluated for each study that presented capacity measurements for more than fifteen cells. Dispersion was obtained by calculating the ratio of the standard deviation to the mean capacity measurement value. High dispersion can be caused by significant heterogeneity between cell capacities or by an outlier cell with capacities far from the mean value of the other cells \cite{schuster2015lithium,vitols2018cell}. In this study, an outlier cell is defined as a cell with a capacity that is more than 1.5 standard deviation away from the mean. Figure~\ref{fig:dispersions} shows the dispersion and the number of outlier cells in the studied articles.

\newpage

The figure shows that the majority of tested retired batteries have a capacity spread below 2\% without outlier cells. However, 20\% of studies have a spread above 4\% with up to 12\% of their cells with an outlier capacity. This dispersion is several orders of magnitude greater than that of fresh batteries and it show the need for specific management strategies.

\subsection{Fast performance assessment}

This section presents the procedure that could be used to assess rapidly the state of health of a retired battery. 

This ideal procedure is made of four steps : defect detection, degradation mechanisms evaluation, performance assessment and dispatch to recycling or a second life in an adapted system. Defect detection is a preliminary step that must be conducted to ensure that the battery is safe and functional. Short-circuit and mechanical damages are some of the defects that can occur to a battery and affect its safety and performance. Degradation mechanisms evaluation consist in assessing the ongoing aging mechanisms that have been triggered during batteries first life. There are a variety of mechanisms that can lead to different effect on the second life of batteries. These two first steps of the procedure have received limited attention in research as the existing experimental techniques are not adapted especially because these steps should be done at pack level in field conditions. Some research teams are trying to adapt existing non destructive techniques to lithium-ion batteries~\cite{kawaura2016operando,roscher2018method,montoya2024quantitative}. However they remain pioneer work that are far to be ready to test retired batteries of electric vehicles found in junkyards and that should be tested at ambient temperature. Most of research efforts focus on performance assessment and dispatch of batteries to recycling or reuse. This section will present the different approaches adopted and their accuracy.

\begin{figure}[h!]
    \centering
    \includegraphics[width=\columnwidth]{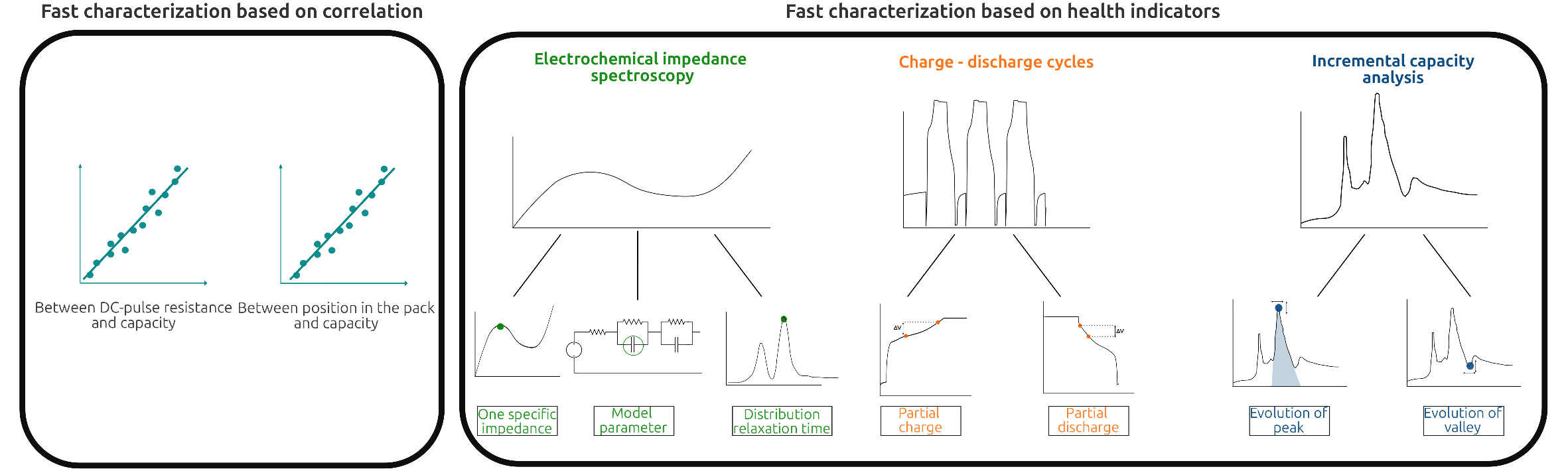}
    \caption{Most common health indicators investigated}
    \label{fig:health_indicators}
\end{figure}

In the context of retired batteries, the duration and cost of the characterization are crucial for ensuring the economic viability of the second life battery market. Contrarily to fresh batteries which performance is known to be maximal, retired batteries have unknown histories and performances. To limit the cost and duration of the performance evaluation, different approaches exist in the literature. Fig.~\ref{fig:health_indicators} presents the most common approaches.

A common approach consist in evaluating the correlation between the capacity and resistance measurement as resistance test could be less expensive and faster than capacity measurements. In the case of a strong correlation, one measure of capacity or resistance would be sufficient to evaluate both metrics. The table~\ref{table:correlation_q_r} shows the correlation found between capacity and resistance in different studies about retired batteries.

\begin{table}[h!]
\caption{Results on correlation of resistance and capacity of retired batteries}
\centering
\begin{tabular}{|c|c|c|}
\hline
   \textbf{No correlation} & \shortstack{\textbf{Low or }\\ \textbf{moderate correlation}}& \shortstack{\textbf{Strong}\\ \textbf{correlation}}   \\ \hline

\cite{bauma18, liao2017performance, tong2017demonstration,schuster2016correlation,wang2023state} & \cite{moreira2022fast,drallmeier2022challenges,tran2022used,popp2024fast}
   & \cite{bauma18,ramirez2023statistical, chang2022experimental,guenther2024second,braco2020experimental,lv2022investigation}  \\ \hline

\end{tabular}
\label{table:correlation_q_r}
\end{table}

The table shows that there is no consensus on this question. Different studies have offered disparate findings on the validity of this assumption The correlation seems to exist for only some of the tested batteries. In \cite{bauma18}, authors indicated that the correlation became more pronounced after aging. This controversy suggest that to date it is not possible to use correlation between capacity and resistance as a reliable tool. Further studies are necessary to understand the science behind this correlation. Another common approach consist in reducing the test time by investigating the correlation of cells performance with their positions in a retired pack. Understanding performance variations in a pack is important for identifying potential weak spots that could appear due to a poor thermal management. A cell at a weak spot would perform worse than its surroundings. As a first step, characterization of the retired module could be limited to assessing this cell, as it is well known that its performance will limit that of the entire module. In \cite{schuster2015lithium,hassini2025}, authors showed through experimental and statistical studies that such weak spot could not be found in the tested retired packs.

To complement the correlation-based approaches, a growing research topic consist in investigating the correlation of health indicators with the capacity. An health indicator is a metric that require limited measurement time and equipment and that is strongly correlated to capacity. In general, health indicator are obtained thanks to data-driven approaches using datasets produced through characterization experiments. Characterization experiments are commonly made of three type of measurements : impedance measurement, charge-discharge cycles and low-current test \cite{hassini2023lithium}.  The impedance based approach consists in measuring impedance in the frequency domain thanks to the electrochemical impedance spectrometry. This technique consists in applying a series of sinusoidal electrical constraints to the battery and measuring its response as a function of frequency. Several authors have highlighted the correlation between some impedance measurement and battery state \cite{gasper2022predicting}, other found that the correlation was more evident after processing impedance to fit a model or after determining the distribution relation time \cite{tian2020fractional,najjar2026fast}. A complete charge and discharge cycles approach usually requires two or three hours of test. Partial charge or discharge are then used to reduce the testing time and enable online performance assessment \cite{ali2019online,bavand2022online,agudelo2023battery}. These features are measured by charging or discharging the battery on a predefined and limited range of voltage. This range is selected as one which has a strong correlation to the full capacity. Reducing the voltage range leads to a reduction of testing time so also of the testing cost. This approach is suitable for industrial implementation as it requires limited experimental setup and may be used for online diagnostic across the whole battery life. Another approach consist in processing voltage and current curves to obtain new curves such as incremental capacity analysis (ICA), differential voltage analysis (DVA), probability density function (PDF) or  average Lorenz radius (ALR) \cite{riviere2015lifepo4,agudelo2021application,dubarry2022best}. Features of the curves are tracked and their evolution is analyzed to determined the degradations mechanisms occurring in the battery.

Table~\ref{tab:studies_fast_soh_eis} shows the studies that used electrochemical spectroscopy impedance measurements for fast assessment of the performance of retired batteries.
\newpage
 \begin{table}[h!]
    \centering
     \resizebox{\columnwidth}{!}{
\begin{tabular}{|c|cc|cc|cccc|}
\hline
\multicolumn{9}{|c|}{\textbf{Health indicators based on impedance}}   \\
\hline
\textbf{Ref.} & \multicolumn{2}{c|}{\textbf{Battery}} & \multirow{2}{*}{\textbf{Health indicator}} &  \multirow{2}{*}{\shortstack{ \textbf{Associated}\\ \textbf{data-driven technique} }}& \multicolumn{4}{c|}{\textbf{Accuracy}} \\
 
  \cline{2-3} \cline{6-9}  & \multicolumn{1}{c|}{\textbf{Number}}&   \textbf{Chemistry}  &  & &\multicolumn{1}{c|}{\textbf{MAE}} & \multicolumn{1}{c|}{\textbf{RMSE}} & \multicolumn{1}{c|}{$\mathbf{R^{2}}$} & \textbf{Time} \\ \hline

\hline 

\cite{amuta2025novel}& 9 cells &\shortstack{ LCO/C,\\ NMC/C} & \shortstack{EIS imaginary impedance\\ at 6.6 Hz and SoC 75-85\%}& No & \shortstack{ not\\ evaluated}& \shortstack{ not\\ evaluated} &$>0.98$ & \shortstack{ not\\ evaluated}\\

           \hline
\cite{camboim2024state}& 10 cells &LFP/C  & \shortstack{Set of 7 EIS impedance\\at 7 frequencies and SoC 100\%}& \shortstack{Yes, Multilayer Perceptron\\ neural network} &$<1.9\%$& $<1\%$&$>0.99$&3 min 30s\\
        \hline
\cite{ezpeleta2025rapid} &  2 cells   & NMC/C  & \shortstack{Parameter Rint of \\an electric circuit model}      &  \shortstack{Yes, Model Fitting \\and Random Forest Model } & \shortstack{ not\\ evaluated} & $<5\%$ & \shortstack{ not\\ evaluated} & \shortstack{ not\\ evaluated}\\
 \hline

\cite{fan2022fast} &  4 modules   & LFP/C  & \shortstack{Parameter Q2 of \\an electric circuit model}&  \shortstack{Yes, Model Fitting} & \shortstack{ not\\ evaluated} &  \shortstack{ not\\ evaluated} & \shortstack{ not\\ evaluated} & 20 min\\
 \hline

\cite{fan2025generalized} & cells & NMC/C, LFP/C & Set of 6 EIS indicators & Yes, CatBoost & $<4.7\%$ &  \shortstack{ not\\ evaluated} &  \shortstack{ not\\ evaluated} &  \shortstack{ not\\ evaluated}\\

\hline  
\cite{kehl2021electrical} & 5 modules & NMC/C & Set of 3 EIS indicators & No &  \shortstack{ not\\ evaluated} &  \shortstack{ not\\ evaluated} &  \shortstack{ not\\ evaluated} &  \shortstack{ not\\ evaluated} \\

\hline

\cite{kemeny2026impact} & 1 module & LMO-LNO/C & \shortstack{Parameter Rohm of \\an electric circuit model} & Yes, Model Fitting &  \shortstack{ not\\ evaluated} & $<0.8\%$ &  \shortstack{ not\\ evaluated} &  \shortstack{ not\\ evaluated} \\

\hline

\cite{luo2021rapid} & 9 cells & LFP/C &\shortstack{Parameter Q2 of \\an electric circuit model}  & Yes, Model Fitting & $<1.3\%$ & \shortstack{ not\\ evaluated} &$<0.98$&  5 min\\

\hline

\cite{li2025state} & 14 cells & LCO/C & \shortstack{Set of 3 high frequency real\\ component indicators from EIS } & Yes, DBO-optimized XGBoost & $ <2.7\%$ & $<3.9\%$ &  \shortstack{ not\\ evaluated} & 18 s\\

\hline

\cite{montes2025multi} & 1 module & NMC/C & Set of resistances from current pulses & 
Yes, Feedforward Neural Network  & $<2.2\%$ &  \shortstack{ not\\ evaluated}  &  \shortstack{ not\\ evaluated}  & 10 s\\
\hline

\cite{peng2026fast} &331 cells & Unknown & Set of 8 EIS indicators & \shortstack{Yes, Constrained Variational Autoencoder\\ - First Integer Neighbor Clustering Hierarchy} &  \shortstack{ not\\ evaluated}  &  \shortstack{ not\\ evaluated}  &  \shortstack{ not\\ evaluated}  &  \shortstack{ not\\ evaluated} \\

 \hline  
 \cite{ramirez2023statistical} & 192 cells &LFP/C  & EIS real impedance at 89 Hz & No & \shortstack{ not\\ evaluated}  & \shortstack{ not\\ evaluated}  &$<0.91	$ & \shortstack{ not\\ evaluated} \\
 \hline  

 \cite{schuster2016correlation} & 1438 cells& NMC/C   & EIS real impedance at SoC 50\% & No & \shortstack{ not\\ evaluated} & \shortstack{ not\\ evaluated}&$<0.53$&\shortstack{ not\\ evaluated}\\
\hline  

\cite{tran2022used} & 12 cells & LCO/C  & \shortstack{Parameters R0 and C1 of \\an electric circuit model} & Yes, Model Fitting &\shortstack{ not\\ evaluated}&\shortstack{ not\\ evaluated}& $<0.9$&\shortstack{ not\\ evaluated}\\

\hline

\cite{vasquez2025} & 1 module & NMC/C  & \shortstack{Parameter R1 of \\an electric circuit model} & Yes, Model Fitting &\shortstack{ not\\ evaluated}&\shortstack{ not\\ evaluated}& $<0.8$& 10 min\\
\hline

\cite{wang2023rapid} & 32 cells & LFP/C & Set of 5 indicators from EIS & Yes, Fuzzy Clustering  &\shortstack{ not\\ evaluated} &\shortstack{ not\\ evaluated} &\shortstack{ not\\ evaluated} & 6 min\\

\hline

\cite{zhang2014novel} & 94 cells & Unknown & Set of resistance from current pulses & \shortstack{Yes, Back Propagation\\ Neural Network} &\shortstack{ not\\ evaluated} & $<4.3\%$  &\shortstack{ not\\ evaluated}&\shortstack{ not\\ evaluated}\\

\hline

\cite{zhu2025rapid}& 18 cells & LFP/C & Set of indicators from EIS  & \shortstack{Yes, Back Propagation\\ Neural Network} & $ <1.1\%$ & $<1.3\%$ & \shortstack{ not\\ evaluated} & 8 min\\
\hline







  \hline

    \end{tabular}
}

\caption{Health indicators based on impedance tested on retired batteries.}
    \label{tab:studies_fast_soh_eis}
\end{table}
     
These studies showed that electrochemical impedance spectroscopy is an interesting technique for quickly assessing the performance of retired batteries. It is used in a variety of manner by evaluating a specific impedance, fitting a model or training a data-driven model. Among the 18 studies described, 14 have used data-driven techniques. Results are promising as the techniques have been applied to cells and modules. \cite{camboim2024state} is the benchmark with an error estimation below 1.9\% in less than 3 minutes and 30 seconds. The main limitation of its approach is the need to set the state of charge to 100\%. \cite{li2025state} and \cite{montes2025multi} have also proposed promising approaches combining electrochemical impedance spectroscopy and data-driven techniques. \textit{Future works should extend the evaluation work to pack level considering different temperatures and initial state of charge. Accuracy, duration and cost of the technique should be thoroughly evaluated. The opportunity to deploy the approach in a system should be also studied \cite{bechara2025embedded}.}\\

Table~\ref{tab:studies_fast_soh_partial_charge} show the studies that used indicators from charge-discharge cycles.
\newpage

 \begin{table}[h!]
    \centering
      \resizebox{\columnwidth}{!}{
  \begin{tabular}{|c|cc|cc|cccc|}
\hline
\multicolumn{9}{|c|}{\textbf{Health indicators based on  charge-discharge cycles}}   \\
\hline
\textbf{Ref.} & \multicolumn{2}{c|}{\textbf{Battery}} & \multirow{2}{*}{\textbf{Health indicator}} &  \multirow{2}{*}{\shortstack{ \textbf{Associated}\\ \textbf{data-driven technique} }}& \multicolumn{4}{c|}{\textbf{Accuracy}} \\
 
  \cline{2-3} \cline{6-9}  & \multicolumn{1}{c|}{\textbf{Number}}&   \textbf{Chemistry}  &  & &\multicolumn{1}{c|}{\textbf{MAE}} & \multicolumn{1}{c|}{\textbf{RMSE}} & \multicolumn{1}{c|}{$\mathbf{R^{2}}$} & \textbf{Time} \\ \hline

\cite{ahmei22} & 1 pack & LMO-LNO/C & \shortstack{Discharged capacity\\ between 3.95V and 3.55 V}&No& $<5\%$ & \shortstack{ not\\ evaluated} &\shortstack{ not\\ evaluated} & 10 min\\

\hline
\cite{amuta2023state} & 4 cells &unknown &  \shortstack{ Voltage integral during \\partial charge between 3.85V and 4.2V} &No &  $2.9\%$& $3.4\%$ & $0.97$ &  \shortstack{ not\\ evaluated}\\
\hline

\cite{braco22}& 6 cells & LMO-LNO/C  & \shortstack{Charged capacity\\ between 3.85V and 4.05V }& No  & $<4.3\%$ & $<2.1\%$&  \shortstack{ not\\ evaluated}& \shortstack{ not\\ evaluated} \\
\hline
\cite{cui2024taking} & 1 cell & LMO/C & Set of 6 indicators from  charge-discharge cycles and current pulses & Yes, ElasticNet Regression & $<1.4\%$ & $<1.8\%$ &\shortstack{ not\\ evaluated} & \shortstack{ not\\ evaluated}\\
\hline

\cite{he2025state}& 9 cells & Unknown & \shortstack{Set of 3 indicators\\ from charge-discharge cycles} & \shortstack{Yes,  Multi-Input Metabolic \\Gated Recurrent Unit}& $ <1\%$ & $<1.2\%$& \shortstack{ not\\ evaluated} & \shortstack{ not\\ evaluated}\\

\hline

\cite{huang2025rapid}& 800 cells & LFP/C & \shortstack{Rest dynamic after \\charging-discharging cycles}& Yes, Support Vector Regression & \shortstack{ not\\ evaluated} & $<2.9\%$&\shortstack{ not\\ evaluated}& \shortstack{ not\\ evaluated}\\

\hline

\cite{jiang2024state} & 36 cells & NMC/C & \shortstack{Set of 13 indicators \\from charging-discharging cycles} & Yes, Bayesian Neural Network& $<2.2\%$& $<1.9\%$& \shortstack{ not\\ evaluated}& \shortstack{ not\\ evaluated}\\

\hline

\cite{li2025lifetime} & 84 cells & Unknown & \shortstack{Set of 10 indicators \\from charging-discharging cycles} & Yes, Hierarchical Bayesian Model & $<6.1\%$ & $<9\%$ & \shortstack{ not\\ evaluated}& \shortstack{ not\\ evaluated}\\

\hline

\cite{li2026rapid} & 1648 cells& LFP/C & \shortstack{Voltage and capacity during \\ partial charge between 3.1V and 3.4V} & \shortstack{Yes, hybrid Convolutional Neural Network (CNN),\\ Conv Block Attention Module (CBAM), and \\ Long Short-Term Memory (LSTM) architecture} & \shortstack{ not\\ evaluated} & \shortstack{ not\\ evaluated} & $<0.99$ & \shortstack{ not\\ evaluated}\\

\hline

\cite{michelini2024impact} & 4 cells & LMO-LNO/C & \shortstack{Set of 11 indicators\\ from charging-discharging cycles} & No & \shortstack{ not\\ evaluated} & \shortstack{ not\\ evaluated} & \shortstack{ not\\ evaluated} & \shortstack{ not\\ evaluated}\\

\hline

\cite{najjar2026fast} & 46 cells & LMO-LNO/C & \shortstack{Charged capacity\\ between 3.4V and 3.6 V}& No & \shortstack{ not\\ evaluated} & \shortstack{ not\\ evaluated}& $<0.99$& \shortstack{ not\\ evaluated}\\
\hline

\cite{ni2023rapid} & cells & LFP/C & \shortstack{Charged capacity \\between SoC 0\% and SoC 20\%} & \shortstack{Yes, improved moth–flame optimisation\\ and support vector regression } & \shortstack{ not\\ evaluated} & $<2.7\%$ & \shortstack{ not\\ evaluated}& 6 min\\
\hline
\cite{quinard2019fast} & 24 cells & LMO-LNO/C & \shortstack{Charged capacity\\ between 3.4V and 3.8 V}& No & $<5.1\%$ &\shortstack{ not\\ evaluated} & $<0.7$& 5 min\\

\hline
 \cite{raza2026ev} & 4 cells & NMC/C & \shortstack{Set of 13 indicators for NMC 65Ah\\ cells and 24 indicators 141Ah from charge-discharge cycles, EIS and ICA} & Yes, ElasticNet &  \shortstack{ not\\ evaluated}& $<11\%$ & \shortstack{ not\\ evaluated} & \shortstack{ not\\ evaluated}\\
 \hline

\cite{sun2022quick} & 1 module & LFP/C & Mean Fréchet distance of charge and discharge curve & No &\shortstack{ not\\ evaluated} & $<2.5\%$ & $>0.99$ & 10 min\\

\hline

\cite{xiong2020online} & 4 cells & NMC/C, LFP/C & Charged capacity between 3.95V and 4 V & No & $<1.2\%$ & $<1.7\%$ & $<0.99$ & \shortstack{ not\\ evaluated}\\
\hline

\cite{zhang2025two} & 121 cells & unknown & \shortstack{Evolution of temperature \\ during charge-discharge cycles} & Yes,  Long Short-Term Memory & $<9.5\%$ & \shortstack{ not\\ evaluated}& \shortstack{ not\\ evaluated}& \shortstack{ not\\ evaluated}\\

\hline
\cite{zhang2026developing} & 360 cells & LFPC/C &  \shortstack{Set of 3 indicators from charge-discharge \\cycles and current pulses} & Yes, 2D Convolutional Network & $<3\%$ & \shortstack{ not\\ evaluated}  & \shortstack{ not\\ evaluated} & 18 min\\

\hline

\cite{zhu2025bayesian} & 84 cells & NMC/C & \shortstack{Set of 12 indicators from \\ charging-discharging cycles} & \shortstack{Yes, Bayesian Calibrated \\Physics-Informed Neural Network} & $1\%$ & $<1.5\%$ & $<0.99$ & \shortstack{ not\\ evaluated}\\
\hline

    \end{tabular}
 }

\caption{Health indicators based on  charge-discharge cycles tested on retired batteries.}
    \label{tab:studies_fast_soh_partial_charge}
\end{table}
Charge and discharge cycles are interesting as they can be measured online on a system. They offer an interesting option for quickly evaluating battery state because they can be measured at the cell or assembly level with limited equipment. Their measurement time is also an advantage, as they can be measured in a few minutes. However, test duration are difficult to compare between studies because the initial state of charge and current may differ. The table show that this approach have been used at cell, module and pack level on different battery technologies. A great variety of indicators have been used and 11 of the 19 studies have used data-driven techniques. \cite{sun2022quick} proposed the benchmark method for this approach with a technique that capture in less than 10 minutes the battery state with an error below 2.5\%. In \cite{amuta2023state}, the authors used different equations to evaluate testing time. This is an interesting study as defining a reference time to compare different studies is still necessary. In \cite{liu2023fast}, authors also estimated the energy consumption of the technique. This is an important point to consider as characterization cost should be minimized. \textit{Future works should include an evaluation of the test duration and cost with a clear methodology.}

Table~\ref{tab:studies_fast_soh_ica} shows the studies that used processed current and voltage curves for fast assessment of the performance of retired batteries.
 \newpage

\begin{table}[h!]
    \centering
     \resizebox{\columnwidth}{!}{

\begin{tabular}{|c|cc|cc|cccc|}
\hline
\multicolumn{9}{|c|}{\textbf{Health indicators based on current and voltage curves}}   \\
\hline
\textbf{Ref.} & \multicolumn{2}{c|}{\textbf{Battery}} & \multirow{2}{*}{\textbf{Health indicator}} &  \multirow{2}{*}{\shortstack{ \textbf{Associated}\\ \textbf{data-driven technique} }}& \multicolumn{4}{c|}{\textbf{Accuracy}} \\
 
  \cline{2-3} \cline{6-9}  & \multicolumn{1}{c|}{\textbf{Number}}&   \textbf{Chemistry}  &  & &\multicolumn{1}{c|}{\textbf{MAE}} & \multicolumn{1}{c|}{\textbf{RMSE}} & \multicolumn{1}{c|}{$\mathbf{R^{2}}$} & \textbf{Time} \\ \hline

\cite{chen2022efficient} & 2 cells & LFP/C &  \shortstack{IC (y-axis) peaks  changes in ICA at 3.42V\\ and 3.47V in charge at C/5} &  \shortstack{Yes, K-means++ \\algorithm } &  \shortstack{ not\\ evaluated} &  \shortstack{ not\\ evaluated} &  \shortstack{ not\\ evaluated} &  \shortstack{ not\\ evaluated}\\
     \hline
\cite{de2025diagnostics} & 4 cells & NMC/C & \shortstack{ICA peaks changes (3.5V, 3.65V, 3.85V in charge,\\ 3.4V, 3.55V, 3.9V in discharge) at C/2} & No & $<1.9\%$ & \shortstack{ not\\ evaluated} & \shortstack{ not\\ evaluated} &\shortstack{ not\\ evaluated}\\

\hline

\cite{du2026efficient}&  320 cells & LFP/C &  \shortstack{Set of 3 V(x-axis) peaks changes\\ in ICA at 3.28V, 3.37V and 3.43V} &  \shortstack{Yes, K-means++ algorithm\\ and Aggregation hierarchical clustering algorithm} &  \shortstack{ not\\ evaluated}&  \shortstack{ not\\ evaluated}&  \shortstack{ not\\ evaluated}&  \shortstack{ not\\ evaluated}\\
     
\hline
\cite{huang2022state}& 1 cell & NCA/C &\shortstack{Area below ICA (incremental capacity analysis)\\ peak at 3.4V} & No & \shortstack{ not\\ evaluated} & \shortstack{ not\\ evaluated} & $>0.98$ & \shortstack{ not\\ evaluated} \\
\hline

\cite{huang2022state} & 1 module & LFP/C &\shortstack{Area below ICA (incremental capacity analysis)\\ peak at 3.4V} & No & \shortstack{ not\\ evaluated} & \shortstack{ not\\ evaluated} & $>0.95$ & \shortstack{ not\\ evaluated} \\

\hline

\cite{huang2024flexible}&  108 cells & NMC/C, LFP/C & \shortstack{V (x-axis) peaks changes\\ in ICA at 3.8V} & Yes, random forest& \shortstack{ not\\ evaluated}& \shortstack{ not\\ evaluated}& \shortstack{ not\\ evaluated}& \shortstack{ not\\ evaluated}\\
\hline

\cite{jiang2017recognition} & 4 cells & LFP/C & Area below ICA (incremental capacity analysis) peak at 3.4V & No & $<2.5\%$ & $<3\%$& $<0.91$ &  \shortstack{ not\\ evaluated}\\

\hline

\cite{li2022screening} & 39 cells & LFP/C & Set of 9 indicators from ICA  & \shortstack{Yes, adaptive genetic algorithm-back\\ propagation neural network} & $<3\%$ &\shortstack{ not\\ evaluated} &$<0.93$ & \shortstack{ not\\ evaluated}\\
\hline

\cite{li2024electrochemical} & 2 cells & LFP/C & Changes in PDF (probability density function) peak at 3.4V & No & \shortstack{ not\\ evaluated} & \shortstack{ not\\ evaluated}&$<0.98$& \shortstack{ not\\ evaluated}\\

\hline

\cite{lin2023state} & 1 module & LFP/C& Area below PDF (probability density function) peak at 3.4V & No & \shortstack{ not\\ evaluated} & $<0.96\%$ & \shortstack{ not\\ evaluated} &\shortstack{ not\\ evaluated} \\
\hline
\cite{lin2023state} & 1 cell & NCA/C& Area below PDF (probability density function) peak at 3.4V & No & \shortstack{ not\\ evaluated}  & \shortstack{ not\\ evaluated} & $<0.9999$ &\shortstack{ not\\ evaluated} \\
\hline
\cite{liu2022characterization} & 1 cell & NCA/C & Set of 15 indicators from ICA, DVA and PDF curves & Yes, Principal Component Regression & $<1.2\%$ & $<1.4\%$ &\shortstack{ not\\ evaluated}  & \shortstack{ not\\ evaluated} \\
     
\hline

\cite{schaltz2021incremental} & 6 cells & NMC/C, LMO/C & Set of 8 indicators from ICA  & No &$<8.6\%$&$<1.4\%$& \shortstack{ not\\ evaluated}& \shortstack{ not\\ evaluated}\\

\hline

\cite{she2021battery} & 2 cells & LFP/C & \shortstack{IC (y-axis) peak  changes \\in ICA at unknown voltage}& \shortstack{Yes, Density-Based Spatial\\ Clustering of Applications with Noise}& $<3\%$ & $<2.5\%$ &  \shortstack{ not\\ evaluated} &  \shortstack{ not\\ evaluated}\\

\hline
\cite{vasquez2025} & 1 module& LFP/C& Area below ICA (incremental capacity analysis) peak at 3.4V & No &\shortstack{ not\\ evaluated}  & \shortstack{ not\\ evaluated} & $<0.99$ & 48 min\\
\hline
\cite{wan2026enhanced} & 200 cells & NCA/C & IC (y-axis) peak  changes in ICA at 4V & 
Yes, OPTICS-DBSCAN algorithms & \shortstack{ not\\ evaluated} & \shortstack{ not\\ evaluated} & $<0.98$ & \shortstack{ not\\ evaluated} \\

\hline

\cite{wen2023fast} & 1 module & LFP/C & Variation of ALR (average Lorenz radius) peak at 3.25V–3.35V & No & \shortstack{ not\\ evaluated}  & $0.5\%$& $>0.99$ &\shortstack{ not\\ evaluated} \\
\hline

\cite{xu2026regenerate} & 86 cells & NMC/C & IC (y-axis) peak  changes in ICA at 3.25V & yes, Enhancement convolutional neural network & $ <0.5\%$ & $<1\%$ & $<0.99$ & 27 min\\

\hline
\cite{zhang2021state} & 216 cells & LFP/C & IC (y-axis) peak  changes in ICA at 3.35V & No &\shortstack{ not\\ evaluated}  &\shortstack{ not\\ evaluated} & $<0.96$&\shortstack{ not\\ evaluated}\\

\hline
\cite{zhang2022reliable} & 3 cells & LFP/C & \shortstack{Set of 8 indicators related to peaks\\ changes in ICA at 3.5V, 3.6V, 3.9V and 4.1V }& \shortstack{Yes, particle swarm optimization\\ - broad learning system network} & \shortstack{ not\\ evaluated} & $<0.6\%$ & $<0.99$ & 2 s\\

\hline

\cite{zhang2023state} & 2 cells and 2 modules & LFP/C & Area below ICA (incremental capacity analysis) peak at 3.4V & No & \shortstack{ not\\ evaluated}  & \shortstack{ not\\ evaluated}  & $0.96$ & \shortstack{ not\\ evaluated} \\
\hline
\cite{zhang2023state} & 1 module & NCA/C & Height of ICA (incremental capacity analysis) peak at 3.4V & No & $<1.6\%$ &  \shortstack{ not\\ evaluated}& $<0.98$ &  \shortstack{ not\\ evaluated}\\

\hline

\cite{zhou2023health} & 110 cells & LFP/C & \shortstack{IC (y-axis) peak  changes\\ in ICA at 3.35V} & \shortstack{Yes, ordering points\\ to identify the clustering structure} & \shortstack{ not\\ evaluated} &\shortstack{ not\\ evaluated}& $<0.95$ & \shortstack{ not\\ evaluated}\\

\hline

\cite{zou2023fast} & 17 cells & NMC/C & \shortstack{Set of 4 indicators related \\to peaks changes in ICA at 3.9V} & Yes, Linear Regression & $<1.2\%$ & $1.5\%$ & \shortstack{ not\\ evaluated}& \shortstack{ not\\ evaluated}\\

           \hline
\cite{ma2021study} & 6 modules & LFP/C & Area below ICA (incremental capacity analysis) peak at 3.33V & No &$<6.7\%$ & $<3.1\%$ & \shortstack{ not\\ evaluated} & \shortstack{ not\\ evaluated}  \\

\hline

    \end{tabular}
 }

\caption{Health indicators based on processed current and voltage curves tested on retired batteries.}
    \label{tab:studies_fast_soh_ica}
\end{table}

Processing current and voltage curves is also a popular technique for assessing the performance of retired batteries. This can be explained by the possibility to bridge the material science with electrical characterization technique by evaluating the degradation modes occurring inside the battery. Several authors concluded that processed current and voltage cureves evolutions are in line with post-mortem analysis \cite{cao2024second,wang2023state}. The table ~\ref{tab:studies_fast_soh_ica} presents 22 studies that used this technique.

A diversity of health indicators have been extracted from current and voltages curves. Incremental capacity analysis (ICA) have been the most popular technique, it was applied on different chemistry and at cell and module level. Schaltz et al. showed that ICA features are not the same at cell and pack level \cite{schaltz2021incremental}. Multiple studies have identified the same indicators as relevant. For example, LFP batteries were mostly studied between 3.3V and 3.5V. \cite{zhang2022reliable}
and \cite{xu2026regenerate} have obtained interesting results with error below 1\%. They both combined ICA analysis with data-driven techniques. One important limit of the reviewed studies is the lack of consistency in health indicator definition and evaluation metrics. Each study named and defined the evaluated features with its own wording and methodology. This limit the comparability between studies. This is also true for accuracy assessment, as different indicators were used making difficult any accuracy comparison. In this review, the features were renamed by defining peaks and valleys according to the voltage they appear at.

Table~\ref{tab:studies_fast_soh_other} presents a pioneer  study that used a non destructive and non electrical technique to evaluate the state of retired batteries.

\begin{table}[h!]
    \centering
     \resizebox{\columnwidth}{!}{
\begin{tabular}{|c|cc|cc|cccc|}

\hline
\multicolumn{9}{|c|}{\textbf{Other techniques}}   \\
\hline
\textbf{Ref.} & \multicolumn{2}{c|}{\textbf{Battery}} & \multirow{2}{*}{\textbf{Health indicator}} &  \multirow{2}{*}{\shortstack{ \textbf{Associated}\\ \textbf{data-driven technique} }}& \multicolumn{4}{c|}{\textbf{Accuracy}} \\
 
  \cline{2-3} \cline{6-9}  & \multicolumn{1}{c|}{\textbf{Number}}&   \textbf{Chemistry}  &  & &\multicolumn{1}{c|}{\textbf{MAE}} & \multicolumn{1}{c|}{\textbf{RMSE}} & \multicolumn{1}{c|}{$\mathbf{R^{2}}$} & \textbf{Time} \\ \hline

\hline 

    \cite{montoya2024quantitative} & 13 cells& NMC/C& \shortstack{Mid-band fit from \\ ultrasonic measurement}& No& \shortstack{ not\\ evaluated} & \shortstack{ not\\ evaluated}&$<0.83$&\shortstack{ not\\ evaluated}\\
    \hline

    \end{tabular}
 }

\caption{Health indicators based on other techniques tested on retired batteries.}
    \label{tab:studies_fast_soh_other}
\end{table}

Table~\ref{tab:studies_fast_soh_other} shows that ultrasonic measurement can be a novel technique for evaluating battery state. However, the time and energy consumption of this techniques have not been thoroughly investigated, making it difficult to evaluate their suitability in an industrial context. Further works will be needed to evaluate this technique and to develop new non destructive techniques.

This review have highlighted several findings regarding rapid characterization techniques for retired batteries. Most research efforts focus on performance and degradation mechanisms evaluations with electrical measurements. A diversity of techniques and indicators have been evaluated and identified as promising.  Incremental capacity analysis and electrochemical spectroscopy techniques have been widely studied as they can capture the degradation mechanisms occurring in the battery \cite{wang2023state,she2021battery,cao2024second}. Several gaps remain in the research efforts. First, defect detection in battery packs and cells have received limited interest. Short-circuit, wiring issues or unsafe cells are still difficult to detect using the existing electrical techniques. New experimental techniques will need to be developed. Most research efforts have also focused on laboratory measurements to evaluate the performance based on a single indicator most commonly the capacity. To dispatch the retired batteries towards reuse or recycling,the decision should be based on a battery passport \cite{hassini2024battery}. The set of performance indicators available should be compared to the performance needs of a set of possible reuse applications. These usage needs would be a sort of second life visa.
The review also showed that the diversity of batteries tested led to a diversity of relevant indicators. It seems that not a single technique or approach can evaluate all battery models. There is a need for a database that would associate each battery model with its relevant health indicators and their validity domain (temperature, state of charge...). 

From a methodological perspective, the time and equipment necessary to conduct these techniques remain major concerns with regard to industrial development. The duration and cost of fast characterization techniques have received little attention despite their significant impact on the profitability of the second-life battery market. More efforts should be done to define common metrics and methodologies to evaluate the different health indicators.
In general, a more cumulative approach for research would be valuable. Most studies present techniques without considering previous results from published articles. This leads to great variability in feature names, experimental conditions, and datasets used for accuracy evaluation. In this article, health indicators and their performance are compared using consistent metrics that enable comparison and help set a benchmark for future studies. The authors hope that this structured methodology will be reused in future studies to encourage more cumulative research. In the future, it would be also interesting to evaluate all health indicators and characterization techniques based on a set of common public and open dataset. This good practice would enable comparison of different results based on the same input \cite{hassini2023lithium} .

\begin{figure}[h!]
    \centering
    \includegraphics[width=\linewidth]{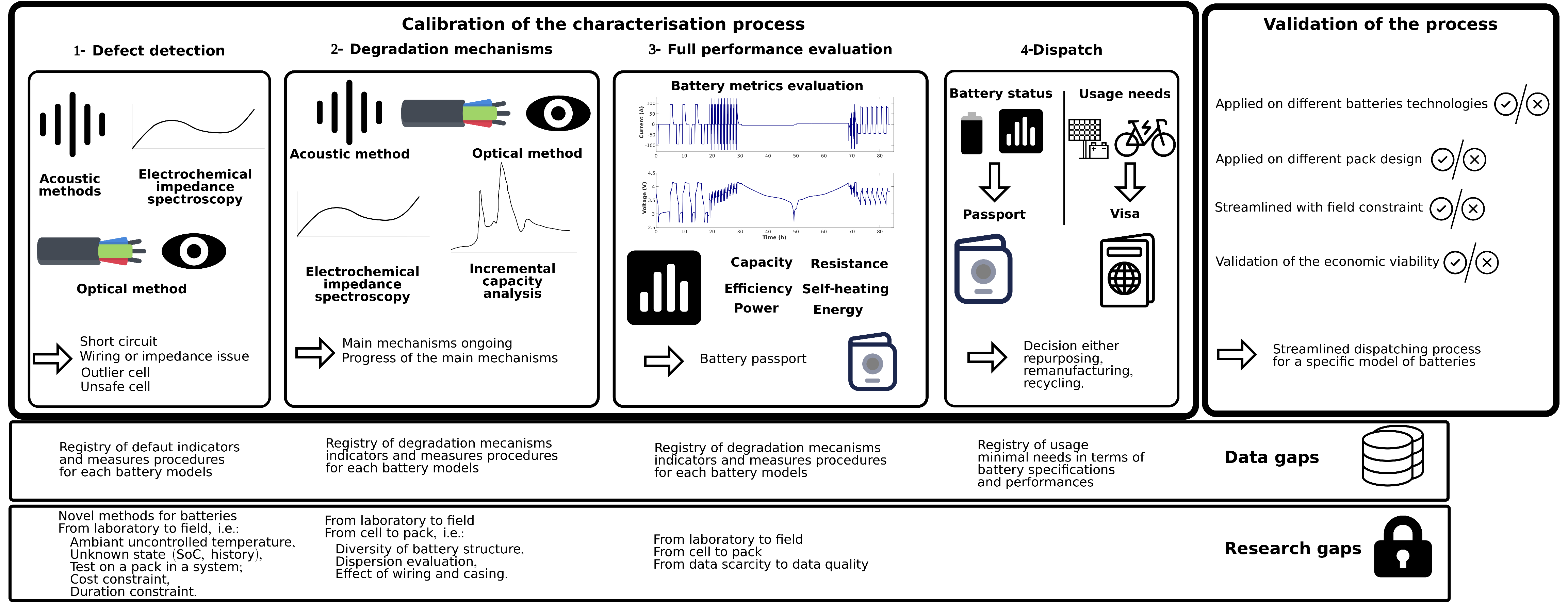}
    \caption{Ideal procedure for the evaluation of the suitability of a retired battery to reuse.}
    \label{fig:fast_carac}
\end{figure}

Figure~\ref{fig:fast_carac} provide a synthesis of an ideal procedure for an evaluation of the suitability of a retired battery to reuse and the existing gaps in current research.

\clearpage

\section{Aging of reused batteries} \label{sec:reusing_batteries}

Previous sections have described the performance of retired batteries. This section explores another major research question : the degradation of a reused battery.

\subsection{Open research questions}

The question of how reused battery degradation evolves is an important research question that affect the economic viability and environmental impact. In 2018, Martinez-Laserna et al. highlighted a debate that existed on this subject in the scientific community \cite{martinez2018}. At that time, several authors were considering that reused batteries would age with a common pattern. They agreed to consider that all reused batteries would degrade in same manner but these authors disagree on this manner. Some authors such as Ahmadi et al. or Wieland et al. consider that aging would slows down during the second life \cite{ahmadi2014,wieland2017}. In these works based on simulations, the authors make the assumption that the reuse of batteries in a second life application will generally be reused in a low stress use, which should result in a slowing down of their degradation rate. Other authors, such as Sathre et al., consider that a battery used in a second-life application will experience accelerated degradation \cite{sathre2015}. Debnath et al. even quantify that the acceleration will be of the order of 3.3\%/year compared with the first life \cite{debnath2014}. In this work, which is also based on simulation, the authors hypothesize that the change in use of batteries will lead to the appearance of new aging mechanisms or the appearance of interactions between these mechanisms. In their view, these new phenomena will accelerate the aging process.

A great deal of research has also focused on the sudden acceleration of battery degradation. In the literature on battery aging, this phenomenon is known as the ‘aging knee’ and is the subject of considerable research. A 2020 standard defines the knee as a rapid decrease in a battery's capacity at the end of its life \cite{ieee20}. In a review of the literature on the subject, Attia et al. show that this phenomenon has been observed in several test campaigns. The authors explain that this phenomenon is the result of a complex and poorly understood interaction between the degradation mechanisms affecting batteries \cite{attia22}. They show that certain battery technologies and certain experimental conditions tend to favor the appearance of such a knee. The article also points out that the modeling and prediction of this phenomenon remain uncertain and essentially highlights the fact that further research is needed to better understand this phenomenon. An important research gap to fill is to determine what conditions tend to favor accelerated aging or knee apparition. 

Another topic of attention consists in identifying the best usage and management strategies to ensure a long and safe lifespan for reused batteries. Reused batteries are sourced from a wide range of origins, they experienced different first life, with diverse charging habits and usage conditions. These differences result in differences of degradation mechanisms triggered and evolution in the safety behavior.  An important research gap to fill is to assess the influence of degradation mechanisms and safety behavior \cite{shahjalal2022review,hu2022review,azizighalehsari2024empowering}. In this section, authors aim to contribute to answering the following research questions :
\begin{itemize}
    \item How reused batteries degrade?
    \item How retired batteries should be reused and managed to ensure a satisfying lifespan?
    \item How battery aging affects their safety?
\end{itemize}

\subsection{General observations}

\begin{figure}[h!]
    \centering
    \includegraphics[width=\columnwidth]{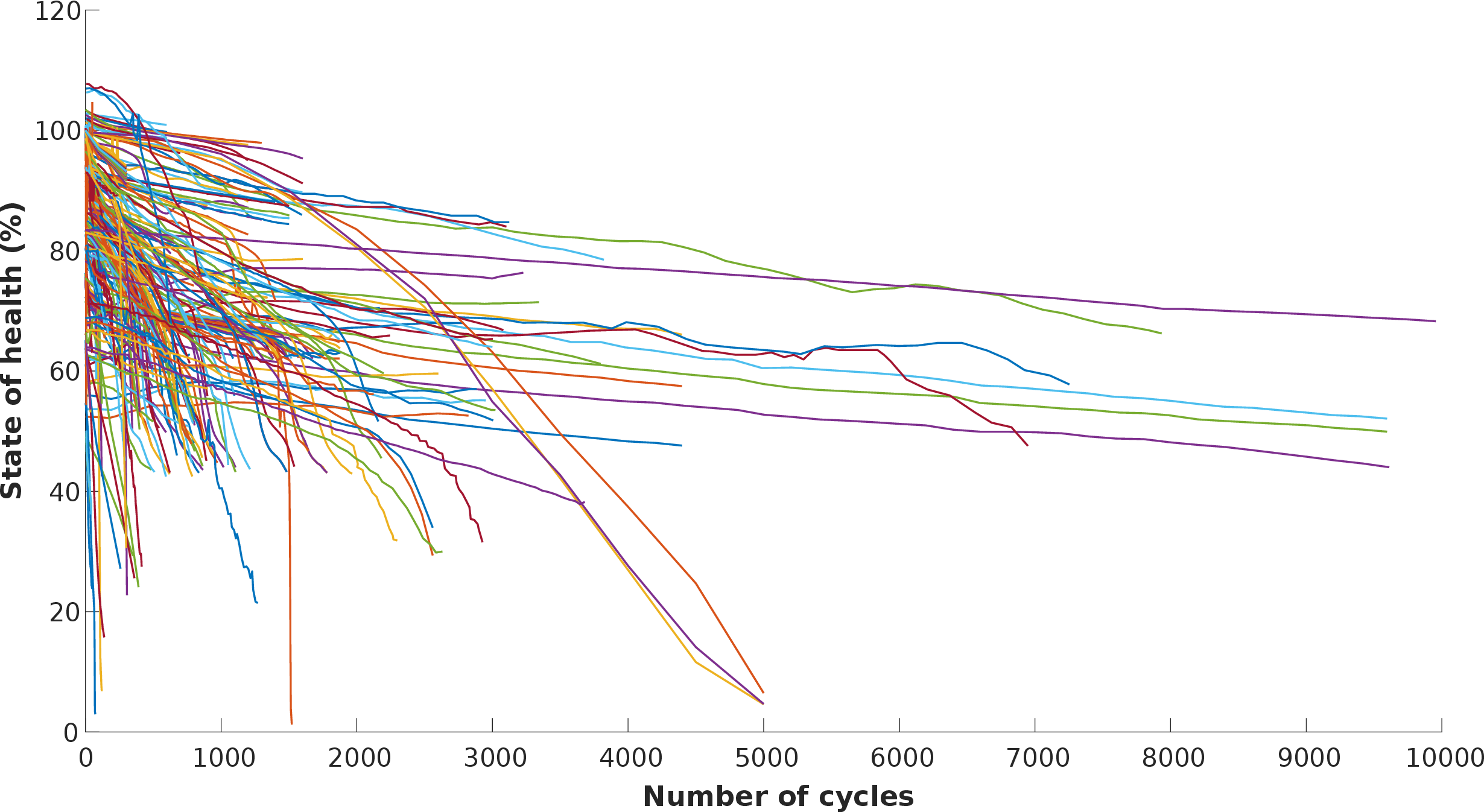}
    \caption{Degradation curves of reused batteries}
    \label{fig:degradation_speeds}
\end{figure}

For this review, 36 articles presenting aging studies on retired batteries were analyzed. In total, the degradation of 892 retired battery cells have been tracked. Fig.~\ref{fig:degradation_speeds} gathers all degradation curves.

From this figure, it appears that no common pattern can be identified for degradation curves. Retired batteries may have very different lifespans and degradation rate. It is important to keep in mind that batteries were tested in laboratories for research projects that have limited budget and duration. This push experimentalist to shorten the aging campaign by choosing experimental conditions known to accelerate battery degradation. Despite this fact, many batteries have reached several thousands of cycles and state of health below 60\%. This is encouraging, as it justifies that reuse is relevant from a durability perspective. The figure also suggests that no behavior change occurs after any threshold of performance loss. These results denies the common assumption that after a specific state of health threshold aging would accelerate. These data also show that no general correlation can be identified between state of health at the end of first life and duration of the second life. In the next sections, the most meaningful analysis are presented. Additional analysis have been conducted to determine if experimental conditions could explain all degradations patterns. This was not the case. Also, analysis on a specific chemistry or battery model did not provide meaningful insights. Some of these analysis are available in appendix.

\subsection{Focus on short lifespan studies}

\begin{figure}[h!]
    \centering
    \includegraphics[width=\columnwidth]{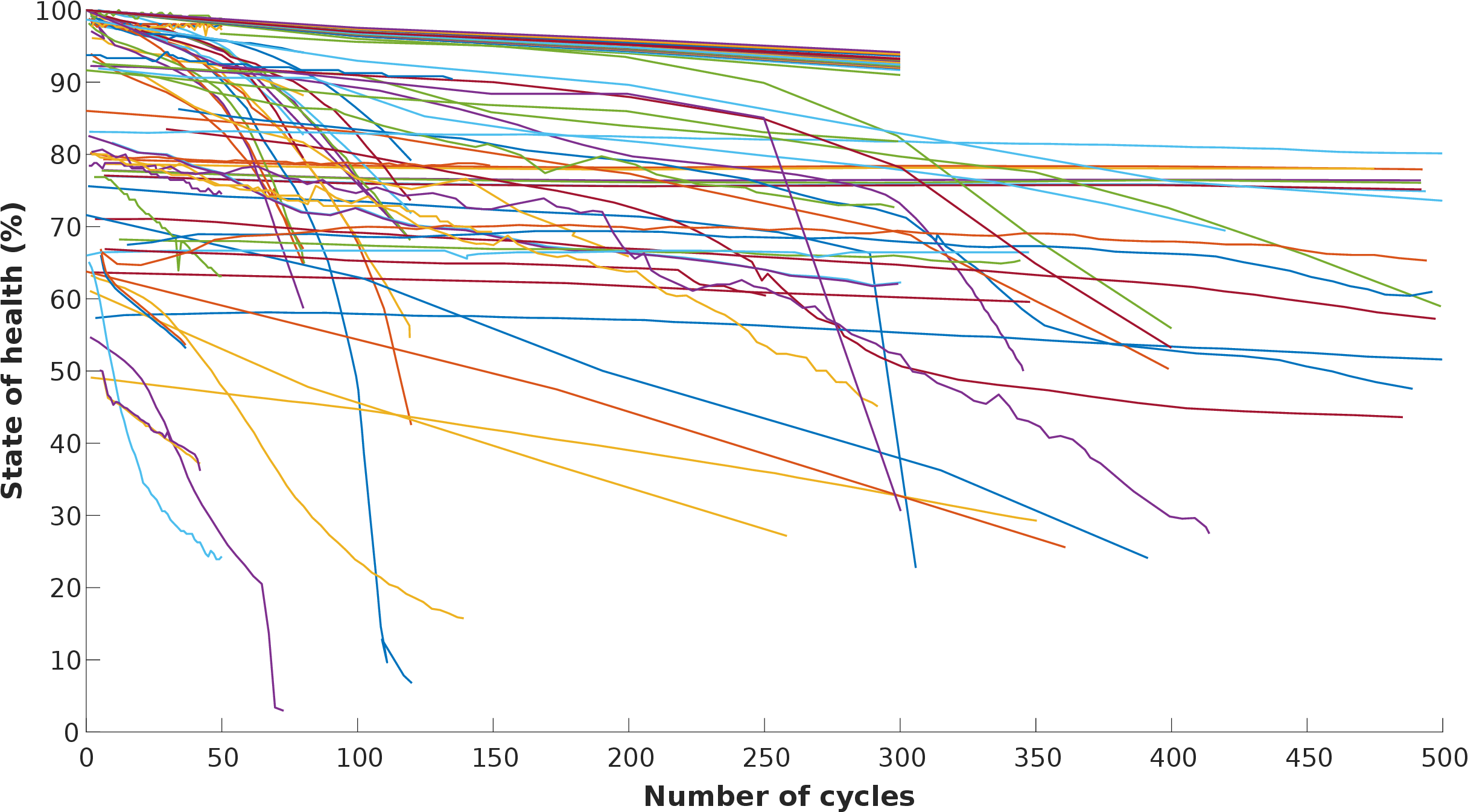}
    \caption{Degradation curves of reused batteries with short lifespan}
    \label{fig:low_lifespan}
\end{figure}

This section aims to identify the factors that would lead to a short lifespan. Figure~\ref{fig:low_lifespan} shows the batteries that have been used for less than 500 cycles in the second life.

\begin{figure}[h!]
    \centering
    \includegraphics[width=\linewidth]{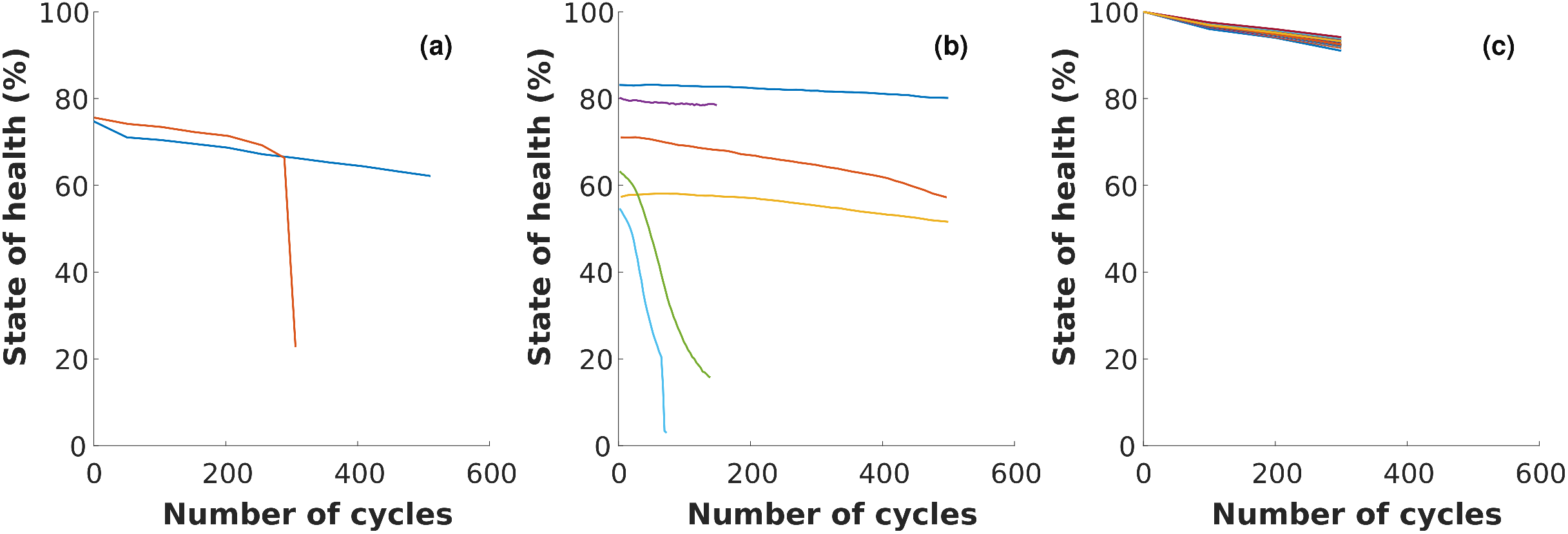}
    \caption{(a) Attidekou et al. \cite{attidekou2020methodologies}, (b) Lv et al. \cite{lv2022investigation}, (c) Baumann et al. \cite{bauma18}}
    \label{fig:interesting_plots_short_lifespan}
\end{figure}

These batteries had low lifespans for different reasons. The section~\ref{section:exp_conditions_short_lifespan} describe the experimental conditions of all cells with short lifespan. In few studies, the short lifespan can be explain by the desire of experimentalist to limit the aging campaign to a certain number of cycles \cite{bauma18}. This explains why several reused batteries have reached a limited number of cycles, despite their homogeneous, slow and linear degradation. Most batteries have experienced a short lifespan because of the experimental conditions chosen. The combination of multiple experimental conditions known to accelerate the degradation have led to reduced lifespan. In several studies, cells with the same initial condition experienced very different second life due to the chosen aging conditions. The figure also shows the absence of correlation between initial state of health and durability.
Fig.~\ref{fig:interesting_plots_short_lifespan} show the degradation curves obtained in three studies. These studies were chosen for the interest of their results.
\newpage

Fig. (a) shows the degradation curves presented in \cite{attidekou2020methodologies}. The table~\ref{tab:low_lifespan} shows that in this study the cells were aged in harsh conditions. Cells were overcharged at a current rate of 0.8C. The two cells were charged with different profiles. The cell that experienced the knee was charged with a constant-current and a constant-voltage phase. This constant-voltage at a high state of charge seems to have triggered the knee as the other cell was charged with a constant current profile only and did not experience the knee. This suggest that the aging knee can be eliminated by avoiding constant voltage phase at a full charge state.

Fig. (b) shows the results from \cite{lv2022investigation}. In this study, 3 of the 6 cells experienced a particularly short lifespan. These three cells were the one charged and discharged at the higher current rate (0.9C in charge and discharge). The three others cells were charged at lower current rate (0.45 in charge and discharge) and they experienced linear and slow degradation. One the cell had an initial state of health of 57\% and reached 52\% after 500 cycles. This example demonstrates that even deeply degraded batteries can be reused and can experience slow and linear degradation if the aging conditions are adapted.

Fig. (c) illustrates the degradation curves from \cite{bauma18}. In this study, all cells were aged in similar conditions. This explains why their degradation was homogeneous. In this study, the short lifespan is a choice of experimentalist that decide to stop the campaign after 300 cycles.

In general, it can be concluded that short lifespan are the consequence of a combination of harsh experimental conditions. To avoid the aging knee, aging conditions should be reasonable to avoid an excessive aging acceleration.

\subsection{Focus on long lifespan studies}

\begin{figure}[h!]
    \centering
    \includegraphics[width=\columnwidth]{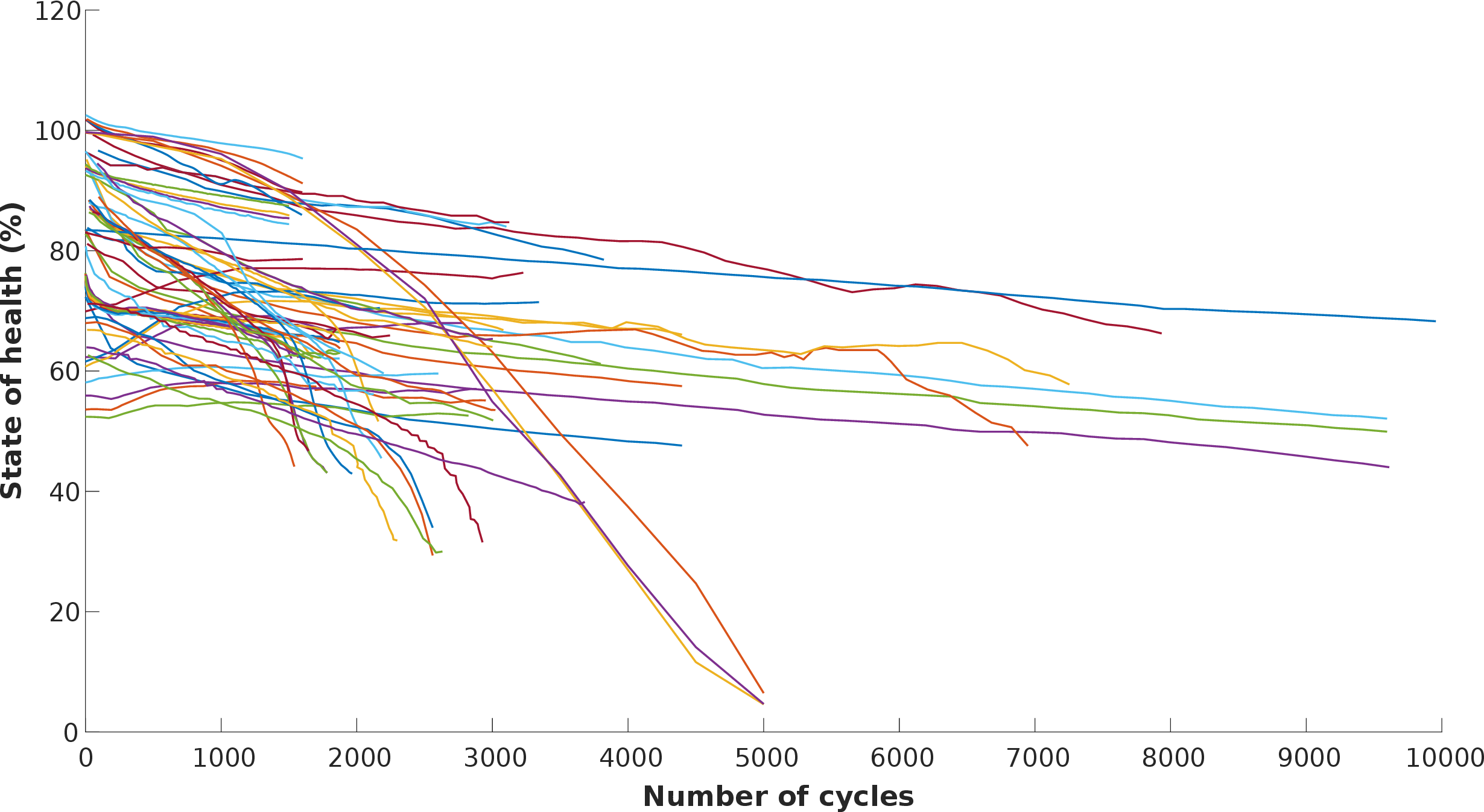}
    \caption{Degradation curves of reused batteries with long lifespan}
    \label{fig:high_lifespan}
\end{figure}

This section aims to identify the factors that would lead to a long lifespan. Figure~\ref{fig:high_lifespan} shows the batteries that have been used for more than 1500 cycles in the second life.

\newpage

\begin{figure}[h!]
    \centering
    \includegraphics[width=\linewidth]{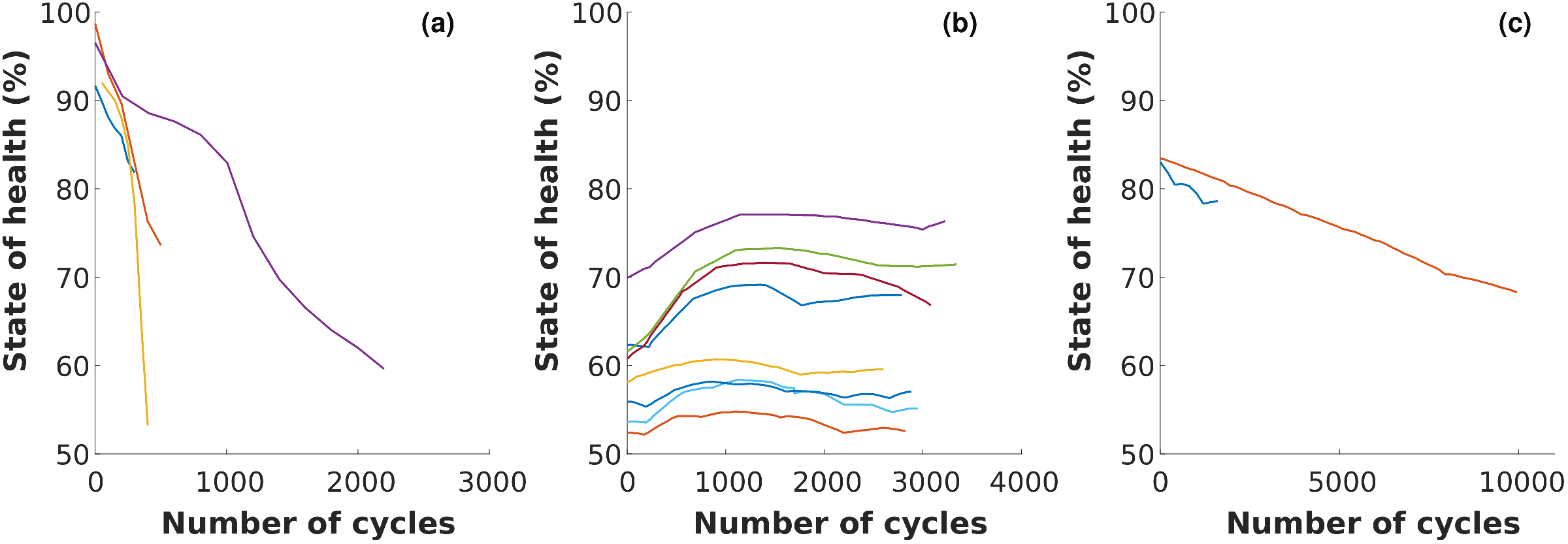}
    \caption{(a) Zhang et al. \cite{zhang2023state}, (b) Cui et al. \cite{cui2024adaptive}, (c) Xu et al. \cite{xu2019study}}
    \label{fig:interesting_plots_long_lifespan}
\end{figure}

This figure shows that long-lifespan batteries have diverse degradation behaviors. The section \ref{section:exp_conditions_long_lifespan} describes the experimental conditions experienced by these cells. The table of experimental conditions shows that using a second-life battery at room temperature with limited current rates or avoiding an extreme state of charge generally improves battery durability. This figure demonstrates that second life can last more than 9000 cycles in accelerated conditions. A large number of the batteries experienced a linear degradation. It is interesting to note that reusing a battery with a state of health lower than 80\% can result in a long second life. Fig.~\ref{fig:interesting_plots_long_lifespan} shows the degradation curves obtained in three studies. These studies were chosen for the interest of their results.

Fig. (a) shows the degradation curves from \cite{zhang2023state}. In this aging campaign, the batteries were charged in CC-CV and discharge in CC between SoC 0 and 100\%. This study illustrate the effect of the current rate. The cell with the highest current rate (2C) experienced a short lifespan (400 cycles). Two cells tested at C/5 experienced a slow and linear degradation. Finally, the cell tested at C/2 reached more than 2000 cycles. The difference between these three last cells is surprising and show that reused cell aged in similar conditions can experience very different degradation.

Fig. (b) presents the degradation curves collected from the study  \cite{cui2024adaptive}. In this work, batteries experienced capacities growth over 3000 cycles. They were aged with full charge but at room temperature. The temperature variation may explain this surprising behavior.

Fig. (c) shows the longest second life aging campaign from \cite{xu2019study}. In this study, cells reached nearly 10,000 aging cycles. They were used at room temperature, within a limited state of charge range (45-55\%), and underwent cycles representative of frequency regulation. These conditions closely resemble real usage, which is encouraging for the viability of second-life batteries.


\subsection{Focus on dispersion evolution}

\begin{figure}[h!]
    \centering
    \includegraphics[width=\columnwidth]{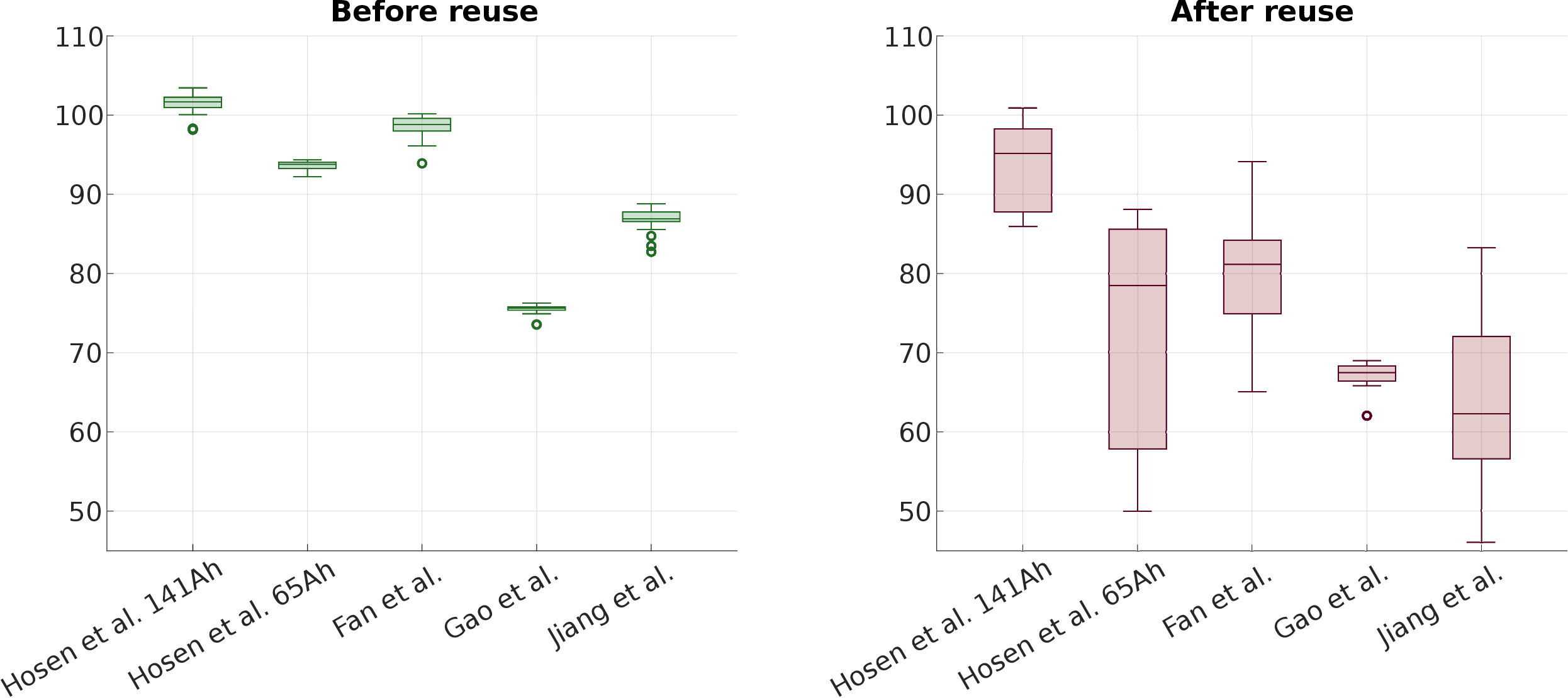}
    \caption{Evolution of dispersion in four aging studies on reused batteries.}
    \label{fig:dispersion}
\end{figure}

The evolution of performance dispersion among the different elements of a battery pack is also a point of interest. Since the performance of a battery assembly depends on its most degraded elements and the heterogeneity between elements, it is desirable to maintain the highest possible performance and homogeneity levels. Several authors have noted that heterogeneity tends to increase with age \cite{schuster2015lithium,bauma18,hassini2026ageing}. This requires careful consideration of performance spread for reused batteries. This review considers 4 studies that tested more than 15 retired batteries: \cite{hosen2026data,jiang2024state,fan2022fast,gao2024comprehensive}. Dispersion is calculated as the ratio of the standard deviation of the capacity of the retired batteries to their mean value. Figure~\ref{fig:dispersion} shows the evolution of dispersion before and after the aging campaign.

The figure shows that dispersion increased significantly for all studies. This highlights the importance of considering performance spread and outliers in retired batteries.

\subsection{Focus on laboratory studies } \label{sec:fresh_batteries}

\begin{figure}[h!]
    \centering
    \includegraphics[width=\linewidth]{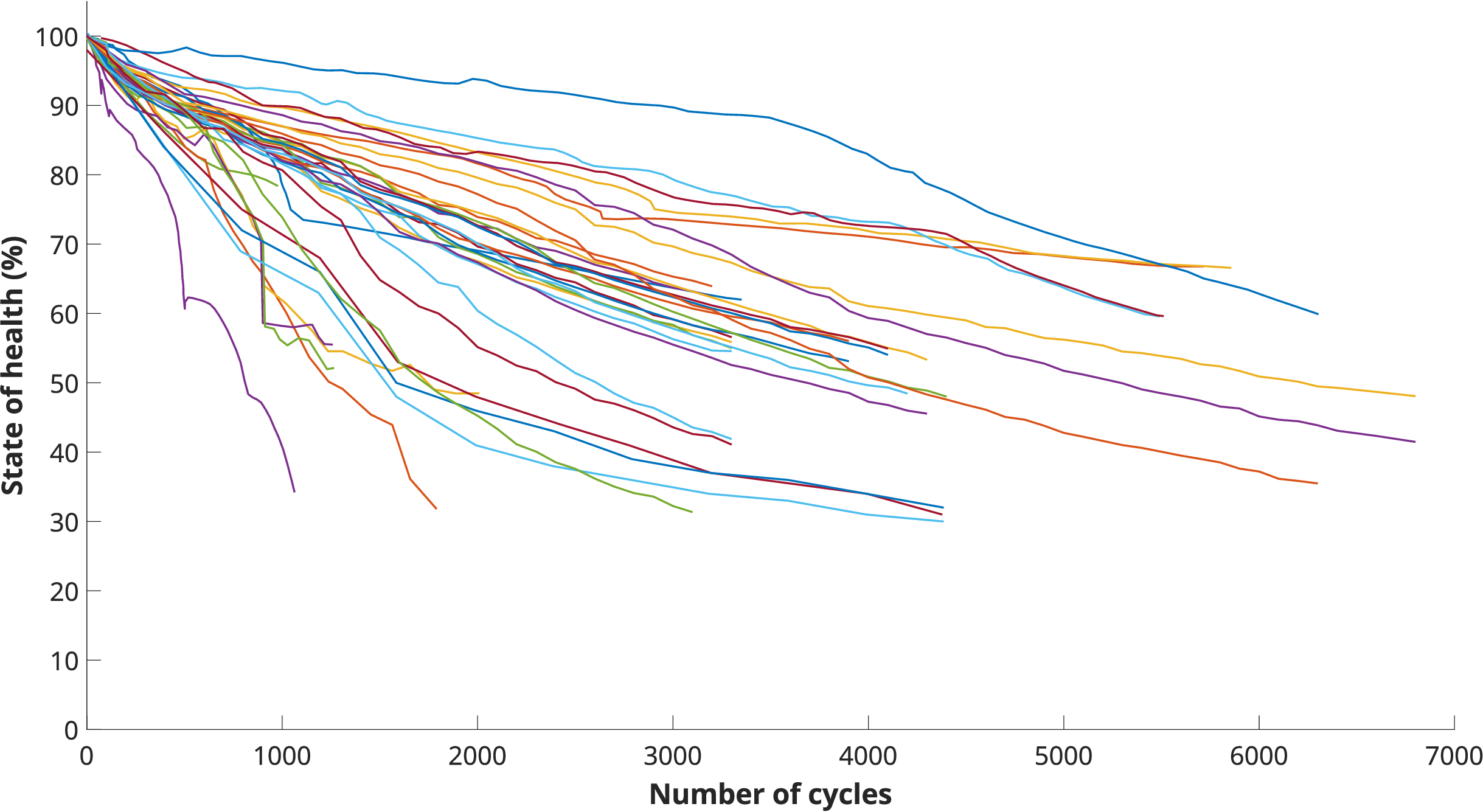}
    \caption{Effect of deep degradation}
    \label{fig:fresh_batteries_low_soh}
\end{figure}

The reviewed aging studies all suffer from the lack of information on their first life history. This section presents complementary studies conducted in laboratories. They have been selected for the deep degradation states they reached or for the usage change that were set up. Figure~\ref{fig:fresh_batteries_low_soh} shows degradation curves of 5 studies that conducted aging campaign  down to state of health 35\% \cite{ezeigwe2026degradation,kebede2023optimal,ohrelius2025polarization,wang2019study,wheeler2025impact}.

These degradation curves are in good agreement with the aging of retired batteries. Degradation curves have similar shapes than between state of health 100\% and 80\%. The degradation mostly follows a linear or square root of time shape. This suggest that existing aging models could be reused and extended for deeper degradation states. Results from \cite{wheeler2025impact,wheeler2025ageing} are particularly interesting as three cells were aged down to 35\% state of health for more than 5000 cycles at 50°C. A post mortem study also showed that this accelerated aging did not triggered dendritic formation or mechanical deformation \cite{wheeler2024postmortem}.

\begin{figure}[h!]
    \centering
    \includegraphics[width=\linewidth]{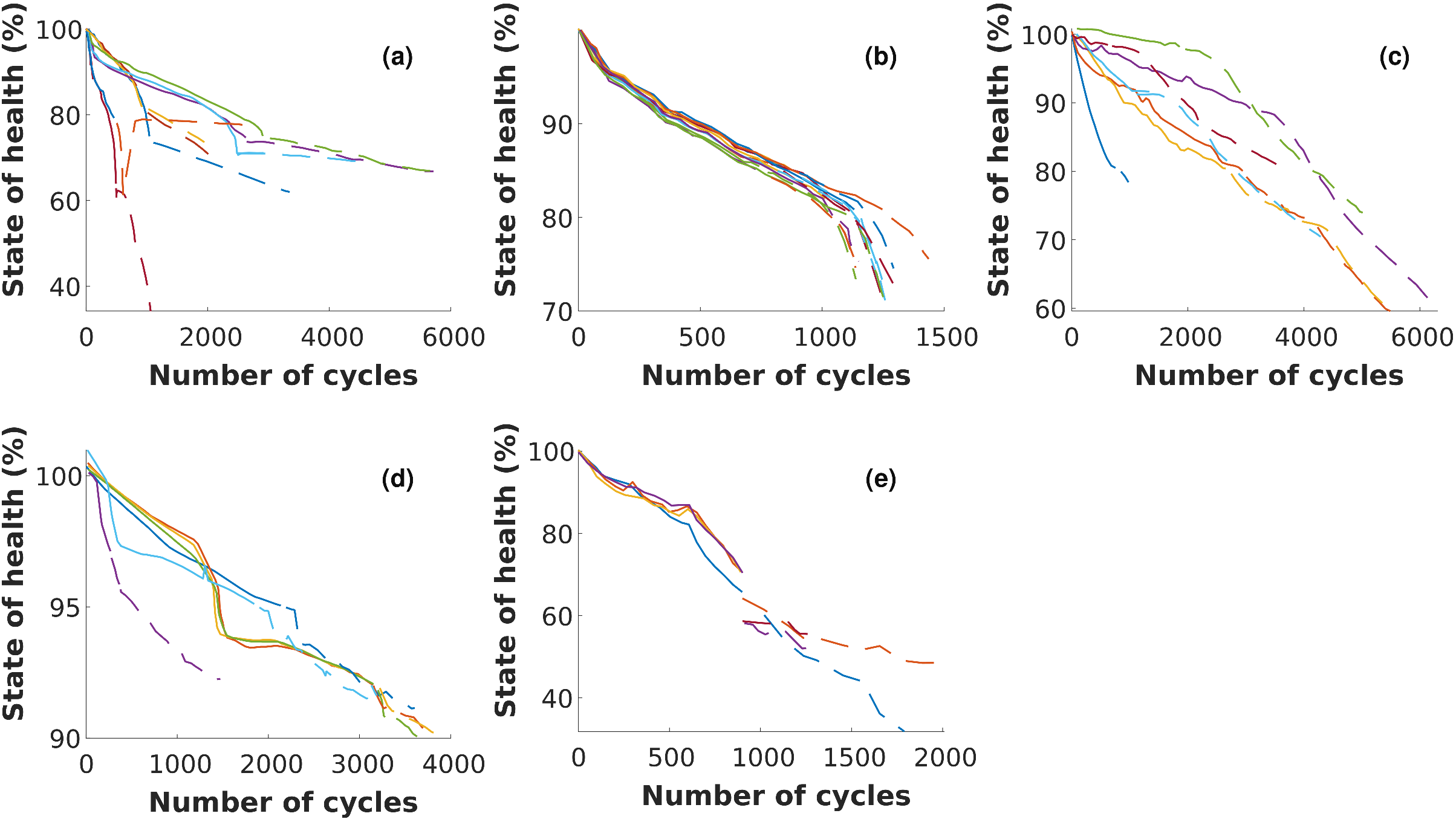}
    \caption{Effect of usage change (a) Ezeigwe et al. \cite{ezeigwe2026degradation}, (b) Fan et al. \cite{fan2026consistency}, (c) Kebede et al. \cite{kebede2023optimal}, (d) Moy et al. \cite{moy2024second} and (e) Ohrelius et al. \cite{ohrelius2025polarization} }
    \label{fig:fresh_batteries_usage_changes}
\end{figure}

Fig.~\ref{fig:fresh_batteries_usage_changes} shows degradations curves from 5 studies that set up usage change during the aging campaign.

Fig. (a) shows the degradation curves from \cite{ezeigwe2026degradation}. In this aging campaign, the first life was conducted at different temperature (5, 25, 45°C) and current rates (1, 1.5C) with a full state of charge window (SoC 0-100\%). The second life was done at C/4 and between SoC 10 and 70\% for all cells except one that aged at C/2 and C/4 between SoC 0 and 100\%. This last cell experienced a degradation that followed the trend of the first usage. The current reduction helped to extend the cell lifespan by 200 cycles. For other cells, a clear reduction of the degradation speed appeared. This suggest that a usage change can be used to change a degradation dynamic.

Fig. (b) presents the results from \cite{fan2026consistency}. In this article, the effect of the switching point from one usage to another is studied. For all cells, charged are done in CC-CV and discharge in CC between SoC 0 and 100\%. The current rates and state of health thresholds are used as variables to evaluate different transitions between first and second usage. The experimental results show that when cells degrade linearly they tends to behave consistently despite a change in their aging conditions. However, if the degradation is nonlinear they tend to be affected according by their new operating conditions. Authors conclude that historical operating conditions subtly influence battery degradation during reuse and that usage change should be preferably done during a linear degradation of the battery.

Fig. (c), (d) and (e) present the degradation curves respectively from \cite{kebede2023optimal}, \cite{moy2024second} and \cite{ohrelius2025polarization}. In these studies, first life was conducted with automotive cycles while second life was a stationary usage. Experimental results indicate that the condition of the cells at the end of their first life tend to influence the performance and further aging in the 2nd life.

\newpage

\subsection{Focus on influence of aging on safety}

Retired batteries raise also concerns due to the lack of understanding of the relation between degradation and safety. As batteries age, their electrical, physical, and chemical characteristics change significantly. This means that the reliability and safety of batteries can be affected positively or negatively by the evolution of aging. The influence of aging on battery safety is a relatively understudied research topic due to the extensive and costly experimental equipment required. This section presents 27 articles related to this important topic.

\begin{figure}[h!]
    \centering
    \includegraphics[width=0.9\columnwidth]{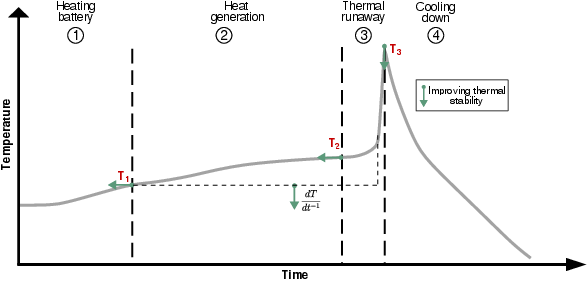}
    \caption{Evolution of temperature during a thermal runaway caused by a thermal abuse.}
    \label{fig:thermal_runaway_steps}
\end{figure}

Ensuring battery safety consists of avoiding the thermal runaway phenomenon. Thermal runaway is caused by abusive conditions and leads to battery destruction, as well as heat production and possible gas emissions \cite{guenther2024second}. Thermal runaway can be broken down into four stages, which are structured around three key temperatures: T1, T2, and T3, as well as their evolution \cite{feng2020,tran2022review}. Figure~\ref{fig:thermal_runaway_steps} shows these four stages and the temperature evolution during thermal runaway.

The first step is an increase in battery temperature. Temperature T1 indicates when the battery begins to heat up on its own. This temperature is known as the self-heating temperature (T1). This reaction is caused by the interaction between the solid electrolyte interface (SEI) and the electrolyte. The excess heat generated comes from parasitic reactions taking place within the cell. Kuntz et al. point out that avoiding dissolving the separator is the key to battery safety \cite{kuntz2023}. T1 is therefore an indicator of the cell's thermal stability. Temperature T2 is the temperature at which thermal runaway occurs. From this point on, the temperature begins to rise rapidly. During this phase, the velocity increases by several orders of magnitude until it reaches the maximum temperature, temperature T3. Wang et al. indicate that thermal runaway takes several minutes, providing time to implement protective measures for people and systems \cite{wang2020}. Improving a battery's thermal stability involves raising temperatures T1 and T2 while lowering temperature T3.

Several authors have demonstrated the link between a battery's thermal stability and the degradation mechanisms that affect it. Table~\ref{table:utilisation_stabilité}  shows the effect of different aging conditions on the thermal stability of batteries.

\begin{table}[h!]
\centering
\begin{tabular}{ p{3cm}| p{4cm}| p{5cm}| p{3cm}| } 

\cline{2-4}
  & \textbf{Improves thermal stability} & \textbf{Degrades thermal stability}& \textbf{No or limited effect}  \\ \cline{1-4}
\multicolumn{1}{|c|}{\shortstack{\textbf{High temperature (SEI)}}}  & \cite{kuntz2023,kabra2025,timke2024impact,jesus2026impact}& \cite{liu2023,zhang2024_1}&\cite{waldmann2017,borner2017}   \\ \cline{1-4}
\multicolumn{1}{|c|}{\shortstack{\textbf{Low temperature (Plating)}}}  &  \cite{wang2025research}  & \cite{kuntz2023,waldmann2017,borner2017,carter2019,friesen2016,gerosa2025,kabra2025,zhang2024_2,liu2023,zhao2026evolution,jesus2026impact}&   \\ \cline{1-4}

\multicolumn{1}{|c|}{\textbf{Calendar}}& \cite{zhang2016}  & \cite{abada2018}  &    \\ \cline{1-4}
\multicolumn{1}{|c|}{\textbf{Overcharge}}& \cite{wang2020}  & \cite{liu2020,leonhardt2026effects}   &   \\ \cline{1-4}
\multicolumn{1}{|c|}{\shortstack{\textbf{Short-circuit or overcurrent}}} &  & \cite{gerosa2025} &   \\ \cline{1-4}
\multicolumn{1}{|c|}{\textbf{Low pressure}}  &   & \cite{xie2020}   &    \\ \cline{1-4}

\end{tabular}
\caption{Effect of usages on thermal stability of aged batteries}

\label{table:utilisation_stabilité}
\end{table}

The effect of lithium plating is especially noticeable. This phenomenon is favored by low temperatures, high states of charge, and high charge current rates. However, other factors are also important. For example, low temperatures may not lead to lithium plating if certain current levels or states of charge are not reached. From a safety standpoint, lithium plating lead to two phenomena that affect thermal stability: the dendritic growth of lithium and exothermic reactions of lithium with the electrolyte \cite{waldmann2017}. 
Table~\ref{table:conception_stabilité}  shows the effect of design on the thermal stability of a cell.

\begin{table}[h!]

\begin{tabular}{ p{4cm}| p{4cm}| p{5cm}| p{3cm}| } 
\cline{2-4}
  & \shortstack{\textbf{Improves thermal stability}} & \shortstack{\textbf{Degrades thermal stability}}& \shortstack{\textbf{No or limited effect}}   \\ \cline{1-4}
\multicolumn{1}{|c|}{\textbf{Power cell}}         & Fresh \cite{kuntz2023} & Cold aging \cite{kuntz2023}   &    \\ \cline{1-4}
\multicolumn{1}{|c|}{\textbf{Energy cell}}& Cold aging \cite{kuntz2023}   & Fresh \cite{kuntz2023}     &   \\ \cline{1-4}
\multicolumn{1}{|c|}{\textbf{Silicone at negative electrode}} & & \cite{kuntz2023}     &    \cite{cruz2025evaluating} \\ \cline{1-4}

\end{tabular}
\caption{Effect of battery design on thermal stability}

\label{table:conception_stabilité}
\end{table}

Kuntz et al. have demonstrated that distinguishing between power and energy cells can reveal two different safety behaviors \cite{kuntz2023}. Power cells are defined as having large, thick electrodes that enable them to deliver high currents. They had better thermal stability when new but degraded after cold aging. Energy cells, on the other hand, have smaller, thinner electrodes that enable them to deliver higher energy densities. These cells were less stable when new but improved with cold aging. The authors also demonstrated that silicone at the negative electrode could degrade thermal stability.

This section summarizes the main findings of the literature on the safety of second-life batteries. It appears that the thermal stability of aged batteries is mainly affected by the degradation mechanisms that have impacted them. Since aged batteries have degraded performance, the energy they dissipate during a runaway event is lower than that of a new battery. To improve the safety of second-life batteries, abusive use and aging conditions that encourage lithium plating should be avoided. Regarding SEI growth and abusive overcharging, different and sometimes contradictory results have been obtained depending on the study. \textit{Further research is therefore required}.

\section{Synthesis of review insights}

This section gather the key take away of this review about reused batteries.

\subsection{Structuring battery research}
The review of more than 150 studies have revealed several limits of current research practices in the battery field. 

A first observation is the lack of a common vocabulary related to the "second life" of batteries. This term is polysemous and designate a wide variety of battery types. In Section~\ref{sec:definitions}, we proposed a clarified vocabulary for designating the batteries at different steps of their life cycle. 

The lack of open science practices is another important limitation of research in the battery field. In this study, only 7\% of the reviewed articles published experimental data alongside their analyses. Additionally, several articles published data in figures that did not facilitate reading and data extraction. We recommend to share experimental dataset alongside the study or at the very least to design figures and tables that enable data extraction.

This article also highlights the lack of a common method for evaluating the performance of characterization techniques. This makes difficult any comparison of results from different studies. Due to these limitations, cumulative research in the battery field is hardly possible in 2026. Initiatives that bring together research teams from around the world are necessary to develop open, incremental, and reproducible research projects in the field of batteries.

From a more technical point of view, we suggest to limit the use of the state of health (SoH) indicator to a specific scope. From our perspective, the SoH should be considered only as an insightful and important indicator for tracking performance of batteries during use. In laboratories, it should remain a reference to determine the level of degradation of the tested battery. For in field studies, it is also a relevant piece of information for online performance tracking and predictive maintenance. However, we would also like to stress the fact that state of health capacity is not relevant for defining end-of-life threshold. Performance fade related to the usage such as maximal acceptable autonomy loss for electric vehicle should be favored. Favoring such definition would lead to the fact that measuring capacity is also an insufficient method for assessing performance in make decision related to batteries end-of-life and reuse. A more complete set of indicators should be measured to assess if a battery should be reused or not. Battery passport seems to be a more complete document that could help in the decision-making.

\subsection{Choosing retired batteries}

Section~\ref{sec:choosing_batteries} of the review focused on the performance of retired batteries. The state of health of more than 10000 retired batteries was evaluated. The mean value was 79.9\% and the median 84.5\%. A Weibull distribution was fitted to the dataset collected. These findings suggest that the common definition of a "second life" based on a 20 or 30\% capacity threshold does not align with experimental results. 

Also the dispersion of retired batteries is several orders of magnitude higher than that of new batteries. This highlight the importance of developing effective thermal and balancing strategies to enhance battery durability.

A study of performance assessment techniques revealed that using correlations between performance and cell position in a battery pack or between capacity and resistance is not a reliable methodology for reducing the characterization time of retired batteries. Instead, looking for health indicators highly correlated with state of health is the most promising approach for selecting batteries for reuse. The benchmark studies were able to evaluate the SoH with an error below 2\% in less than 4 minutes.

This review also emphasizes the need to develop a more holistic approach to select batteries that can be reused. In 2026, the evaluation of defects and degradation mechanisms have received limited interest while they have key roles for the viability of batteries reuse. The need for developing a combination of techniques that could evaluate different dimensions of the battery performance is also clear. The concept of battery passport and visa are valuable tool for making an informed decision about possible reuse.

\subsection{Aging of reused batteries}

Section~\ref{sec:reusing_batteries} presented results related to aging of reused batteries. 36 articles and more than 890 cells were studied. This section demonstrated that batteries can be reused for several thousand cycles and beyond an 80\% state of health. 

The battery second life does not lead to specific degradation pattern. There is a diversity of degradation speed and shapes. Experimental conditions are the primary factors to explain a short or a long lifespan. History in the first life seems to also have an effect.

Aging acceleration, also known as the aging knee, generally occurs due to a combination of accelerating factors. To ensure a long lifespan for reused batteries, the current rate and state of charge should be adapted to the measured capacity of the batteries. The authors also showed that the degradation rate could be controlled by changing experimental conditions.

This review also emphasized the importance of dispersion in reused batteries. Aging increases heterogeneity among these batteries. This necessitates developing adapted thermal and balancing management strategies for second-life batteries. Finally, the effect of aging on safety was shown to be related to degradation mechanisms in the battery. Safety risks can be minimized by avoiding conditions that accelerate aging.

In a nutshell, a control of experimental conditions and an adequate battery management strategy lead to a long, linear and safe aging during the second life.

\subsection{Directions for future research}

This review also outlines directions for future battery research.

This article have shown that second life batteries could have different meaning. In this review, we propose to define second life batteries as the ones that underwent a usage and/or structure change. A usage change may be a first life in a vehicle and a reuse in a stationary application or in a smaller vehicle for instance. While a structure change could be a change of the number of module in the battery, a modification of the serie-parallel configuration or a change in BMS or thermal management systems for instance. For such batteries, major research questions are how a usage or structure change may affect battery degradation and dispersion. A second category of batteries are the deeply degraded ones. As the threshold of 80\% state of health have been popular over the last decades, authors suggest to define batteries that are more aged than this threshold as deeply degraded ones. For these batteries, major research questions are how deep degradation affects degradation speed and dispersion. It is important to note that a battery may be both a second life one and a deeply degraded but it is not always the case. Authors hope that separating both concepts would be helpful to avoid any confusion between these two battery types. Hopefully, this would be beneficial for producing coherent research results in the field.

This article also emphasizes the importance of distinguishing between the various applications of state of health. Currently, health status is used to track performance, stop aging experiments, and select retired batteries suitable for reuse. However, the authors argue that this is not relevant. Experiment design and reuse potential evaluation should be based on a more comprehensive performance evaluation. Two important tools that should be used systematically are a battery passport and an evaluation of degradation mechanisms and battery defects. While state of health is relevant for performance tracking, it should be augmented by including different performance metrics and by considering dispersion and outliers. This observation underscores the necessity of novel experimental techniques that can be employed at the cell assembly level to evaluate performance dispersion among cells.

Due to its impact on battery durability, the study of dispersion and the evolution of outliers should also be a more popular research topic. Considering this characteristic of retired batteries is consequently an important topic for future research. For example, developing experimental techniques to rapidly and efficiently assess the spread among cells is an important consideration. In addition, considering the heterogeneity of performance when defining the state of health of a battery module or pack is paramount. The definition of end-of-life should also consider cell heterogeneity, as it is currently common to assume that all cells reach the performance threshold simultaneously. Studies predicting state of health and remaining useful life would benefit from considering this. The ability to model dispersion and incorporate it into energy management studies is a promising new research topic that should be promoted in the future. Developing BMS, thermal, and balancing techniques to limit the influence of dispersion is also an important research topic.

Considering the diversity of battery technologies, there is a clear need to structure the existing studies on each battery model. The development of a registry of experimental studies for each battery model could help to determine health indicators and aging experimental conditions that are suitable for this specific technology. This would also lead to a more cumulative research approach as experimentalist would avoid reproducing an existing result.

Finally, this article examined the reuse of batteries from an electrical engineering perspective. However, major gaps in battery reuse include the capability to eco-design future generations of batteries and the development of logistics and industrial techniques to collect and dismantle the growing number of retired batteries worldwide. To gain a comprehensive understanding of the topic, new meta-reviews could examine battery reuse from various perspectives, such as logistics, economics, and safety.

\section{Conclusion and outlooks}

In summary, this study clarified the current state of knowledge regarding retired batteries. It provided a detailed overview of experimental studies on the performance and degradation of retired batteries in a second life. As the electric vehicle and battery market develops and expands, studies like those summarized in this analysis need to be conducted and repeated.

The dataset produced to conduct this review is shared under open license : \\

\newpage 

\appendix

\section{Equations}

\label{eq:weibull}
The MATLAB function fitdist was used to fit a distribution function to the experimental data \cite{fitdist}. A generalized extreme value distribution was chosen. Its probability density function is based on three parameters $k$, $\beta$ and $\mu$ and is expressed as follows:
\begin{equation}
f(k,\sigma,\mu) = \left\{
    \begin{array}{ll}
       \frac{k}{\sigma}(\frac{\mu}{\sigma})^{k-1}e^{-(\frac{\mu}{\sigma})^k} &, \mu \geq 0,\\
       0, & \mu<0.
    \end{array}
\right\}
\end{equation}

with : 
\begin{itemize}
	\item [$\bullet$] $k$ is the shape parameter of the distribution,
	\item [$\bullet$] $\sigma$ is the scale parameter of the  distribution,
    \item [$\bullet$] $\mu$ is the location parameter of the  distribution.

\end{itemize}


\section{Experimental conditions for short lifespan studies} \label{section:exp_conditions_short_lifespan}
This table presents the experimental conditions of the aging campaigns that led to short lifespan ($<500$ cycles). The number X of cells that experienced short lifespan and the total number Y of cells tested are indicated as (X/Y) in the reference column. The experimental conditions in bold are accelerating factors as they are conditions that are known to accelerate aging. 

\begin{table}[h!]
    \centering
      \resizebox{\columnwidth}{!}{
    \begin{tabular}{cccccccc}
       \textbf{Reference}&\shortstack{\textbf{Battery} \\ \textbf{Chemistry - Format - Qnom}} & \textbf{T°}& \shortstack{\textbf{Initial}\\\textbf{degradation}}& \shortstack{\textbf{Charge} \\ \textbf{Type, C-rate, }$\mathbf{SoC_{max}(\%)}$} & \shortstack{\textbf{Discharge} \\ \textbf{Type, C-rate, }$\mathbf{SoC_{min}(\%)}$} & \shortstack{\textbf{Final}\\ \textbf{SoH/cycle}}  \\
       \hline

\multirow{6}{*}{ \shortstack{Chen et al. \cite{chen2022efficient} \\(6/6)}} &  \multirow{6}{*}{LFP/C-Cylindrical-1.1Ah} &  \multirow{6}{*}{?}     &79\%& \multirow{6}{*}{CC, \textbf{0.9}, \textbf{100}}& \multirow{6}{*}{CC, -0.9, \textbf{0}}& 78\%/492\\
        &  &   & \textbf{79\%}&  &  & 79\%/497 \\
        &  &   & \textbf{78\%}&  &  & 76\%/492 \\
        &  &   & \textbf{78\%}&  &  & 76\%/492 \\
        &  &   & \textbf{77\%}&  &  & 75\%/494 \\
        &  &   & \textbf{77\%}&  &  & 75\%/492 \\
\hline
\multirow{16}{*}{ \shortstack{Fan et al. \cite{fan2022fast} \\(16/16)}} &  \multirow{16}{*}{\shortstack{LFP/C-Cylindrical-15.5Ah}} &  \multirow{16}{*}{25}&99\%& \multirow{16}{*}{CC-CV, \textbf{1, 100}}& \multirow{16}{*}{Pulse CC, -0.7, \textbf{0}}& 7\%/120\\
        &  &   & 99\%&  &  & 43\%/120 \\
        &  &   & 100\%&  &  & 55\%/120 \\
        &  &   & 100\%&  &  & 64\%/91 \\

        &  &   & 100\%&  &  & 68\%/120 \\
        &  &   & 100\%&  &  & 72\%/120 \\
        &  &   & 100\%&  &  & 73\%/120 \\
        
        &  &   & 100\%&  &  & 79\%/120 \\
        &  &   & 94\%&  &  & 67\%/80 \\
        &  &   & 96\%&  &  & 88\%/80 \\

        &  &   & 97\%&  &  & 59\%/80 \\
        &  &   & 98\%&  &  & 65\%/80 \\
        &  &   & 98\%&  &  & 83\%/80 \\

        &  &   & 99\%&  &  & 80\%/80 \\
        &  &   & 98\%&  &  & 94\%/80 \\
        &  &   & 99\%&  &  & 65\%/80 \\

\hline
\multirow{1}{*}{ \shortstack{Haram et al. \cite{haram2023second} (1/1)}} &  \multirow{1}{*}{LMO+LNO/C-Pouch-65Ah} &  \multirow{1}{*}{?}&\textbf{50\%}& \multirow{1}{*}{?}& \multirow{1}{*}{?}& 30\%/350\\

\hline
\multirow{1}{*}{ \shortstack{Huang et al. \cite{huang2022state} (1/1)}} &  \multirow{1}{*}{LFP/C-Cylindrical-40Ah} &  \multirow{1}{*}{25}&97\%& \multirow{1}{*}{CC-CV, \textbf{2, 100}}& \multirow{1}{*}{CC, \textbf{-2, 0}}& 56\%/400\\

\hline
\multirow{1}{*}{ \shortstack{Huang et al. \cite{huang2022state} (1/1)}} &  \multirow{1}{*}{NCA/C-Cylindrical-5Ah} &  \multirow{1}{*}{25}&82\%& \multirow{1}{*}{CC-CV, 0.5, \textbf{100}}& \multirow{1}{*}{CC, -0.5, \textbf{0}}& 63\%/300\\

\hline
\multirow{2}{*}{ \shortstack{Jiang et al. \cite{jiang2024state} \\(2/36)}} &  \multirow{2}{*}{NMC/C-Cylindrical-2.6Ah} &  \multirow{2}{*}{25}&84\%& \multirow{1}{*}{CC-CV, \textbf{1}, 60}& \multirow{1}{*}{CC, \textbf{-2}, 30}& 44\%/485\\
        &  &   & 86\%&CC-CV, \textbf{1}, 60  &  CC, \textbf{-2}, 30& 48\%/489 \\

\hline
\multirow{1}{*}{\shortstack{Li et al. \cite{li2024electrochemical} (1/2)}} &  \multirow{1}{*}{LFP/C-Prismatic-25Ah} &  \multirow{1}{*}{\textbf{55}}&86\%& \multirow{1}{*}{CC, \textbf{1, 100}}& \multirow{1}{*}{CC, \textbf{-3, 0}}& 50\%/400\\

\hline
\multirow{1}{*}{ \shortstack{Lin et al.\cite{lin2023state} (1/1)}} &  \multirow{1}{*}{NCA/C-Cylindrical-4.8Ah} &  \multirow{1}{*}{25}&91\%& \multirow{1}{*}{CC, \textbf{1, 100}}& \multirow{1}{*}{CC, -0.5, \textbf{0}}& 66\%/200\\

\hline
\multirow{2}{*}{ \shortstack{Liu et al. \cite{liu2022characterization} \\(2/2)}} &  \multirow{2}{*}{NCA/C+Si-Cylindrical-4.8Ah} &  \multirow{2}{*}{25}&83\%& \multirow{2}{*}{CC, 0.5, \textbf{100}}& \multirow{2}{*}{CC, -0.5, \textbf{0}}& 62\%/300\\
        &  &   & 93\%&  &  & 73\%/300 \\

\hline
\multirow{6}{*}{ \shortstack{Lv et al. \cite{lv2022investigation} \\(6/6)}} &  \multirow{6}{*}{\shortstack{LFP/C\\Cylindrical\\5.5Ah}} &  \multirow{6}{*}{25}&83\%& \multirow{3}{*}{CC,0.45,\textbf{100}}& \multirow{3}{*}{CC, -0.45, \textbf{0}}& 80\%/500\\
        &  &   & \textbf{71\%}&  &  & 57\%/497 \\
        &  &   & \textbf{57\%}&  &  & 52\%/500 \\

        &  &   & 80\%&CC, \textbf{0.9, 100}  & CC, -0.9, \textbf{0} & 78\%/149 \\
        &  &   & \textbf{63\%}&CC, \textbf{0.9, 100} & CC, -0.9, \textbf{0} & 16\%/140 \\   
        &  &   & \textbf{55\%}&CC, \textbf{0.9, 100}  & CC, -0.9, \textbf{0} & 3\%/73 \\

\hline
\multirow{4}{*}{ \shortstack{Tran et al. \cite{tran2022used} \\(4/12)}} &  \multirow{4}{*}{LCO/C-Cylindrical-2.15Ah} &  \multirow{4}{*}{22}&66\%& \multirow{4}{*}{CC-CV, \textbf{0.8, 100}}& \multirow{4}{*}{CC, -0.8, \textbf{0}}& 53\%/37\\
        &  &   & 67\%&  &  & 54\%/37 \\
        &  &   & 50\%&  &  & 37\%/42 \\
        &  &   & 50\%&  &  & 37\%/42 \\

\hline

   \end{tabular}
}
    \caption{Studies with short lifespan reused batteries}
    \label{tab:low_lifespan}
\end{table}

\begin{table}[h!]
    \centering
      \resizebox{\columnwidth}{!}{
    \begin{tabular}{cccccccc}
       \textbf{Reference}&\shortstack{\textbf{Battery} \\ \textbf{Chem.-Format-Qnom}} & \textbf{T°}& \shortstack{\textbf{Initial}\\\textbf{degradation}}& \shortstack{\textbf{Charge} \\ \textbf{Type, C-rate, }$\mathbf{SoC_{max}(\%)}$} & \shortstack{\textbf{Discharge} \\ \textbf{Type, C-rate, }$\mathbf{SoC_{min}(\%)}$} & \shortstack{\textbf{Final}\\ \textbf{SoH/cycle}}  \\

\hline
\multirow{2}{*}{\shortstack{ Xiong et al. \cite{xiong2020online} \\(2/3)}} &  \multirow{2}{*}{NMC/C-Cylindrical-2.4Ah} &  \multirow{2}{*}{?}&80\%& \multirow{2}{*}{CC-CV, 0.5, \textbf{100}}& \multirow{1}{*}{CC, -0.5, \textbf{0}}& 45\%/292\\
        &  &   & 80\%&  & CC, -1, \textbf{0} & 28\%/414 \\

        \hline
\multirow{2}{*}{\shortstack{ Zhang et al. \cite{zhang2021state} \\(2/2)}} &  \multirow{2}{*}{NCA/C+Si-Cylindrical-4.8Ah} &  \multirow{2}{*}{25}&92\%& \multirow{1}{*}{CC-CV, 0.2, \textbf{100}}& \multirow{1}{*}{CC, -0.2, \textbf{0}}& 82\%/300\\
        &  &   & 99\%& CC-CV, 0.5, \textbf{100} & CC, -0.5, \textbf{0} & 73\%/500 \\
 \hline
\multirow{1}{*}{\shortstack{ Zhang  et al. \cite{zhang2021state} (1/2)}} &  \multirow{1}{*}{LFP/C-Cylindrical-40Ah} &  \multirow{1}{*}{25}&92\%& \multirow{1}{*}{CC-CV, 0.2, \textbf{100}}& \multirow{1}{*}{CC, \textbf{-2, 0}}& 53\%/400\\

\hline

\multirow{2}{*}{ \shortstack{Attidekou  et al. \cite{attidekou2020methodologies} \\(2/2)}} &  \multirow{2}{*}{LMO+LNO/C-Pouch-66Ah} &  \multirow{2}{*}{25}&100\%& \multirow{1}{*}{CC, \textbf{0.8, $>$100}}& \multirow{1}{*}{CC-CV, -0.8, \textbf{10}}& 87\%/500\\
        &  &   & 100\%& CC-CV, \textbf{0.8, $>$100} & CC-CV, -0.8, \textbf{10} & 30\%/300 \\

\hline

\multirow{1}{*}{ \shortstack{Hosen  et al. \cite{hosen2026data} (1/32)}} &  \multirow{1}{*}{?-Pouch-65Ah} &  \multirow{1}{*}{25}&92\%& \multirow{1}{*}{?, \textbf{0.8}, 90}& \multirow{1}{*}{?, -0.8, 20}& 50\%/345\\

\hline
  \multirow{2}{*}{ \shortstack{Ma  et al. \cite{ma2015investigation} (2/6)}} &  \multirow{2}{*}{LMO/C-?-90Ah} &  \multirow{2}{*}{40}&92\%& \multirow{2}{*}{CC-CV, \textbf{0.3, 100}}& \multirow{1}{*}{Driving, -0.3, 50}& 59\%/500\\     
 &  &   & 92\%&  & \multirow{1}{*}{Driving, -0.3, 20} & 70\%/420 \\

\hline
  \multirow{4}{*}{ \shortstack{Perez  et al. \cite{perez2025novel} (4/8)}} &  \multirow{4}{*}{LMO-LNO/C-Pouch-66Ah} &  \multirow{3}{*}{45}&61\%& \multirow{1}{*}{?, \textbf{1.5, 100}}& \multirow{1}{*}{?, -1.5, 50}& 27\%/258\\     
 &  &   & 64\%& \multirow{1}{*}{?, 0.3, \textbf{100}} & \multirow{1}{*}{?, -0.3, 50} & 70\%/420 \\
 &  &   & 72\%& \multirow{1}{*}{?, \textbf{0.7, 100}} & \multirow{1}{*}{?, -0.7, 50} & 24\%/390 \\
 &  & 25  & 64\%& \multirow{1}{*}{?, \textbf{0.7}, 55} & \multirow{1}{*}{?, -0.7, 45} & 60\%/350 \\

\hline
\multirow{24}{*}{\shortstack{ Baumann et al. \cite{bauma18} \\(24/24)}} &  \multirow{24}{*}{LMO/C-Prismatic-50Ah} &  \multirow{24}{*}{35}&100\%& \multirow{24}{*}{CC, 0.5, \textbf{95}}& \multirow{24}{*}{CC, -1, \textbf{5}}& 91\%/300\\
        &  &   & 100\%&  &  & 92\%/300 \\
        &  &   & 100\%&  &  & 92\%/300 \\
        &  &   & 100\%&  &  & 92\%/300 \\
        &  &   & 100\%&  &  & 92\%/300 \\
        &  &   & 100\%&  &  & 92\%/300 \\
        &  &   & 100\%&  &  & 92\%/300 \\
        &  &   & 100\%&  &  & 92\%/300 \\
        &  &   & 100\%&  &  & 92\%/300 \\
        &  &   & 100\%&  &  & 92\%/300 \\
        &  &   & 100\%&  &  & 92\%/300 \\
        &  &   & 100\%&  &  & 92\%/300 \\
        &  &   & 100\%&  &  & 93\%/300 \\
        &  &   & 100\%&  &  & 93\%/300 \\
        &  &   & 100\%&  &  & 93\%/300 \\
        &  &   & 100\%&  &  & 93\%/300 \\
        &  &   & 100\%&  &  & 93\%/300 \\
        &  &   & 100\%&  &  & 94\%/300 \\
        &  &   & 100\%&  &  & 94\%/300 \\
        &  &   & 100\%&  &  & 94\%/300 \\
        &  &   & 100\%&  &  & 94\%/300 \\
        &  &   & 100\%&  &  & 92\%/300 \\
        &  &   & 100\%&  &  & 92\%/300 \\
        &  &   & 100\%&  &  & 93\%/300 \\

\hline

\multirow{3}{*}{ \shortstack{Vasquez et al.\cite{vasquez2025} \\(3/3)}} &  \multirow{3}{*}{LFP/C-?-4.5Ah} &  \textbf{40}&\textbf{68\%}& \multirow{3}{*}{CC, \textbf{1, 100}}& \multirow{3}{*}{CC, -1, \textbf{0}}& 65\%/345\\
        &  &   20& \textbf{66\%}&  &  & 66\%/347 \\
        &  &   \textbf{50}& \textbf{67\%}&  &  & 60\%/250 \\
\hline
\multirow{2}{*}{\shortstack{Vasquez et al. \cite{vasquez2025} \\(2/2)}} &  \multirow{2}{*}{NMC/C-?-60Ah} &  \multirow{2}{*}{20}&68\%& \multirow{2}{*}{CC-CV, 0.2, \textbf{100}}& \multirow{2}{*}{CC, -0.2, \textbf{0}}& 61\%/500\\
        &  &   & 67\%&  &  & 65\%/495 \\

\hline
    \end{tabular}

 }
\end{table}

\clearpage

\section{Experimental conditions for long lifespan studies}\label{section:exp_conditions_long_lifespan}
This table presents the experimental conditions of the aging campaigns that led to long lifespan ($>1500$ cycles). The number X of cells that experienced long lifespan and the total number Y of cells tested are indicated as (X/Y) in the reference column. The experimental conditions in bold are accelerating factors as they are conditions that are known to accelerate aging.

\begin{table}[h!]
    \centering
      \resizebox{\columnwidth}{!}{

    \begin{tabular}{cccccccc}
       \textbf{Reference}&\shortstack{\textbf{Battery} \\ \textbf{Chem.-Format-Qnom}} & \textbf{T°}& \shortstack{\textbf{Initial}\\\textbf{degradation}}& \shortstack{\textbf{Charge} \\ \textbf{Type, C-rate, }$\mathbf{SoC_{max}(\%)}$} & \shortstack{\textbf{Discharge} \\ \textbf{Type, C-rate, }$\mathbf{SoC_{min}(\%)}$} & \shortstack{\textbf{Final}\\ \textbf{SoH/cycle}} \\
       \hline
\multirow{4}{*}{ \shortstack{Cao et al. \cite{cao2024second} \\(4/4)}} &  \multirow{4}{*}{LFP/C-Prismatic-107Ah} &  \multirow{3}{*}{23}     &99\%& \multirow{4}{*}{CC, \textbf{0.6}, \textbf{100}}& \multirow{4}{*}{CC, -0.95,\textbf{0}}& 66\%/7935\\
        &  &   & 97\%&  &  & 78\%/3822 \\
        &  &   & 86\%&  &  & 58\%/7254 \\
        &  &  \textbf{50} & 89\%&  &  & 48\%/6948 \\
\hline

\multirow{8}{*}{ \shortstack{Cui et al. \cite{cui2024adaptive} \\(8/8)}} &  \multirow{8}{*}{\shortstack{LMO/C-Pouch-32.5Ah}} &  \multirow{8}{*}{room}&\textbf{62\%}& \multirow{8}{*}{CC, 0.5, \textbf{100}}& \multirow{8}{*}{CC, -0.75, \textbf{0}}& 68\%/2788\\
        &  &   & \textbf{52\%}&  &  & 53\%/2824 \\
        &  &   & \textbf{58\%}&  &  & 60\%/2601 \\
        &  &   & \textbf{70\%}&  &  & 76\%/3228 \\
        &  &   & \textbf{62\%}&  &  & 71\%/3343 \\
        &  &   & \textbf{54\%}&  &  & 55\%/2951 \\
        &  &   & \textbf{61\%}&  &  & 67\%/3081 \\
        &  &   & \textbf{56\%}&  &  & 57\%/2883 \\

\hline
\multirow{8}{*}{ \shortstack{Jiang et al. \cite{jiang2024state}\\ (8/36)}} &  \multirow{8}{*}{\shortstack{NMC/C-Cylindrical-2.6Ah}} &  \multirow{8}{*}{25}&87\%& \multirow{6}{*}{CC-CV, \textbf{1}, 75}& \multirow{2}{*}{CC, \textbf{-2}, 35}& 44\%/1540\\

       &  &   & 87\%&  &  & 52\%/2159 \\
       &  &   & 87\%&  & \multirow{4}{*}{CC, \textbf{-2}, 45} & 52\%/2159 \\

       &  &   & 86\%&  &  & 43\%/1781 \\

       &  &   & 88\%&  &  & 46\%/2182 \\
       &  &   & 88\%&  &  & 52\%/2345 \\

       &  &   & 86\%&\multirow{2}{*}{CC-CV, \textbf{1}, 70}  & \multirow{2}{*}{CC, \textbf{-2}, 50} & 47\%/1645 \\
       &  &  & 88\%&  & & 43\%/1965 \\

\hline
\multirow{1}{*}{ \shortstack{Sun et al. \cite{sun2022quick} (1/1)}} &  \multirow{1}{*}{\shortstack{LFP/C-Cylindrical-40Ah}} &  \multirow{1}{*}{25}&93\%& \multirow{1}{*}{CC, \textbf{2}, 80}& \multirow{1}{*}{CC, \textbf{-2}, 30}& 61\%/3798\\

\hline
\multirow{1}{*}{ \shortstack{Xiong et al. \cite{xiong2020online} (1/1)}} &  \multirow{1}{*}{\shortstack{LFP/C-Cylindrical-2.6Ah}} &  \multirow{1}{*}{room}&81\%& \multirow{1}{*}{CC-CV, \textbf{1, 100}}& \multirow{1}{*}{CC, -1, \textbf{0}}& 56\%/2125\\

\hline

\multirow{2}{*}{ \shortstack{Xu et al. \cite{xu2019study} \\(2/2)}} &  \multirow{2}{*}{\shortstack{LFP/C-Pouch-22Ah}} &  \multirow{1}{*}{25}&83\%& \multirow{1}{*}{CC, 0.5, \textbf{100}}& \multirow{1}{*}{CC, -0.5, \textbf{0}}& 79\%/1601\\

 &  &  & 83\%&\shortstack{Stationnary}, 1, 55  & \shortstack{Stationnary}, -1, 45 & 68\%/9955 \\

\hline

\multirow{4}{*}{ \shortstack{Zhang et al. \cite{zhang2020performance} \\(4/4)}} &  \multirow{4}{*}{\shortstack{LFP/C-Cylindrical-40Ah}} &  \multirow{4}{*}{25}&89\%& \multirow{1}{*}{CC-CV, \textbf{1}, 80}& \multirow{1}{*}{CC, \textbf{-1}, 30}& 54\%/3023\\

   &  &  & 92\%&CC-CV, 0.3, \textbf{100}  &CC, -0.3, \textbf{0} & 64\%/2999 \\
   &  &  & 94\%&CC-CV, 0.2, \textbf{100}  &CC, -0.2, \textbf{0} & 65\%/3000 \\
   &  &  & 86\%& CC, \textbf{2, 100}      &CC, \textbf{-2, 0 }  & 51\%/3000 \\
\hline

\multirow{1}{*}{ \shortstack{Zhang et al. \cite{zhang2021state} (1/2)}} &  \multirow{1}{*}{\shortstack{LFP/C-Cylindrical-40Ah}} &  \multirow{1}{*}{25}&96\%& \multirow{1}{*}{CC-CV, 0.5, \textbf{100}}& \multirow{1}{*}{CC, -0.5, \textbf{0}}& 60\%/2198\\

\hline

\multirow{6}{*}{ \shortstack{Braco et al. \cite{braco2020experimental} \\(6/6)}} &  \multirow{6}{*}{\shortstack{LMO+LNO/C-Pouch-66Ah}} &  \multirow{6}{*}{25}&\textbf{71\%}& \multirow{6}{*}{CC-CV, \textbf{1, 100}}& \multirow{6}{*}{CC, -1, \textbf{0}}& 32\%/2927 \\

       &  &  & \textbf{69\%}&  & & 34\%/2560 \\
       &  &  & \textbf{68\%}&  & & 29\%/2560 \\
       &  &  & \textbf{67\%}&  & & 32\%/2298 \\
       &  &  & \textbf{64\%}&  & & 38\%/3681 \\
       &  &  & \textbf{62\%}&  & & 30\%/2631 \\

\hline

\multirow{7}{*}{ \shortstack{Hosen et al. \cite{hosen2026data} \\(7/32)}} &  \multirow{7}{*}{\shortstack{?-Pouch-65Ah}} &  \multirow{7}{*}{25}&102\%& \multirow{1}{*}{?, 1, 80}& \multirow{1}{*}{?, -1, 20}& 95\%/1600 \\

&  &  & 102\%&\multirow{1}{*}{?, 0.3, 80}  & \multirow{1}{*}{?, -0.3, 20}& 90\%/1600 \\

&  &  & 102\%&\multirow{1}{*}{?, 0.3, 90}  & \multirow{1}{*}{?, -0.3, \textbf{10}}& 86\%/1600 \\
&  &  & 102\%&\multirow{1}{*}{?, 0.3, 90}  & \multirow{1}{*}{?, -0.3, \textbf{10}}& 87\%/1500 \\
&  &  & 102\%&\multirow{1}{*}{?, 1, 90}  & \multirow{1}{*}{?, -1, 20}& 92\%/1600 \\

&  &  & 102\%&\multirow{1}{*}{?, 0.8, 80}  & \multirow{1}{*}{?, -0.8, 20}& 85\%/1500 \\

&  &  & 94\%&\multirow{1}{*}{?, 0.4, 80}  & \multirow{1}{*}{?, -0.4, 20}& 84\%/1500 \\

\hline

   \end{tabular}
}
    \caption{Studies with long lifespan reused batteries}
    \label{tab:high_lifespan}
\end{table}

\begin{table}[h!]
    \centering
      \resizebox{\columnwidth}{!}{

    \begin{tabular}{cccccccc}
       \textbf{Reference}&\shortstack{\textbf{Battery} \\ \textbf{Chem.-Format-Qnom}} & \textbf{T°}& \shortstack{\textbf{Initial}\\\textbf{degradation}}& \shortstack{\textbf{Charge} \\ \textbf{Type, C-rate, }$\mathbf{SoC_{max}(\%)}$} & \shortstack{\textbf{Discharge} \\ \textbf{Type, C-rate, }$\mathbf{SoC_{min}(\%)}$} & \shortstack{\textbf{Final}\\ \textbf{SoH/cycle}} \\
\hline

\multirow{15}{*}{ \shortstack{Gao et al. \cite{gao2024comprehensive} \\(15/24)}} &  \multirow{15}{*}{\shortstack{NMC/C-Pouch-66Ah}} &  \multirow{15}{*}{?}&\textbf{76\%}& \multirow{15}{*}{?}& \multirow{15}{*}{?}& 65\%/1870 \\
       &  &  & 76\%&  & & 64\%/1880 \\
       &  &  & 74\%&  & & 62\%/1680 \\
       &  &  & 75\%&  & & 62\%/1500 \\
       &  &  & 75\%&  & & 62\%/1520 \\
       &  &  & 76\%&  & & 62\%/1500 \\
       &  &  & 76\%&  & & 63\%/1880 \\
       &  &  & 76\%&  & & 62\%/1870 \\
       &  &  & 76\%&  & & 67\%/1860 \\
       &  &  & 76\%&  & & 62\%/1620 \\
       &  &  & 76\%&  & & 62\%/1640 \\
       &  &  & 76\%&  & & 65\%/1870 \\
       &  &  & 76\%&  & & 62\%/1570 \\
       &  &  & 76\%&  & & 62\%/1500 \\
       &  &  & 76\%&  & & 63\%/1880 \\

\hline

\multirow{2}{*}{ \shortstack{Fan et al. \cite{fan2025generalized} \\(2/4)}} &  \multirow{2}{*}{\shortstack{NCA/C-Cylindrical-2.75Ah}} &  \multirow{2}{*}{25}&\textbf{71\%}& \multirow{2}{*}{CC, 0.5, \textbf{100}}& \multirow{2}{*}{Driving, -1, \textbf{0}}& 96\%/3120 \\
       &  &  & 96\%&  & & 84\%/3100 \\

\hline

\multirow{2}{*}{ \shortstack{Jiang et al. \cite{jiang2017recognition} \\(2/6)}} &  \multirow{2}{*}{\shortstack{LFP/C-Prismatic-60Ah}} &  \multirow{2}{*}{25}&\textbf{81\%}& \multirow{2}{*}{Stationnary, 1.25, 60}& \multirow{2}{*}{Stationnary, \textbf{-1.25}, 37}& 66\%/2240 \\
       &  &  & 84\%&  & & 70\%/2170 \\

\hline

\multirow{1}{*}{ \shortstack{Zhang et al. \cite{zhang2023state} \\(1/4)}} &  \multirow{1}{*}{\shortstack{LFP/C-Cylindrical-40Ah}} &  \multirow{1}{*}{25}&\textbf{96\%}& \multirow{1}{*}{CC-CV, 0.5, 100}& \multirow{1}{*}{CC, \textbf{-0.5}, 0}& 60\%/2200 \\

\hline
   \end{tabular}
}
\end{table}

\clearpage
\section{Effect of accelerating factors} \label{section:effect_accelerating_factors}

An analysis of experimental definitions helps to understand the above results. Table~\ref{tab:accelerating factors} shows the definition of accelerating factors used for this study.
\begin{table}[h!]
    \centering
    \begin{tabular}{cc}
    \hline
        Temperature & $<10^{\circ}C$ or $>30^{\circ}C$ \\
         
        $|Idch_{rate}|$ & $>1$ \\
         
        $|Ich_{rate}|$  & $>0.5$ \\
         
        $SoC_{min}$  & $<20$ \\
        
        $SoC_{max}$  & $>90$ \\
            \hline

    \end{tabular}
    \caption{Accelerating factors}
    \label{tab:accelerating factors}
\end{table}

\begin{figure}[h!]
    \centering
    \includegraphics[width=\columnwidth]{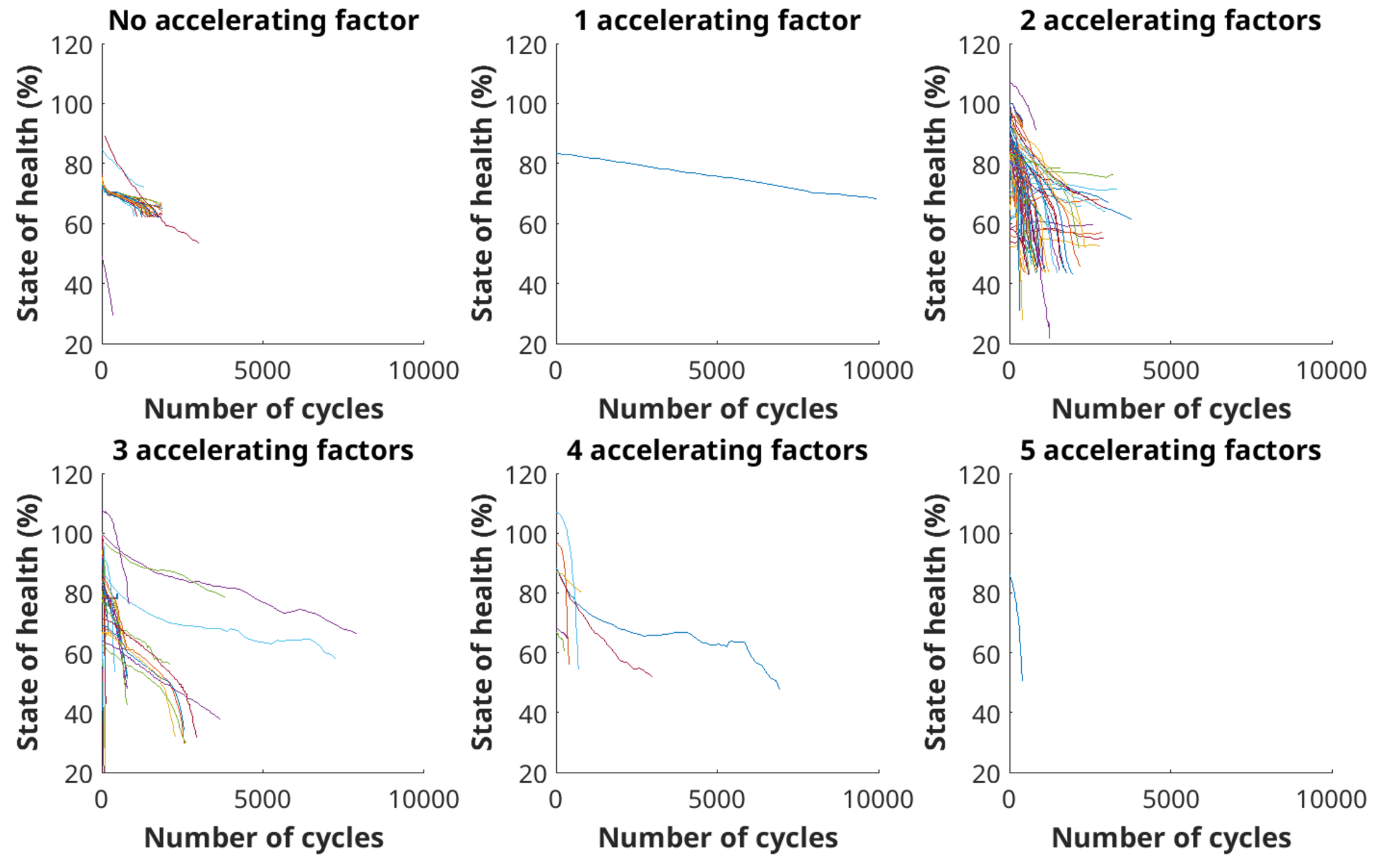}
    \caption{Influence of number of accelerating factors on degradation of reused batteries}
    \label{fig:accelerating_factors_influence}
\end{figure}

Figure~\ref{fig:accelerating_factors_influence} shows the degradation curves as a function of the number of accelerating factors used by experimenters.\\

This figure shows that no direct correlation can be made between the number of accelerating factors and reused battery lifespan. However, it suggests that a linear degradation occurs for batteries aged with one or zero accelerating factors. Faster degradation occurs when more accelerating factors, and aging knee is mostly occurring when 3 or more accelerating factors are used simultaneously. This result suggests that, when aged batteries are tested, attention should be given to ensure that the number of accelerating factors is limited. Also, current rate and state of charge range should be updated with measured capacity, as using nominal values may lead to faster degradation or even lead to abuse testing.

 In several studies where only one accelerating factor was used, the lifespan was also limited. This may be explained by the battery's initial life, as several authors have shown that degradation mechanisms occurring during the first life affect degradation during the second life~\cite{coron2020impact,tasnim2024assessing,wheeler2024postmortem}. Another possibility is that performance degradation was not sufficiently considered in the design of the experiments. Theoretical current rates and states of charge are commonly defined based on nominal capacity~\cite{hassini2023lithium}. For retired batteries, the actual battery capacity is often below the nominal value. This creates a significant discrepancy between the theoretical current rate used to design the experiment and the actual current rate perceived by the battery. For example, if a retired battery has a nominal capacity of 10 Ah and a real capacity of 8 Ah, the theoretical current rate of 1 C would be perceived by the battery as 1.25 C. It is also known that aging leads to an increase in resistance. This also leads to a discrepancy between the expected state of charge range and the range perceived by the battery. Accelerated aging and the appearance of knees in degradation curves may be caused by this experimental design problem. This raises a major question about the representativeness of battery aging experimentation design.  Further investigation on this matter is necessary before concluding that battery reuse is not viable.



%

 \bibliographystyle{elsarticle-num} 
 \bibliography{bibliography}

@article{sun2017,
  title={Tracing global lithium flow: A trade-linked material flow analysis},
  author={Sun, Xin and Hao, Han and Zhao, Fuquan and Liu, Zongwei},
  journal={Resources, Conservation and Recycling},
  volume={124},
  pages={50--61},
  year={2017},
  publisher={Elsevier}
}

@article{ohrelius2025polarization,
  title={Polarization effects and aging characterization of lithium-ion batteries for second life applications},
  author={Ohrelius, Mathilda and Lindbergh, G{\"o}ran and Lindstr{\"o}m, Rakel Wreland},
  journal={Journal of Power Sources},
  volume={655},
  pages={237931},
  year={2025},
  publisher={Elsevier}
}

@article{agudelo2023battery,
  title={Battery state-of-health estimation based on multiple charge and discharge features},
  author={Agudelo, Brian Ospina and Zamboni, Walter and Postiglione, Fabio and Monmasson, Eric},
  journal={Energy},
  volume={263},
  pages={125637},
  year={2023},
  publisher={Elsevier}
}

@article{fan2026consistency,
  title={Consistency sorting of retired lithium-ion batteries: From the perspective of maximizing remaining useful discharge},
  author={Fan, Wenjun and Wang, Xueyuan and Yuan, Yongjun and Zhou, Xiao and Jiang, Bo and Qian, Long and Wei, Xuezhe and Dai, Haifeng},
  journal={Applied Energy},
  volume={402},
  pages={127046},
  year={2026},
  publisher={Elsevier}
}

@article{tian2020fractional,
  title={Fractional order battery modelling methodologies for electric vehicle applications: Recent advances and perspectives},
  author={Tian, JinPeng and Xiong, Rui and Shen, WeiXiang and Sun, FengChun},
  journal={Science China Technological Sciences},
  volume={63},
  number={11},
  pages={2211--2230},
  year={2020},
  publisher={Springer}
}

@article{zhao2026evolution,
  title={Evolution of safety risks during battery degradation: mechanisms and implications},
  author={Zhao, Yiwen and Deng, Junjun and Sun, Zhenyu and Lin, Ni and Li, Weihan},
  journal={Journal of Energy Chemistry},
  year={2026},
  publisher={Elsevier}
}

@article{leonhardt2026effects,
  title={Effects of ageing on thermal stability and gas emissions in lithium-ion cells regarding second-life applications},
  author={Leonhardt, Robert and Lecocq, Amandine and von Nidda, Jonas Krug and Part, Florian and Jandric, Aleksander and Prenner, Stefanie and Berthaud, Maxime and Rabette, Cl{\'e}ment and Marlair, Guy and Schmidt, Anita and others},
  journal={Energy Reports},
  volume={15},
  pages={109240},
  year={2026},
  publisher={Elsevier}
}

@article{timke2024impact,
  title={Impact of State of Health (SOH) on the Thermal Safety of Lithium Ion Cells for Long 1st Life and 2nd Life Applications},
  author={Timke, Bennet and Winter, Martin and Niehoff, Philip},
  journal={Journal of The Electrochemical Society},
  volume={171},
  number={10},
  pages={100526},
  year={2024},
  publisher={IOP Publishing}
}

@article{ezeigwe2026degradation,
  title={Degradation of a Lithium-Ion Battery Cell for Enhanced First and Second Life: Effects of Temperature, Orientation, C-Rate and State of Charge},
  author={Ezeigwe, Ejikeme Raphael and Woll, Sivert A and Erichsen, Lene TB and Solberg, Simon BB and Hughes, Gareth M and Du, Wenjia and Lamb, Jacob J and Wind, Julia and Lian, Torleif and Shearing, Paul R and others},
  journal={Batteries},
  volume={12},
  number={4},
  pages={121},
  year={2026},
  publisher={MDPI}
}

@article{heymans2014economic,
  title={Economic analysis of second use electric vehicle batteries for residential energy storage and load-levelling},
  author={Heymans, Catherine and Walker, Sean B and Young, Steven B and Fowler, Michael},
  journal={Energy Policy},
  volume={71},
  pages={22--30},
  year={2014},
  publisher={Elsevier}
}

@article{pilali2026techno,
  title={Techno-Economic--Environmental Assessment of PV--Second-Life Battery Systems with PCM-Assisted Thermal Management in Low Peak-Load Residential Grids of Iran},
  author={Pilali, Ebrahim and Shafiei, Safiye and Salimi, Mohsen and Amidpour, Majid and Soltani, M},
  journal={Results in Engineering},
  pages={109038},
  year={2026},
  publisher={Elsevier}
}

@article{seger2023storage,
  title={A storage degradation model of Li-ion batteries to integrate ageing effects in the optimal management and design of an isolated microgrid},
  author={Seger, Pedro VH and Rigo-Mariani, R{\'e}my and Thivel, Pierre-Xavier and Riu, Delphine},
  journal={Applied Energy},
  volume={333},
  pages={120584},
  year={2023},
  publisher={Elsevier}
}

@article{chaud2025techno,
  title={Techno-economic impacts of battery replacement for different EV usage patterns},
  author={Chaud, Martin and German, Ronan and Hittinger, Eric and Bouscayrol, Alain and Castex, Elodie},
  journal={IEEE Open Journal of Vehicular Technology},
  volume={6},
  pages={1650--1660},
  year={2025},
  publisher={IEEE}
}

@article{bavand2022online,
  title={Online estimations of Li-ion battery SOC and SOH applicable to partial charge/discharge},
  author={Bavand, Amin and Khajehoddin, S Ali and Ardakani, Masoud and Tabesh, Ahmadreza},
  journal={IEEE Transactions on transportation electrification},
  volume={8},
  number={3},
  pages={3673--3685},
  year={2022},
  publisher={IEEE}
}

@article{roscher2018method,
  title={Method and measurement setup for battery state determination using optical effects in the electrode material},
  author={Roscher, Valentin and Riemschneider, Karl-Ragmar},
  journal={IEEE Transactions on Instrumentation and Measurement},
  volume={67},
  number={4},
  pages={735--744},
  year={2018},
  publisher={IEEE}
}

@article{harper2019recycling,
  title={Recycling lithium-ion batteries from electric vehicles},
  author={Harper, Gavin and Sommerville, Roberto and Kendrick, Emma and Driscoll, Laura and Slater, Peter and Stolkin, Rustam and Walton, Allan and Christensen, Paul and Heidrich, Oliver and Lambert, Simon and others},
  journal={nature},
  volume={575},
  number={7781},
  pages={75--86},
  year={2019},
  publisher={Nature Publishing Group UK London}
}

@article{hertel2024towards,
  title={Towards a green electromobility transition: a systematic review of the state of the art on electric vehicle battery systems disassembly},
  author={Hertel, Dominik and Braeunig, Gerald and Thuerer, Matthias},
  journal={Journal of Manufacturing Systems},
  volume={74},
  pages={387--396},
  year={2024},
  publisher={Elsevier}
}

@Misc{USABC1996,
  author       = {{USABC}},
  editor       = {USABC},
  title        = {USABC Electric Vehicle Battery Test Procedures Manual, Revision 2. Tech. rep.},
  year         = {1996},
}

@inproceedings{jeanneret2025ageing,
  title={Ageing aware cell balancing in electric vehicle batteries},
  author={Jeanneret, Bruno and Redondo-Iglesias, Eduardo},
  booktitle={38th Electric Vehicle Symposium (EVS38)},
  pages={5--p},
  year={2025}
}

@article{ali2019online,
  title={Online remaining useful life prediction for lithium-ion batteries using partial discharge data features},
  author={Ali, Muhammad Umair and Zafar, Amad and Nengroo, Sarvar Hussain and Hussain, Sadam and Park, Gwan-Soo and Kim, Hee-Je},
  journal={Energies},
  volume={12},
  number={22},
  pages={4366},
  year={2019},
  publisher={MDPI}
}

@inproceedings{riviere2015lifepo4,
  title={LiFePO4 battery state of health online estimation using electric vehicle embedded incremental capacity analysis},
  author={Riviere, Elie and Venet, Pascal and Sari, Ali and Meniere, Frederic and Bultel, Yann},
  booktitle={2015 IEEE Vehicle Power and Propulsion Conference (VPPC)},
  pages={1--6},
  year={2015},
  organization={IEEE}
}

@TechReport{ipcc2022,
  author      = {IPCC},
  title       = {Climate Change 2022: Mitigation of Climate Change. Contribution of Working Group III to the Sixth Assessment Report of the Intergovernmental Panel on Climate Change},
  institution = {IPCC},
  year        = {2022},
}

@article{piombo2025novel,
  title={A novel application-aware retired lithium-ion batteries regrouping approach to enable improved second life},
  author={Piombo, Gabriele and Niri, Mona Faraji and Marco, James},
  journal={Journal of Cleaner Production},
  volume={489},
  pages={144526},
  year={2025},
  publisher={Elsevier}
}

@inproceedings{zhang2025prediction,
  title={Prediction and Data Analysis of Decommissioned Quantity of New Energy Vehicle Traction Batteries through Weibull Model},
  author={Zhang, Yizhen and Hu, Song and Zhao, Fengyun and Wang, Pan},
  booktitle={2025 IEEE International Conference on Electronics, Energy Systems and Power Engineering (EESPE)},
  pages={707--713},
  year={2025},
  organization={IEEE}
}

@article{zhao2025lstm,
  title={LSTM-SVM-Weibull modeling for decommissioning amount prediction of power batteries based on attention mechanism and ISPBO algorithm},
  author={Zhao, Mengna and Chen, Shiping},
  journal={Applied Intelligence},
  volume={55},
  number={5},
  pages={305},
  year={2025},
  publisher={Springer}
}

@article{hu2025mathematical,
  title={Mathematical Model and Prediction Analysis of Automobile Power Battery Withdrawal Considering Weibull Distribution},
  author={Hu, Song and Jiang, Xiaotong and Zhao, Fengyun and Wang, Jingyi and Wang, Rui and Wu, Meng},
  journal={Mathematics and Computer Science: Research Updates Vol. 3},
  pages={100--110},
  year={2025},
  publisher={BP International}
}

@article{qaadan2025weibull,
  title={Weibull Distribution-Informed Neural Networks (WDINNs): a probabilistic framework for enhanced degradation prediction in LIBs},
  author={Qaadan, Sahar and Alshare, Aiman and Popp, Alexander and Alazrai, Rami and Daoud, Mohammad I and Ali, Mostafa Z and Schmuelling, Benedikt},
  journal={Ionics},
  pages={1--18},
  year={2025},
  publisher={Springer}
}

@article{zhu2025towards,
  title={Towards a circular supply chain for retired electric vehicle batteries: A systematic literature review},
  author={Zhu, He and Hu, Jiayao and Yang, Ying},
  journal={International Journal of Production Economics},
  pages={109556},
  year={2025},
  publisher={Elsevier}
}

@article{dubarry2020perspective,
  title={Perspective on state-of-health determination in lithium-ion batteries},
  author={Dubarry, Matthieu and Baure, George and Anse{\'a}n, David},
  journal={Journal of Electrochemical Energy Conversion and Storage},
  volume={17},
  number={4},
  pages={044701},
  year={2020},
  publisher={American Society of Mechanical Engineers}
}

@article{lee2021lithium,
  title={Lithium battery SOH monitoring and an SOC estimation algorithm based on the SOH result},
  author={Lee, Jong-Hyun and Lee, In-Soo},
  journal={Energies},
  volume={14},
  number={15},
  pages={4506},
  year={2021},
  publisher={MDPI}
}

@article{huang2017online,
  title={An online SOC and SOH estimation model for lithium-ion batteries},
  author={Huang, Shyh-Chin and Tseng, Kuo-Hsin and Liang, Jin-Wei and Chang, Chung-Liang and Pecht, Michael G},
  journal={Energies},
  volume={10},
  number={4},
  pages={512},
  year={2017},
  publisher={MDPI}
}

@article{rochefort2025forecasting,
  title={Forecasting the remaining useful life (RUL) of metallized film capacitors affected by the self-healing phenomenon},
  author={Rochefort, Claire and Venet, Pascal and Clerc, Guy and Sari, Ali and Mitova, Radoslava and Wang, Miao-Xin},
  journal={IEEE Transactions on Dielectrics and Electrical Insulation},
  year={2025},
  publisher={IEEE}
}

@article{el2019lifetime,
  title={Lifetime prediction of lithium-ion capacitors based on accelerated aging tests},
  author={El Ghossein, Nagham and Sari, Ali and Venet, Pascal},
  journal={Batteries},
  volume={5},
  number={1},
  pages={28},
  year={2019},
  publisher={MDPI}
}

@article{ayob2022soc,
  title={SOC, SOH and RUL estimation for supercapacitor management system: Methods, implementation factors, limitations and future research improvements},
  author={Ayob, Afida and Ansari, Shaheer and Lipu, Molla Shahadat Hossain and Hussain, Aini and Saad, Mohamad Hanif Md},
  journal={Batteries},
  volume={8},
  number={10},
  pages={189},
  year={2022},
  publisher={MDPI}
}

@article{jouin2016estimating,
  title={Estimating the end-of-life of PEM fuel cells: Guidelines and metrics},
  author={Jouin, Marine and Bressel, Mathieu and Morando, Simon and Gouriveau, Rafael and Hissel, Daniel and P{\'e}ra, Marie-C{\'e}cile and Zerhouni, Noureddine and Jemei, Samir and Hilairet, Mickael and Bouamama, Belkacem Ould},
  journal={Applied energy},
  volume={177},
  pages={87--97},
  year={2016},
  publisher={Elsevier}
}

@article{hua2022review,
  title={A review on lifetime prediction of proton exchange membrane fuel cells system},
  author={Hua, Zhiguang and Zheng, Zhixue and Pahon, Elodie and P{\'e}ra, Marie-C{\'e}cile and Gao, Fei},
  journal={Journal of Power Sources},
  volume={529},
  pages={231256},
  year={2022},
  publisher={Elsevier}
}

@inproceedings{vitols2018cell,
  title={Cell capacity dispersion analysis based battery pack design},
  author={Vitols, Kristaps and Grinfogels, Edgars and Nikonorovs, Deniss},
  booktitle={2018 IEEE 6th Workshop on Advances in Information, Electronic and Electrical Engineering (AIEEE)},
  pages={1--5},
  year={2018},
  organization={IEEE}
}

@article{al2022techno,
  title={Techno-economic feasibility of retired electric-vehicle batteries repurpose/reuse in second-life applications: A systematic review},
  author={Al-Alawi, Mohammed Khalifa and Cugley, James and Hassanin, Hany},
  journal={Energy and Climate Change},
  volume={3},
  pages={100086},
  year={2022},
  publisher={Elsevier}
}

@article{junior2023towards,
  title={Towards a business model for second-life batteries: Barriers, opportunities, uncertainties, and technologies},
  author={J{\'u}nior, Carlos Ant{\^o}nio Rufino and Sanseverino, Eleonora Riva and Gallo, Pierluigi and Koch, Daniel and Kotak, Yash and Schweiger, Hans-Georg and Zanin, Hudson},
  journal={Journal of Energy Chemistry},
  volume={78},
  pages={507--525},
  year={2023},
  publisher={Elsevier}
}

@techreport{cready2003technical,
  title={Technical and economic feasibility of applying used EV batteries in stationary applications},
  author={Cready, Erin and Lippert, John and Pihl, Josh and Weinstock, Irwin and Symons, Philip},
  year={2003},
  institution={Sandia National Lab.}
}

@article{faessler2021stationary,
  title={Stationary, second use battery energy storage systems and their applications: A research review},
  author={Faessler, Bernhard},
  journal={Energies},
  volume={14},
  number={8},
  pages={2335},
  year={2021},
  publisher={MDPI}
}

@article{drevon2017intercoder,
  title={Intercoder reliability and validity of WebPlotDigitizer in extracting graphed data},
  author={Drevon, Daniel and Fursa, Sophie R and Malcolm, Allura L},
  journal={Behavior modification},
  volume={41},
  number={2},
  pages={323--339},
  year={2017},
  publisher={Sage Publications Sage CA: Los Angeles, CA}
}

@article{wraalsen2021circular,
  title={Circular business models for lithium-ion batteries-Stakeholders, barriers, and drivers},
  author={Wr{\aa}lsen, Benedikte and Prieto-Sandoval, Vanessa and Mejia-Villa, Andres and O'Born, Reyn and Hellstr{\"o}m, Magnus and Faessler, Bernhard},
  journal={Journal of Cleaner Production},
  volume={317},
  pages={128393},
  year={2021},
  publisher={Elsevier}
}

@article{kampker2023identification,
  title={Identification of challenges for second-life battery systems—A literature review},
  author={Kampker, Achim and Heimes, Heiner Hans and Offermanns, Christian and Vienenk{\"o}tter, Janis and Frank, Merlin and Holz, Daniel},
  journal={World Electric Vehicle Journal},
  volume={14},
  number={4},
  pages={80},
  year={2023},
  publisher={MDPI}
}

@article{zhao2021review,
  title={A review on battery market trends, second-life reuse, and recycling},
  author={Zhao, Yanyan and Pohl, Oliver and Bhatt, Anand I and Collis, Gavin E and Mahon, Peter J and R{\"u}ther, Thomas and Hollenkamp, Anthony F},
  journal={Sustainable Chemistry},
  volume={2},
  number={1},
  pages={167--205},
  year={2021},
  publisher={MDPI}
}

@article{baars2021,
  title={Circular economy strategies for electric vehicle batteries reduce reliance on raw materials},
  author={Baars, Joris and Domenech, Teresa and Bleischwitz, Raimund and Melin, Hans Eric and Heidrich, Oliver},
  journal={Nature Sustainability},
  volume={4},
  number={1},
  pages={71--79},
  year={2021},
  publisher={Nature Publishing Group UK London}
}

@article{cheng2022,
  title={Sustainability evaluation of second-life battery applications in grid-connected PV-battery systems},
  author={Cheng, Ming and Ran, Aihua and Zheng, Xueling and Zhang, Xuan and Wei, Guodan and Zhou, Guangmin and Sun, Hongbin},
  journal={Journal of Power Sources},
  volume={550},
  pages={232132},
  year={2022},
  publisher={Elsevier}
}

@article{bridge2022,
  title={Towards the lithium-ion battery production network: Thinking beyond mineral supply chains},
  author={Bridge, Gavin and Faigen, Erika},
  journal={Energy Research \& Social Science},
  volume={89},
  pages={102659},
  year={2022},
  publisher={Elsevier}
}

@inproceedings{wolfs2010economic,
  title={An economic assessment of “second use” lithium-ion batteries for grid support},
  author={Wolfs, Peter},
  booktitle={2010 20th Australasian Universities Power Engineering Conference},
  pages={1--6},
  year={2010},
  organization={IEEE}
}

@techreport{burke2009performance,
    author = {Burke, Andrew},
    title = {Performance, charging, and second-use considerations for lithium batteries for plug-in electric vehicles},
    institution = {UC Davis Institute of Transportation},
    year = {2009}
}

@article{casals2017electric,
  title={Electric vehicle battery reuse: Preparing for a second life},
  author={Casals, Lluc Canals and Garc{\'\i}a, Beatriz Amante and Cremades, L{\'a}zaro V},
  journal={Journal of Industrial Engineering and Management},
  volume={10},
  number={2},
  pages={266--285},
  year={2017}
}

@article{cusenza2019reuse,
  title={Reuse of electric vehicle batteries in buildings: An integrated load match analysis and life cycle assessment approach},
  author={Cusenza, Maria Anna and Guarino, Francesco and Longo, Sonia and Mistretta, Marina and Cellura, Maurizio},
  journal={Energy and Buildings},
  volume={186},
  pages={339--354},
  year={2019},
  publisher={Elsevier}
}

@article{rallo2020economic,
  title={Economic analysis of the disassembling activities to the reuse of electric vehicles Li-ion batteries},
  author={Rallo, H{\'e}ctor and Benveniste, G and Gestoso, I and Amante, B},
  journal={Resources, conservation and recycling},
  volume={159},
  pages={104785},
  year={2020},
  publisher={Elsevier}
}

@article{lander2023breaking,
  title={Breaking it down: A techno-economic assessment of the impact of battery pack design on disassembly costs},
  author={Lander, Laura and Tagnon, Chris and Nguyen-Tien, Viet and Kendrick, Emma and Elliott, Robert JR and Abbott, Andrew P and Edge, Jacqueline S and Offer, Gregory J},
  journal={Applied Energy},
  volume={331},
  pages={120437},
  year={2023},
  publisher={Elsevier}
}

@article{bai2024techno,
  title={Techno-economic assessment of isolated micro-grids with second-life batteries: A reliability-oriented iterative design framework},
  author={Bai, Hanyu and Lei, Shunbo and Geng, Sijia and Hu, Xiaosong and Li, Zhaojian and Song, Ziyou},
  journal={Applied Energy},
  volume={364},
  pages={123068},
  year={2024},
  publisher={Elsevier}
}

@article{chauhan2024energy,
  title={Energy storage potential of used electric vehicle batteries for supporting renewable energy generation in India},
  author={Chauhan, Rajat and Santran, Ram and Obrecht, Matevz and Singh, Rhythm},
  journal={Energy for Sustainable Development},
  volume={81},
  pages={101513},
  year={2024},
  publisher={Elsevier}
}

@article{hassini2023lithium,
  title={Lithium--ion battery data: from production to prediction},
  author={Hassini, Marwan and Redondo-Iglesias, Eduardo and Venet, Pascal},
  journal={Batteries},
  volume={9},
  number={7},
  pages={385},
  year={2023},
  publisher={MDPI}
}

@article{rohatgi2021webplotdigitizer,
  title={WebPlotDigitizer (version 4.2)[Computer software]},
  author={Rohatgi, A},
  journal={Pacifica, CA,[Updated August, 2021]. Available from: https://automeris. io/webplotdigitizer},
  year={2021}
}

@article{martinez2018,
  title={Battery second life: Hype, hope or reality? A critical review of the state of the art},
  author={Martinez-Laserna, Egoitz and Gandiaga, I{\~n}igo and Sarasketa-Zabala, Elixabet and Badeda, Julia and Stroe, D-I and Swierczynski, Maciej and Goikoetxea, Ander},
  journal={Renewable and Sustainable Energy Reviews},
  volume={93},
  pages={701--718},
  year={2018},
  publisher={Elsevier}
}

@Article{ahmadi2014,
  author  = {Ahmadi, Leila and Fowler, Michael and Young, Steven and al.},
  title   = {Energy efficiency of Li-ion battery packs re-used in stationary power applications},
  journal = {Sustainable Energy Technol. Assess.},
  year    = {2014},
  volume  = {8},
  pages   = {9--17},
  issn    = {2213-1388},
}

@InProceedings{wieland2017,
  author    = {Wieland, Marco and Gerhard, Sebastian and Schmidt, Achim},
  title     = {Model-based Lifetime Analysis of 2\textsuperscript{nd}-life Lithium-Ion Battery Storage Systems for Stationary Applications},
  booktitle = {NEIS},
  year      = {2017},
  location  = {Wiesbaden, Germany},
  isbn      = {978-3-658-15028-0},
  pages     = {175--181},
}

@Article{sathre2015,
  author    = {Sathre, Roger and Scown, Corinne D and Kavvada, Olga and al.},
  title     = {Energy and climate effects of second-life use of electric vehicle batteries in California through 2050},
  journal   = {J. Power Sources},
  year      = {2015},
  volume    = {288},
  pages     = {82--91},
  issn      = {0378-7753},
}

@Article{debnath2014,
  author    = {Debnath, Uttam Kumar and Ahmad, Iftekhar and Habibi, Daryoush},
  title     = {Quantifying economic benefits of second life batteries of gridable vehicles in the smart grid},
  journal   = {International Journal of Electrical Power \& Energy Systems},
  year      = {2014},
  volume    = {63},
  pages     = {577--587},
  issn      = {0142-0615},
}

@Article{ieee20,
  author  = {IEEE},
  title   = {{IEEE} Recommended Practice for Sizing Lead-Acid Batteries for Stationary Applications},
  journal = {IEEE Std 485-2020},
  year    = {2020},
  pages   = {1--69},
}

@Article{attia22,
  author    = {Attia, Peter M and Bills, Alexander and Planella, Ferran Brosa and al.},
  title     = {“Knees” in lithium-ion battery aging trajectories},
  journal   = {J. Electrochem. Soc.},
  year      = {2022},
  volume    = {169},
  number    = {6},
  pages     = {060517},
  publisher = {IOP Publishing},
}

@Article{paul13,
  author    = {Paul, Sebastian and Diegelmann, Christian and Kabza, Herbert and al.},
  title     = {Analysis of ageing inhomogeneities in lithium-ion battery systems},
  journal   = {J. Power Sources},
  year      = {2013},
  volume    = {239},
  pages     = {642--650},
  publisher = {Elsevier},
}

@Article{rumpf17,
  author    = {Rumpf, Katharina and Naumann, Maik and Jossen, Andreas},
  title     = {Experimental investigation of parametric cell-to-cell variation and correlation based on 1100 commercial lithium-ion cells},
  journal   = {Journal of Energy Storage},
  year      = {2017},
  volume    = {14},
  pages     = {224--243},
  publisher = {Elsevier},
}

@article{haram2021,
  title={Feasibility of utilising second life EV batteries: Applications, lifespan, economics, environmental impact, assessment, and challenges},
  author={Haram, Mohammed Hussein Saleh Mohammed and Lee, Jia Woon and Ramasamy, Gobbi and Ngu, Eng Eng and Thiagarajah, Siva Priya and Lee, Yuen How},
  journal={Alexandria Engineering Journal},
  volume={60},
  number={5},
  pages={4517--4536},
  year={2021},
  publisher={Elsevier}
}

@article{li2021,
  title={On the feature selection for battery state of health estimation based on charging--discharging profiles},
  author={Li, Yuanyuan and Stroe, Daniel-Ioan and Cheng, Yuhua and Sheng, Hanmin and Sui, Xin and Teodorescu, Remus},
  journal={Journal of Energy Storage},
  volume={33},
  pages={102122},
  year={2021},
  publisher={Elsevier}
}

@article{deng2021,
  title={A rest-time-based prognostic model for remaining useful life prediction of lithium-ion battery},
  author={Deng, Liming and Shen, Wenjing and Wang, Hongfei and Wang, Shuqiang},
  journal={Neural Computing and Applications},
  volume={33},
  pages={2035--2046},
  year={2021},
  publisher={Springer}
}

@article{zhou2020,
  title={Hybrid genetic algorithm method for efficient and robust evaluation of remaining useful life of supercapacitors},
  author={Zhou, Yanting and Wang, Yanan and Wang, Kai and Kang, Le and Peng, Fei and Wang, Licheng and Pang, Jinbo},
  journal={Applied Energy},
  volume={260},
  pages={114169},
  year={2020},
  publisher={Elsevier}
}

@article{severson2019,
  title={Data-driven prediction of battery cycle life before capacity degradation},
  author={Severson, Kristen A and Attia, Peter M and Jin, Norman and Perkins, Nicholas and Jiang, Benben and Yang, Zi and Chen, Michael H and Aykol, Muratahan and Herring, Patrick K and Fraggedakis, Dimitrios and others},
  journal={Nature Energy},
  volume={4},
  number={5},
  pages={383--391},
  year={2019},
  publisher={Nature Publishing Group UK London}
}

@article{wildfeuer2023,
  title={Experimental degradation study of a commercial lithium-ion battery},
  author={Wildfeuer, Leo and Karger, Alexander and Ayg{\"u}l, Deniz and Wassiliadis, Nikolaos and Jossen, Andreas and Lienkamp, Markus},
  journal={Journal of Power Sources},
  volume={560},
  pages={232498},
  year={2023},
  publisher={Elsevier}
}

@Article{saxen15,
  author    = {Saxena, Samveg and Le Floch, Caroline and MacDonald, Jason and al.},
  title     = {Quantifying EV battery end-of-life through analysis of travel needs with vehicle powertrain models},
  journal   = {J. Power Sources},
  year      = {2015},
  volume    = {282},
  pages     = {265--276},
  publisher = {Elsevier},
}

@Article{arrin21,
  author       = {Arrinda, Mikel and Oyarbide, Mikel and Macicior, Haritz and al.},
  title        = {Application dependent end-of-life threshold definition methodology for batteries in electric vehicles},
  journal      = {MDPI},
  journaltitle = {Batteries},
  year         = {2021},
  volume       = {7},
  number       = {1},
  pages        = {12},
}

@Article{casal19a,
  author    = {Canals Casals, Lluc and Rodr{\'\i}guez, Marta and Corchero, Cristina and al.},
  title     = {Evaluation of the End-of-Life of Electric Vehicle Batteries According to the State-of-Health},
  journal   = {World Electric Vehicle Journal},
  year      = {2019},
  volume    = {10},
  number    = {4},
  month     = oct,
  pages     = {63},
  issn      = {2032-6653},
  publisher = {MDPI AG},
}

@Article{hossa19,
  author  = {Hossain, Eklas and Murtaugh, Darren and Mody, Jaisen and al.},
  title   = {A Comprehensive Review on Second-Life Batteries: Current State, Manufacturing Considerations, Applications, Impacts, Barriers Potential Solutions, Business Strategies, and Policies},
  pages   = {73215--73252},
  volume  = {7},
  journal = {IEEE Access},
  year    = {2019},
}

@article{anselma2021,
  title={Battery state-of-health sensitive energy management of hybrid electric vehicles: Lifetime prediction and ageing experimental validation},
  author={Anselma, Pier Giuseppe and Kollmeyer, Phillip and Lempert, Jeremy and Zhao, Ziyu and Belingardi, Giovanni and Emadi, Ali},
  journal={Applied Energy},
  volume={285},
  pages={116440},
  year={2021},
  publisher={Elsevier}
}

@article{salim2019,
  title={Drivers, barriers and enablers to end-of-life management of solar photovoltaic and battery energy storage systems: A systematic literature review},
  author={Salim, Hengky K and Stewart, Rodney A and Sahin, Oz and Dudley, Michael},
  journal={Journal of cleaner production},
  volume={211},
  pages={537--554},
  year={2019},
  publisher={Elsevier}
}

@article{gomez2023,
  title={Optimal Design of Photovoltaic Domestic Installations Considering Second-Purpose Batteries},
  author={G{\'o}mez-Gonz{\'a}lez, Manuel and Tostado-V{\'e}liz, Marcos and Valverde, Manuel and Jurado, Francisco},
  journal={IEEE Access},
  year={2023},
  publisher={IEEE}
}

@article{xu2020,
  title={Future material demand for automotive lithium-based batteries},
  author={Xu, Chengjian and Dai, Qiang and Gaines, Linda and Hu, Mingming and Tukker, Arnold and Steubing, Bernhard},
  journal={Communications Materials},
  volume={1},
  number={1},
  pages={99},
  year={2020},
  publisher={Nature Publishing Group UK London}
}

@article{chai2021,
  title={An evaluation framework for second-life EV/PHEV battery application in power systems},
  author={Chai, Songjian and Xu, Ning Zhou and Niu, Ming and Chan, Ka Wing and Chung, Chi Yung and Jiang, Hui and Sun, Yuxin},
  journal={IEEE access},
  volume={9},
  pages={152430--152441},
  year={2021},
  publisher={IEEE}
}

@article{fallah2024,
  title={Exploring the state of health of electric vehicle batteries at end of use; hierarchical waste flow analysis to determine the recycling and reuse potential},
  author={Fallah, Narjes and Fitzpatrick, Colin},
  journal={Journal of Remanufacturing},
  volume={14},
  number={1},
  pages={155--168},
  year={2024},
  publisher={Springer}
}

@article{ma2024,
  title={Pathway decisions for reuse and recycling of retired lithium-ion batteries considering economic and environmental functions},
  author={Ma, Ruifei and Tao, Shengyu and Sun, Xin and Ren, Yifang and Sun, Chongbo and Ji, Guanjun and Xu, Jiahe and Wang, Xuecen and Zhang, Xuan and Wu, Qiuwei and others},
  journal={Nature Communications},
  volume={15},
  number={1},
  pages={7641},
  year={2024},
  publisher={Nature Publishing Group UK London}
}

@article{le2020,
  title={Comparative life cycle assessment of lithium-ion battery chemistries for residential storage},
  author={Le Varlet, Thomas and Schmidt, Oliver and Gambhir, Ajay and Few, Sheridan and Staffell, Iain},
  journal={Journal of Energy storage},
  volume={28},
  pages={101230},
  year={2020},
  publisher={Elsevier}
}

@article{marques2019,
  title={Comparative life cycle assessment of lithium-ion batteries for electric vehicles addressing capacity fade},
  author={Marques, Pedro and Garcia, Rita and Kulay, Luiz and Freire, Fausto},
  journal={Journal of Cleaner Production},
  volume={229},
  pages={787--794},
  year={2019},
  publisher={Elsevier}
}

@InProceedings{abdel17,
  author    = {Abdel-Monem, Mohamed and Hegazy, Omar and Omar, Noshin and al.},
  title     = {Lithium-ion Batteries: Comprehensive Technical Analysis of Second-Life Batteries for Smart Grid Applications},
  booktitle = {IEEE ECCE},
  pages={P--1},
  year      = {2017},
  location  = {Warsaw, Poland},
  address   = {Warsaw, Poland},
}

@article{amuta2023state,
  title={State of Health Assessment of Spent Lithium--Ion Batteries Based on Voltage Integral during the Constant Current Charge},
  author={Amuta, Ote and Kowal, Julia},
  journal={Batteries},
  volume={9},
  number={11},
  pages={537},
  year={2023},
  publisher={MDPI}
}

@article{attidekou2020methodologies,
  title={Methodologies for large-size pouch lithium-ion batteries end-of-life gateway detection in the second-life application},
  author={Attidekou, Pierrot S and Milojevic, Zoran and Muhammad, Musbahu and Ahmeid, Mohamed and Lambert, Simon and Das, Prodip K},
  journal={Journal of The Electrochemical Society},
  volume={167},
  number={16},
  pages={160534},
  year={2020},
  publisher={IOP Publishing}
}

@InProceedings{hassini2025,
author={Hassini, Marwan and Mintsa-Eya, Colette and Redondo-Iglesias, Eduardo and Venet, Pascal},
title={Influence of Cell Position on the Capacity of Retired Batteries: Experimental and Statistical Studies},
booktitle={IECON},
year={2025},
organization={IEEE},
location={Madrid, Spain},
pages={1-5},
}

@InProceedings{azizi22,
  author       = {Azizighalehsari, Seyedreza and Venugopal, Prasanth and Singh, Deepak Pratap and al.},
  title        = {Performance Evaluation of Retired Lithium-ion Batteries for Echelon Utilization},
  booktitle    = {IECON},
  year         = {2022},
  organization = {IEEE},
  location     = {Brussels, Belgium},
  pages        = {1--6},
  address      = {Brussels, Belgium},
}

@Article{bauma18,
  author    = {Baumann, Michael and Wildfeuer, Leo and Rohr, Stephan and al.},
  title     = {Parameter variations within Li-Ion battery packs--Theoretical investigations and experimental quantification},
  journal   = {Journal of Energy Storage},
  year      = {2018},
  volume    = {18},
  pages     = {295--307},
  publisher = {Elsevier},
}

@Article{braco22,
  author    = {Braco, Elisa and San Martin, Idoia and Sanchis, Pablo and al.},
  title     = {Health indicator selection for state of health estimation of second-life lithium-ion batteries under extended ageing},
  journal   = {Journal of Energy Storage},
  year      = {2022},
  volume    = {55},
  pages     = {105366},
  publisher = {Elsevier},
}

@article{braco2020experimental,
  title={Experimental assessment of cycling ageing of lithium-ion second-life batteries from electric vehicles},
  author={Braco, Elisa and San Mart{\'\i}n, Idoia and Berrueta, Alberto and Sanchis, Pablo and Urs{\'u}a, Alfredo},
  journal={Journal of Energy Storage},
  volume={32},
  pages={101695},
  year={2020},
  publisher={Elsevier}
}

@inproceedings{brand2012aging,
  title={Ageing inhomogeneity of long-term used BEV-batteries and their reusability for 2nd-life applications},
  author={Brand, Martin J and Quinger, Daniel and Walder, Georg and Jossen, Andreas and Lienkamp, Markus},
  booktitle={Electric Vehicle Symposium and Exhibition (EVS 26)},
  year={2012},
  pages={1-7},
}

@inproceedings{burke2013life,
  title={Life cycle testing of lithium batteries for fast charging and second-use applications},
  author={Burke, Andrew and Miller, Marshall},
  booktitle={2013 World Electric Vehicle Symposium and Exhibition (EVS27)},
  pages={1--10},
  year={2013},
  organization={IEEE}
}

@article{camboim2024state,
  title={State of health estimation of second-life batteries through electrochemical impedance spectroscopy and dimensionality reduction},
  author={Camboim, Marcelo Miranda and Moreira, Aghatta Cioquetta and Maria de F{\'a}tima, NC and Beck, Raul Fernando and Arioli, Vitor Torquato and Omae, Camila and Ding, Hongwu},
  journal={Journal of Energy Storage},
  volume={78},
  pages={110063},
  year={2024},
  publisher={Elsevier}
}

@inproceedings{moreira2022fast,
  title={A Fast and Automatic Methodology for Selecting Lithium-ion Cells for Second-Life Batteries},
  author={Moreira, Aghatta Cioquetta and Camboim, Marcelo Miranda and Beck, Raul Fernando and Maria de Fatima, NC and Arioli, Vitor Torquato and Moura, Jonathan and Omae, Camila and Ding, Hongwu},
  booktitle={2022 IEEE PES Innovative Smart Grid Technologies Conference Europe (ISGT-Europe)},
  pages={1--5},
  year={2022},
  organization={IEEE}
}

@article{ahmei22,
  author    = {Ahmeid, Mohamed and Muhammad, Musbahu and Lambert, Simon and al.},
  title     = {A rapid capacity evaluation of retired electric vehicle battery modules using partial discharge test},
  journal   = {Journal of Energy Storage},
  year      = {2022},
  volume    = {50},
  pages     = {104562},
  publisher = {Elsevier}
}

@article{cao2024second,
  title={Second-Life Assessment of Commercial LiFePO4 Batteries Retired from EVs},
  author={Cao, Zhi and Gao, Wei and Fu, Yuhong and Turchiano, Christopher and Vosoughi Kurdkandi, Naser and Gu, Jing and Mi, Chris},
  journal={Batteries},
  volume={10},
  number={9},
  pages={306},
  year={2024},
  publisher={MDPI}
}

@inproceedings{castro2022air,
  title={From the Air to the Ground: An Experimental Approach to Assess LiPo Batteries for a Second Life},
  author={Castro, Julio and Flores, Carolina and Gonzalez, Diego and Quintero, Vanessa and Perez, Aramis},
  booktitle={2022 Prognostics and Health Management Conference (PHM-2022 London)},
  pages={489--494},
  year={2022},
  organization={IEEE}
}

@article{chang2022experimental,
  title={Experimental assessment of the discharge characteristics of multi-type retired lithium-ion batteries in parallel for echelon utilization},
  author={Chang, Long and Ma, Chen and Zhang, Yulong and Li, Hongyu and Xiao, Linjing},
  journal={Journal of Energy Storage},
  volume={55},
  pages={105539},
  year={2022},
  publisher={Elsevier}
}

@article{chen2017degradation,
  title={A degradation-based sorting method for lithium-ion battery reuse},
  author={Chen, Hao and Shen, Julia},
  journal={PLoS One},
  volume={12},
  number={10},
  pages={e0185922},
  year={2017},
  publisher={Public Library of Science San Francisco, CA USA}
}

@article{chen2022efficient,
  title={An efficient regrouping method of retired lithium-ion iron phosphate batteries based on incremental capacity curve feature extraction for echelon utilization},
  author={Chen, Zuhang and Deng, Yelin and Li, Honglei and Liu, Weiwei},
  journal={Journal of Energy Storage},
  volume={56},
  pages={105917},
  year={2022},
  publisher={Elsevier}
}

@article{chung2021charge,
  title={Charge and discharge profiles of repurposed LiFePO4 batteries based on the UL 1974 standard},
  author={Chung, Hsien-Ching},
  journal={Scientific Data},
  volume={8},
  number={1},
  pages={165},
  year={2021},
  publisher={Nature Publishing Group UK London}
}

@article{drallmeier2022challenges,
  title={Challenges of a Fast Diagnostic to Inform Screening of Retired Batteries},
  author={Drallmeier, Joseph A and Wong, Clement and Solbrig, Charles E and Siegel, Jason B and Stefanopoulou, Anna G},
  journal={IFAC-PapersOnLine},
  volume={55},
  number={24},
  pages={185--190},
  year={2022},
  publisher={Elsevier}
}

@article{fan2022fast,
  title={A fast screening and recombinant method based on short-time pulse discharge and electrochemical impedance spectroscopy for decommissioned power batteries},
  author={Fan, Luo and Haihong, Huang and Haixin, Wang},
  journal={Journal of Electrochemical Energy Conversion and Storage},
  volume={19},
  number={2},
  pages={021020},
  year={2022},
  publisher={American Society of Mechanical Engineers}
}

@article{gao2022annual,
  title={Annual operating characteristics analysis of photovoltaic-energy storage microgrid based on retired lithium iron phosphate batteries},
  author={Gao, Yan and Cai, Yingling and Liu, Chenglin},
  journal={Journal of Energy Storage},
  volume={45},
  pages={103769},
  year={2022},
  publisher={Elsevier}
}

@article{garg2020development,
  title={Development of recycling strategy for large stacked systems: Experimental and machine learning approach to form reuse battery packs for secondary applications},
  author={Garg, Akhil and Yun, Liu and Gao, Liang and Putungan, Darwin Barayang},
  journal={Journal of Cleaner Production},
  volume={275},
  pages={124152},
  year={2020},
  publisher={Elsevier}
}

@article{gu2023quick,
  title={A quick and intelligent screening method for large-scale retired batteries based on cloud-edge collaborative architecture},
  author={Gu, Xin and Li, Jinglun and Zhu, Yuhao and Wang, Yue and Mao, Ziheng and Shang, Yunlong},
  journal={Energy},
  volume={285},
  pages={129342},
  year={2023},
  publisher={Elsevier}
}

@article{haram2023second,
  title={Second Life EV Batteries: Technical Evaluation, Design Framework, and Case Analysis},
  author={Haram, Mohammed Hussein Saleh Mohammed and Sarker, Md Tanjil and Ramasamy, Gobbi and Ngu, Eng Eng},
  journal={IEEE Access},
  year={2023},
  publisher={IEEE}
}

@article{huang2022state,
  title={State of health estimation of lithium-ion batteries based on the regional frequency},
  author={Huang, Shaotang and Liu, Cuicui and Sun, Huiqin and Liao, Qiangqiang},
  journal={Journal of Power Sources},
  volume={518},
  pages={230773},
  year={2022},
  publisher={Elsevier}
}

@article{jiang2017recognition,
  title={Recognition of battery aging variations for LiFePO4 batteries in 2nd use applications combining incremental capacity analysis and statistical approaches},
  author={Jiang, Yan and Jiang, Jiuchun and Zhang, Caiping and Zhang, Weige and Gao, Yang and Guo, Qipei},
  journal={Journal of Power Sources},
  volume={360},
  pages={180--188},
  year={2017},
  publisher={Elsevier}
}

@article{cui2024taking,
  title={Taking second-life batteries from exhausted to empowered using experiments, data analysis, and health estimation},
  author={Cui, Xiaofan and Khan, Muhammad Aadil and Pozzato, Gabriele and Singh, Surinder and Sharma, Ratnesh and Onori, Simona},
  journal={Cell Reports Physical Science},
  volume={5},
  number={5},
  year={2024},
  publisher={Elsevier}
}

@inproceedings{cui2024adaptive,
  title={Adaptive Health Monitoring of Second-Life Batteries},
  author={Cui, Xiaofan and Khan, Muhammad Aadil and Singh, Surinder and Sharma, Ratnesh and Onori, Simona},
  booktitle={2024 American Control Conference (ACC)},
  pages={1--6},
  year={2024},
  organization={IEEE}
}

@article{dang2024optimal,
  title={Optimal configuration of retired battery reconfigurable network considering switching losses},
  author={Dang, Jian and Xiao, Dou and Zhang, Xu and Jia, Rong and Jiao, Yang},
  journal={Journal of Energy Storage},
  volume={101},
  pages={113735},
  year={2024},
  publisher={Elsevier}
}

@article{du2021state,
  title={State-of-charge estimation for second-life lithium-ion batteries based on cell difference model and adaptive fading unscented Kalman filter algorithm},
  author={Du, Banghua and Yu, Zhang and Yi, Shuhao and He, Yanlin and Luo, Yulin},
  journal={International Journal of Low-Carbon Technologies},
  volume={16},
  number={3},
  pages={927--939},
  year={2021},
  publisher={Oxford University Press}
}

@article{faraji2023accelerated,
  title={Accelerated state of health estimation of second life lithium-ion batteries via electrochemical impedance spectroscopy tests and machine learning techniques},
  author={Faraji-Niri, Mona and Rashid, Muhammad and Sansom, Jonathan and Sheikh, Muhammad and Widanage, Dhammika and Marco, James},
  journal={Journal of Energy Storage},
  volume={58},
  pages={106295},
  year={2023},
  publisher={Elsevier}
}

@article{zhao2022research,
  title={Research on state of health for the series battery module based on the Weibull distribution},
  author={Zhao, Qian and Jiang, Haobin and Chen, Biao and Wang, Cheng and Xu, Shanzhen and Zhu, Jianhui and Chang, Lv},
  journal={Journal of The Electrochemical Society},
  volume={169},
  number={2},
  pages={020523},
  year={2022},
  publisher={IOP Publishing}
}

@article{di2023novel,
  title={A novel methodology to study and compare active energy-balance architectures with dynamic equalization for second-life battery applications},
  author={Di Rienzo, Roberto and Nicodemo, Niccol{\`o} and Verani, Alessandro and Baronti, Federico and Roncella, Roberto and Saletti, Roberto},
  journal={Journal of Energy Storage},
  volume={73},
  pages={108772},
  year={2023},
  publisher={Elsevier}
}

@article{gao2024comprehensive,
  title={Comprehensive study of the aging knee and second-life potential of the Nissan Leaf e+ batteries},
  author={Gao, Wei and Cao, Zhi and Fu, Yuhong and Turchiano, Christopher and Kurdkandi, Naser Vosoughi and Gu, Jing and Mi, Chirs},
  journal={Journal of Power Sources},
  volume={613},
  pages={234884},
  year={2024},
  publisher={Elsevier}
}

@article{gao2024evaluation,
  title={Evaluation of the second-life potential of the first-generation Nissan Leaf battery packs in energy storage systems},
  author={Gao, Wei and Cao, Zhi and Kurdkandi, Naser Vosoughi and Fu, Yuhong and Mi, Chirs},
  journal={eTransportation},
  volume={20},
  pages={100313},
  year={2024},
  publisher={Elsevier}
}

@article{guenther2024second,
  title={Second Life for Lithium-Ion Traction Batteries.},
  author={Guenther, Lea H and Klein, Volker and Loef, Georg and Pohl, Andr{\'e} and Okube, Haimanot and Winterbur, Ralf and R{\"o}wer, Georg and Wegener, Bianca and Goertz, Roland and Bracke, Stefan},
  journal={Sustainability (2071-1050)},
  volume={16},
  number={17},
  year={2024}
}

@article{hassini2024battery,
  title={Battery Passports for Second-Life Batteries: An Experimental Assessment of Suitability for Mobile Applications},
  author={Hassini, Marwan and Redondo-Iglesias, Eduardo and Venet, Pascal},
  journal={Batteries},
  volume={10},
  number={5},
  pages={153},
  year={2024},
  publisher={MDPI}
}

@inproceedings{hassini2023capacity,
  title={Capacity Dispersion and Impact of Outliers in a Second Life Battery},
  author={Hassini, Marwan and Seger, Pedro VH and Redondo-Iglesias, Eduardo and Pelissier, Serge and Venet, Pascal},
  booktitle={2023 IEEE Vehicle Power and Propulsion Conference (VPPC)},
  pages={1--4},
  year={2023},
  organization={IEEE}
}

@article{jiang2024state,
  title={State of Health Estimation for Second-Life Lithium-Ion Batteries in Energy Storage System With Partial Charging-Discharging Workloads},
  author={Jiang, Yiyue and Ke, Yuqi and Yang, Fangfang and Ji, Jinchen and Peng, Weiwen},
  journal={IEEE Transactions on Industrial Electronics},
  year={2024},
  publisher={IEEE}
}

@article{jung2023rapid,
  title={Rapid Estimation of Battery Storage Capacity through Multiple Linear Regression},
  author={Jung, Chulwon and Choi, Woongchul},
  journal={Batteries},
  volume={9},
  number={8},
  pages={424},
  year={2023},
  publisher={MDPI}
}

@article{kebede2023optimal,
  title={Optimal sizing and lifetime investigation of second life lithium-ion battery for grid-scale stationary application},
  author={Kebede, Abraham Alem and Hosen, Md Sazzad and Kalogiannis, Theodoros and Behabtu, Henok Ayele and Assefa, Marta Zemedu and Jemal, Towfik and Ramayya, Venkata and Van Mierlo, Joeri and Coosemans, Thierry and Berecibar, Maitane},
  journal={Journal of Energy Storage},
  volume={72},
  pages={108541},
  year={2023},
  publisher={Elsevier}
}

@article{kehl2021electrical,
  title={Electrical characterization of li-ion battery modules for second-life applications},
  author={Kehl, Daniel and Jennert, Torben and Lienesch, Frank and Kurrat, Michael},
  journal={Batteries},
  volume={7},
  number={2},
  pages={32},
  year={2021},
  publisher={MDPI}
}

@article{lacap2021development,
  title={Development and demonstration of microgrid system utilizing second-life electric vehicle batteries},
  author={Lacap, Joseph and Park, Jae Wan and Beslow, Lucas},
  journal={Journal of Energy Storage},
  volume={41},
  pages={102837},
  year={2021},
  publisher={Elsevier}
}

@article{lai2019rapid,
  title={A rapid screening and regrouping approach based on neural networks for large-scale retired lithium-ion cells in second-use applications},
  author={Lai, Xin and Qiao, Dongdong and Zheng, Yuejiu and Ouyang, Minggao and Han, Xuebing and Zhou, Long},
  journal={Journal of cleaner production},
  volume={213},
  pages={776--791},
  year={2019},
  publisher={Elsevier}
}

@article{li2017lifetime,
  title={Lifetime test design for second-use electric vehicle batteries in residential applications},
  author={Li, He and Alsolami, Mohammed and Yang, Shaobing and Alsmadi, Yazan M and Wang, Jin},
  journal={IEEE Transactions on Sustainable Energy},
  volume={8},
  number={4},
  pages={1736--1746},
  year={2017},
  publisher={IEEE}
}

@article{li2020fast,
  title={A fast classification method of retired electric vehicle battery modules and their energy storage application in photovoltaic generation},
  author={Li, Xinzhou and Zhang, Lizhong and Liu, Yi and Pan, Aiqiang and Liao, Qiangqiang and Yang, Xiu},
  journal={International Journal of Energy Research},
  volume={44},
  number={3},
  pages={2337--2344},
  year={2020},
  publisher={Wiley Online Library}
}

@article{li2022screening,
  title={Screening of retired lithium-Ion batteries using incremental capacity charging curve-based residual capacity estimation method for facilitating sustainable circular lithium-ion battery system},
  author={Li, Honglei and Cong, Liang and Ma, Huazheng and Liu, Weiwei and Deng, Yelin and Kong, Shuai},
  journal={Journal of Manufacturing Science and Engineering},
  volume={144},
  number={2},
  pages={021003},
  year={2022},
  publisher={American Society of Mechanical Engineers}
}

@article{li2024electrochemical,
  title={Electrochemical-thermal behaviors of retired power lithium-ion batteries during high-temperature and overcharge/over-discharge cycles},
  author={Li, Caisheng and Liu, Xianqing and Wang, Changhong and Ye, Lisheng and Wu, Tingting and Liang, Zhixuan and Zhang, Zejie and Zeng, Ying and Li, Kaizhe},
  journal={Case Studies in Thermal Engineering},
  volume={61},
  pages={104898},
  year={2024},
  publisher={Elsevier}
}

@article{liao2017performance,
  title={Performance assessment and classification of retired lithium ion battery from electric vehicles for energy storage},
  author={Liao, Qiangqiang and Mu, Miaomiao and Zhao, Shuqi and Zhang, Lizhong and Jiang, Tao and Ye, Jilei and Shen, Xiaowang and Zhou, Guoding},
  journal={International Journal of Hydrogen Energy},
  volume={42},
  number={30},
  pages={18817--18823},
  year={2017},
  publisher={Elsevier}
}

@article{lin2023state,
  title={State of health estimation of lithium-ion batteries based on remaining area capacity},
  author={Lin, Zhicheng and Hu, Houpeng and Liu, Wei and Zhang, Zixia and Zhang, Ya and Geng, Nankun and Liao, Qiangqiang},
  journal={Journal of Energy Storage},
  volume={63},
  pages={107078},
  year={2023},
  publisher={Elsevier}
}

@article{lin2023retired,
  title={Retired battery screening based on Markov transition field and Swin transformer},
  author={Lin, Mingqiang and Wu, Jian and Zou, Jianqing and Meng, Jinhao and Wang, Wei and Wu, Ji},
  journal={IEEE Transactions on Transportation Electrification},
  year={2023},
  publisher={IEEE}
}

@article{aeppli2025aging,
  title={Aging behavior of LiFePO4-based battery cells at stack level: A Second-Life cycling study},
  author={Aeppli, Daniel and Hack, Erwin and Held, Marcel},
  journal={Journal of Energy Storage},
  volume={129},
  pages={117135},
  year={2025},
  publisher={Elsevier}
}

@article{beatty2025long,
  title={Long-Term Sweat Testing Dataset for Second-Life Batteries},
  author={Beatty, Matthew and Strickland, Dani and Warren, Joe and Chan, John and Ferreira, Pedro},
  journal={Scientific Data},
  volume={12},
  number={1},
  pages={1068},
  year={2025},
  publisher={Nature Publishing Group UK London}
}

@article{montes2025multi,
  title={Multi-feature data-driven model for fast SOH assessment of batteries after their first life},
  author={Montes, Tom{\'a}s and Etxandi-Santolaya, Maite and Casals, Lluc Canals and Eichman, Josh},
  journal={Mathematics and Computers in Simulation},
  volume={238},
  pages={150--162},
  year={2025},
  publisher={Elsevier}
}

@article{perez2025novel,
  title={A novel aging modeling approach for second-life lithium-ion batteries},
  author={Perez, Ane and San Mart{\'\i}n, Idoia and Sanchis, Pablo and Ursua, Alfredo},
  journal={eTransportation},
  volume={24},
  pages={100400},
  year={2025},
  publisher={Elsevier}
}

@article{attidekou2026rapid,
  title={Rapid prediction of electrolyte-based cell failure in retired Li-ion electric vehicle pouch cells},
  author={Attidekou, Pierrot S and Milojevic, Zoran and Sweeney, Luke and Muhammad, Musbahu and Ahmeid, Mohamed and Lambert, Simon and Slater, Peter R and Anderson, Paul A},
  journal={Journal of Physics: Energy},
  volume={8},
  number={1},
  pages={01LT01},
  year={2026},
  publisher={IOP Publishing}
}

@article{sordi2026degradation,
  title={Degradation of LiFePO4 batteries after a real hybrid-bus extended application: Investigation of ageing phenomena and heterogeneity of performance},
  author={Sordi, G and Trippetta, GM and Luder, D and Berg, S and Li, W and Figgemeier, E and Sauer, DU and Casalegno, A and Rabissi, C},
  journal={Journal of Energy Storage},
  volume={152},
  pages={120653},
  year={2026},
  publisher={Elsevier}
}

@article{liu2022characterization,
  title={Characterization of aging mechanisms and state of health for second-life 21700 ternary lithium-ion battery},
  author={Liu, Cuicui and Wen, Xiankui and Zhong, Jingliang and Liu, Wei and Chen, Jianhong and Zhang, Jiawei and Wang, Zhiqin and Liao, Qiangqiang},
  journal={Journal of Energy Storage},
  volume={55},
  pages={105511},
  year={2022},
  publisher={Elsevier}
}

@phdthesis{de2025diagnostics,
  title={Diagnostics of the state of health and aging prognosis of second-life lithium-ion battery packs for the systemic study of energy microgrids},
  author={de Oliveira Albuquerque, Lucas},
  year={2025},
  school={Universit{\'e} de Toulouse}
}

@article{huang2025rapid,
  title={Rapid State of Health Estimation Strategy for Retired Batteries Based on Resting Voltage Curves},
  author={Huang, Haihong and Liu, Xin and Chang, Wenjing and Wang, Yuhang},
  journal={Batteries},
  volume={11},
  number={2},
  pages={66},
  year={2025},
  publisher={MDPI}
}

@article{yuwono2026characterization,
  title={Characterization of new and used Lithium-Ion electric motorcycle batteries degradation through cycle testing and surface temperature evaluation for second-life applications},
  author={Yuwono, Triyogi and Wikarta, Alief and Andrianto, Surya Putra and Andisetiawan, Anugrah and Hidayat, Romel and Ramdhiny, Masytha and Salsabila, Rifdah Adya and Zulfa, Liyana Labiba and Aftab, Sikandar and others},
  journal={Energy Conversion and Management: X},
  volume={30},
  pages={101640},
  year={2026},
  publisher={Elsevier}
}

@article{yuwono2026performance,
  title={Performance Evaluation and Material Degradation Analysis of Used Batteries of Online Taxi Electric Motorcycles in Indonesia for Second Life Application},
  author={Yuwono, Triyogi and Wikarta, Alief and Andrianto, Surya Putra and Ramdhiny, Masytha and Andisetiawan, Anugrah and Hidayat, Romel and Salsabila, Rifdah Adya and Zhao, Hongyang and others},
  journal={Results in Engineering},
  pages={110503},
  year={2026},
  publisher={Elsevier}
}

@article{peng2026fast,
  title={Fast sorting of retired lithium-ion batteries: An EIS and VAE-FINCH based soft clustering method},
  author={Peng, Weiwen and Wu, Xinhong and Li, Ziyuan and Zhu, Rong and Tan, Xiaojun},
  journal={Journal of Energy Storage},
  volume={151},
  pages={120518},
  year={2026},
  publisher={Elsevier}
}

@article{xu2026regenerate,
  title={Regenerate large-scale retired second-life battery datasets via recovered capacity labels-based deep learning},
  author={Xu, Yuchen and Zeng, Tianxiang and Peng, Weiwen and Tian, Jinpeng and Yi, Xiaojian and Xu, Qizhi and Zhu, Shun-Peng},
  journal={Energy Storage Materials},
  pages={104947},
  year={2026},
  publisher={Elsevier}
}

@article{chang2026novel,
  title={A novel sorting method and second-life utilization strategy for retired lithium-ion batteries},
  author={Chang, Long and Du, Henghui and Wang, Weibin and Zhu, Rui and Li, Changlong and Duan, Bin},
  journal={Journal of Energy Storage},
  volume={147},
  pages={120129},
  year={2026},
  publisher={Elsevier}
}

@article{wan2026enhanced,
  title={An enhanced sorting method for second-life lithium-ion batteries based on two-step clustering},
  author={Wan, Fu and Lin, Yupeng and Yang, Da and Li, Shufan and Liu, Ruiqi and Zhu, Lei and Yin, Wenwei and Chen, Weigen},
  journal={Journal of Cleaner Production},
  volume={544},
  pages={147712},
  year={2026},
  publisher={Elsevier}
}

@article{wang2026adaptive,
  title={Adaptive Dual-Anchor Fusion Framework for Robust SOC Estimation and SOH Soft-Sensing of Retired Batteries with Heterogeneous Aging},
  author={Wang, Hai and Liu, Rui and Guo, Yupeng and Liu, Yijun and Chen, Jiawei and Jiang, Yan and Li, Jianying},
  journal={Batteries},
  volume={12},
  number={2},
  pages={49},
  year={2026},
  publisher={MDPI}
}

@article{du2026efficient,
  title={An efficient screening and clustering strategy based on isolation forest algorithm and agglomerative hierarchical clustering algorithm for echelon utilization of retired lithium-ion batteries},
  author={Du, Qimeng and He, Fang and Jiang, Zhengyi and Yang, Guohui and Zhu, Boyuan and Chen, Siru and Yang, Haiyan and Zhao, Yaomin},
  journal={Sustainable Energy Technologies and Assessments},
  volume={86},
  pages={104883},
  year={2026},
  publisher={Elsevier}
}

@article{raza2026ev,
  title={From EV to stationary energy storage: EIS-based SoH estimation for second life Li-ion batteries},
  author={Raza, Kashif and Samadi, Amir Farbod and Berecibar, Maitane and Hosen, Md Sazzad},
  journal={Journal of Energy Storage},
  volume={141},
  pages={119316},
  year={2026},
  publisher={Elsevier}
}

@article{li2025lifetime,
  title={Lifetime Early Prediction for Second-Life Lithium-Ion Battery: A Hierarchical Bayesian Learning Method},
  author={Li, Ziyuan and Han, Yu and Peng, Weiwen},
  journal={Reliability Engineering \& System Safety},
  pages={111462},
  year={2025},
  publisher={Elsevier}
}

@article{fan2025generalized,
  title={Generalized capacity estimation for retired lithium-ion batteries: Unsupervised electrochemical impedance spectroscopy dimensionality reduction under random retirement conditions},
  author={Fan, Wenjun and Jiang, Bo and Wang, Xueyuan and Zhu, Jiangong and Liu, Haonan and Yuan, Yongjun and Wei, Xuezhe and Dai, Haifeng},
  journal={Energy Storage Materials},
  pages={104836},
  year={2025},
  publisher={Elsevier}
}

@article{zhang2025two,
  title={A two-stage data-driven temperature prediction method considering battery aging for retired batteries and validation of preemptive thermal control},
  author={Zhang, Tao and Ye, Songtao and Xi, Huan and An, Dou},
  journal={Applied Thermal Engineering},
  pages={128666},
  year={2025},
  publisher={Elsevier}
}

@article{zhu2025bayesian,
  title={Bayesian calibrated physics-informed neural networks for second-life battery SOH estimation},
  author={Zhu, Rong and Hu, Jiawen and Peng, Weiwen},
  journal={Reliability Engineering \& System Safety},
  volume={264},
  pages={111432},
  year={2025},
  publisher={Elsevier}
}

@article{li2025state,
  title={State of health estimation for lithium-ion batteries under unknown usage history via distribution of relaxation times analysis and interpretable feature selection},
  author={Li, Xin and Jin, Zhao and Jiang, Weizhen and Tang, Ruoli and Yang, Xiangguo and Zhang, Yan},
  journal={Journal of Energy Storage},
  volume={136},
  pages={118421},
  year={2025},
  publisher={Elsevier}
}

@article{amuta2025novel,
  title={A Novel Efficient Grading of Spent Lithium-Ion Batteries Using Electrochemical Impedance Spectrometry},
  author={Amuta, Ote and Kowal, Julia},
  journal={Batteries},
  volume={11},
  number={11},
  pages={404},
  year={2025},
  publisher={MDPI}
}

@article{ezpeleta2025rapid,
  title={Rapid and Non-Invasive SoH Estimation of Lithium-Ion Cells via Automated EIS and EEC Models},
  author={Ezpeleta, Ignacio and Fern{\'a}ndez, Javier and Gir{\'a}ldez, David and Freire, Lorena},
  journal={Batteries},
  volume={11},
  number={9},
  pages={325},
  year={2025},
  publisher={MDPI}
}

@article{sarker2025performance,
  title={Performance evaluation of second-life ev batteries for off-grid solar energy storage system},
  author={Sarker, Md Tanjil and Al Qwaid, Marran and Ramasamy, Gobbi and Haram, Mohammed Hussein Saleh Mohammed},
  journal={IEEE Access},
  year={2025},
  publisher={IEEE}
}

@article{liu2023fast,
  title={Fast sorting method of retired batteries based on multi-feature extraction from partial charging segment},
  author={Liu, Xingtao and Tang, Qinbin and Feng, Yitian and Lin, Mingqiang and Meng, Jinhao and Wu, Ji},
  journal={Applied Energy},
  volume={351},
  pages={121930},
  year={2023},
  publisher={Elsevier}
}

@article{he2025state,
  title={The state of health estimation of retired lithium-ion batteries using a multi-input metabolic gated recurrent unit},
  author={He, Yu and Pattanadech, Norasage and Sukemoke, Kasiean and Pan, Minling and Chen, Lin},
  journal={Energies},
  volume={18},
  number={5},
  pages={1035},
  year={2025},
  publisher={MDPI}
}

@article{huang2024flexible,
  title={A flexible screening scheme for retired lithium-ion batteries based on novel capacity indicator and random forest algorithm},
  author={Huang, Peng and Zhang, Ying and Kang, Yongzhe and Gu, Pingwei and Duan, Bin and Zhang, Chenghui},
  journal={IEEE Transactions on Transportation Electrification},
  volume={11},
  number={1},
  pages={544--557},
  year={2024},
  publisher={IEEE}
}

@inproceedings{zhu2025rapid,
  title={Rapid Health Diagnosis Method for Retired Lithium-ion Batteries based on Electrochemical Impedance Spectroscopy},
  author={Zhu, Xufan and Han, Wenxuan and Dong, Jiawei and Geng, Guangchao and Jiang, Quanyuan},
  booktitle={2025 10th Asia Conference on Power and Electrical Engineering (ACPEE)},
  pages={2505--2510},
  year={2025},
  organization={IEEE}
}

@article{agudelo2021application,
  title={Application domain extension of incremental capacity-based battery SoH indicators},
  author={Agudelo, Brian Ospina and Zamboni, Walter and Monmasson, Eric},
  journal={Energy},
  volume={234},
  pages={121224},
  year={2021},
  publisher={Elsevier}
}

@article{bechara2025embedded,
  title={Embedded spectroscopy: Potentialities and constraints for onboard battery diagnostics},
  author={Bechara, Charles and Friedrich, Guy and Forgez, Christophe and Cregut, Samuel},
  journal={Power Electronic Devices and Components},
  pages={100118},
  year={2025},
  publisher={Elsevier}
}

@article{najjar2026fast,
  title={Fast \& Reliable End-of-Life Evaluation of Li-Ion Batteries in Electric Vehicles: A Dual Metric Approach for Ageing Heterogeneity Assessment},
  author={Najjar-Giudicelli, Adrien and Lespinats, Sylvain and Guillet, Nicolas and Thivel, Pierre-Xavier},
  journal={Journal of The Electrochemical Society},
  volume={173},
  number={7},
  pages={070517},
  year={2026},
  publisher={IOP Publishing}
}

@article{tao2024generative,
  title={Generative learning assisted state-of-health estimation for sustainable battery recycling with random retirement conditions},
  author={Tao, Shengyu and Ma, Ruifei and Zhao, Zixi and Ma, Guangyuan and Su, Lin and Chang, Heng and Chen, Yuou and Liu, Haizhou and Liang, Zheng and Cao, Tingwei and others},
  journal={Nature Communications},
  volume={15},
  number={1},
  pages={10154},
  year={2024},
  publisher={Nature Publishing Group UK London}
}

@article{li2026rapid,
  title={Rapid Estimation for the Maximum Remaining Capacity of Retired Lithium-Ion Batteries Based on CNN-CBAM-LSTM},
  author={Li, Aqing and Cui, Penghao and Cao, Yifei and Zhou, Peng and Yang, Lei and Bian, Guochen and Shao, Zhendong},
  journal={Batteries},
  volume={12},
  number={4},
  pages={145},
  year={2026},
  publisher={MDPI}
}

@article{hosen2026data,
  title={A Data-Driven Approach to Map the Aging of Two Types of Dismantled Commercial High-Energy NMC Cells},
  author={Hosen, Md Sazzad and Samadi, Amir Farbod and Raza, Kashif and Berecibar, Maitane},
  journal={World Electric Vehicle Journal},
  volume={17},
  number={5},
  pages={244},
  year={2026},
  publisher={MDPI}
}

@article{zhang2026developing,
  title={Developing an efficient hierarchical regrouping method for uncharacterized retired electric vehicle Li-ion batteries based on partial pulse discharge curves},
  author={Zhang, Jun and Tong, Zheming and Ren, Ning and Chen, Xing and Cao, Xiangkun Elvis},
  journal={eTransportation},
  pages={100591},
  year={2026},
  publisher={Elsevier}
}

@article{kemeny2026impact,
  title={Impact of Internal Temperature Stabilisation on Electrochemical Impedance Spectroscopy Based State-of-Health Evaluation of Electric Vehicle Battery Modules for Second-Life Applications},
  author={Kemeny, Martin and Baga, Erik and Mulpuri, Sai Krishna and Venugopal, Prasanth and Mikolasek, Miroslav},
  journal={Batteries \& Supercaps},
  volume={9},
  number={5},
  pages={e202500926},
  year={2026},
  publisher={Wiley Online Library}
}

@article{wheeler2025ageing,
  title={An ageing study of twenty 18650 lithium-ion Graphite/LFP cells in first and second life use},
  author={Wheeler, William and Bultel, Yann and Venet, Pascal and Sari, Ali},
  journal={Scientific Data},
  volume={12},
  number={1},
  pages={392},
  year={2025},
  publisher={Nature Publishing Group UK London}
}

@article{marini2026design,
  title={Design of an embedded hardware platform for cell-level diagnostics in commercial battery modules},
  author={Marini, Gabriele and Colombo, Alessandro and Lanubile, Andrea and Paxton, William A and Onori, Simona},
  journal={IEEE Open Journal of Power Electronics},
  year={2026},
  publisher={IEEE}
}

@article{braco2021experimental,
  title={Experimental assessment of first-and second-life electric vehicle batteries: Performance, capacity dispersion, and aging},
  author={Braco, Elisa and San Mart{\'\i}n, Idoia and Berrueta, Alberto and Sanchis, Pablo and Urs{\'u}a, Alfredo},
  journal={IEEE Transactions on Industry Applications},
  volume={57},
  number={4},
  pages={4107--4117},
  year={2021},
  publisher={IEEE}
}

@inproceedings{locorotondo2020electrical,
  title={Electrical lithium battery performance model for second life applications},
  author={Locorotondo, Edoardo and Cultrera, Vincenzo and Pugi, Luca and Berzi, Lorenzo and Pasquali, Manlio and Andrenacci, Natascia and Lutzemberger, Giovanni and Pierini, Marco},
  booktitle={2020 IEEE International Conference on Environment and Electrical Engineering and 2020 IEEE Industrial and Commercial Power Systems Europe (EEEIC/I\&CPS Europe)},
  pages={1--6},
  year={2020},
  organization={IEEE}
}

@article{coron2020impact,
  title={Impact of lithium-ion cell condition on its second life viability},
  author={Coron, E and Geni{\`e}s, S and Cugnet, M and Thivel, PX},
  journal={Journal of The Electrochemical Society},
  volume={167},
  number={11},
  pages={110556},
  year={2020},
  publisher={IOP Publishing}
}

@article{lucu2020dataa,
  title={Data-driven nonparametric Li-ion battery ageing model aiming at learning from real operation data--Part A: Storage operation},
  author={Lucu, Mattin and Martinez-Laserna, Egoitz and Gandiaga, I{\~n}igo and Liu, Kailong and Camblong, Haritza and Widanage, Widanalage Dhammika and Marco, James},
  journal={Journal of Energy Storage},
  volume={30},
  pages={101409},
  year={2020},
  publisher={Elsevier}
}

@article{luo2021rapid,
  title={Rapid prediction of the state of health of retired power batteries based on electrochemical impedance spectroscopy},
  author={Luo, Fan and Huang, HaiHong and Ni, LiPing and Li, Tan},
  journal={Journal of Energy Storage},
  volume={41},
  pages={102866},
  year={2021},
  publisher={Elsevier}
}

@article{lv2022investigation,
  title={Investigation on the thermo-electric-electrochemical characteristics of retired LFP batteries for echelon applications},
  author={Lv, Youfu and Luo, Weiming and Mo, Ya and Zhang, Guoqing},
  journal={RSC advances},
  volume={12},
  number={22},
  pages={14127--14136},
  year={2022},
  publisher={Royal Society of Chemistry}
}

@article{lyu2022synchronous,
  title={Synchronous state of health estimation and remaining useful lifetime prediction of Li-Ion battery through optimized relevance vector machine framework},
  author={Lyu, Zhiqiang and Wang, Geng and Gao, Renjing},
  journal={Energy},
  volume={251},
  pages={123852},
  year={2022},
  publisher={Elsevier}
}

@article{ma2015investigation,
  title={Investigation of path dependence in commercial lithium-ion cells for pure electric bus applications: Aging mechanism identification},
  author={Ma, Zeyu and Jiang, Jiuchun and Shi, Wei and Zhang, Weige and Mi, Chunting Chris},
  journal={Journal of Power Sources},
  volume={274},
  pages={29--40},
  year={2015},
  publisher={Elsevier}
}

@article{ma2020multilayer,
  title={Multilayer SOH equalization scheme for MMC battery energy storage system},
  author={Ma, Zhan and Gao, Feng and Gu, Xin and Li, Nan and Wu, Qiang and Wang, Xiaohui and Wang, Xiaolong},
  journal={IEEE Transactions on Power Electronics},
  volume={35},
  number={12},
  pages={13514--13527},
  year={2020},
  publisher={IEEE}
}

@article{ma2021study,
  title={Study on Capacity Estimation Methods of Second-Life Application Batteries},
  author={Ma, Linkang and Zhang, Caiping and Wang, Jinyu and Wang, Kairang and Chen, Jie},
  journal={World Electric Vehicle Journal},
  volume={12},
  number={4},
  pages={163},
  year={2021},
  publisher={MDPI}
}

@article{martinez2018technical,
  title={Technical viability of battery second life: A study from the ageing perspective},
  author={Martinez-Laserna, Egoitz and Sarasketa-Zabala, Elixabet and Sarria, Igor Villarreal and Stroe, Daniel-Ioan and Swierczynski, Maciej and Warnecke, Alexander and Timmermans, Jean-Marc and Goutam, Shovon and Omar, Noshin and Rodriguez, Pedro},
  journal={IEEE Transactions on Industry Applications},
  volume={54},
  number={3},
  pages={2703--2713},
  year={2018},
  publisher={IEEE}
}

@article{michelini2024impact,
  title={Impact of Prolonged Electrochemical Cycling on Health Indicators of Aged Lithium-Ion Batteries for a Second-Life Use},
  author={Michelini, Emanuele and H{\"o}schele, Patrick and Ellersdorfer, Christian and Moser, J{\"o}rg},
  journal={IEEE Access},
  year={2024},
  publisher={IEEE Publications}
}

@article{mohsin2022new,
  title={A new lead-acid battery state-of-health evaluation method using electrochemical impedance spectroscopy for second life in rural electrification systems},
  author={Mohsin, Muhammad and Picot, Antoine and Maussion, Pascal},
  journal={Journal of Energy Storage},
  volume={52},
  pages={104647},
  year={2022},
  publisher={Elsevier}
}

@article{wheeler2024postmortem,
  title={Postmortem Analysis of 18650 Graphite/LFP Cells in a Long-Term Aging Study for Second-Life Applications},
  author={Wheeler, William and Bultel, Yann and Venet, Pascal and Sari, Ali and Riviere, Elie},
  journal={Batteries},
  volume={10},
  number={4},
  pages={119},
  year={2024},
  publisher={MDPI}
}

@article{montoya2024quantitative,
  title={Quantitative Ultrasound Spectroscopy for Screening Cylindrical Lithium-Ion Batteries for Second-Life Applications},
  author={Montoya-Bedoya, Simon and Garcia-Tamayo, Esteban and Rohrbach, Daniel and Gaviria-Cardona, Juan Pablo and Martinez-Tejada, Hader V and Planden, Brady and Howey, David A and Florez, Whady F and Valencia, Ra{\'u}l A and Bernal, Miguel},
  journal={Batteries \& Supercaps},
  volume={7},
  number={5},
  pages={e202400002},
  year={2024},
  publisher={Wiley Online Library}
}

@article{kawaura2016operando,
  title={Operando measurement of solid electrolyte interphase formation at working electrode of Li-ion battery by time-slicing neutron reflectometry},
  author={Kawaura, Hiroyuki and Harada, Masashi and Kondo, Yasuhito and Kondo, Hiroki and Suganuma, Yoshitake and Takahashi, Naoko and Sugiyama, Jun and Seno, Yoshiki and Yamada, Norifumi L},
  journal={ACS Applied Materials \& Interfaces},
  volume={8},
  number={15},
  pages={9540--9544},
  year={2016},
  publisher={ACS Publications}
}

@inproceedings{mowri2021modification,
  title={Modification of degradation mechanism identification technique for cell grading},
  author={Mowri, Sadia Tasnim and Barai, Anup and Gupta, Aniruddha and Marco, James},
  booktitle={2021 IEEE Vehicle Power and Propulsion Conference (VPPC)},
  pages={1--7},
  year={2021},
  organization={IEEE}
}

@article{tasnim2024assessing,
  title={Assessing the Impact of First-Life Lithium-Ion Battery Degradation on Second-Life Performance},
  author={Tasnim Mowri, Sadia and Barai, Anup and Moharana, Sanghamitra and Gupta, Aniruddha and Marco, James},
  journal={Energies},
  volume={17},
  number={2},
  pages={501},
  year={2024},
  publisher={MDPI}
}

@article{ni2023rapid,
  title={Rapid estimation of residual capacity for retired LiFePO4 batteries using voltage interval at low state of charge},
  author={Ni, Yulong and Xu, Jianing and Zhu, Chunbo and Zhang, He and Yu, Yuelong and Song, Kai and Wu, Chao},
  journal={Energy Storage Materials},
  volume={55},
  pages={463--478},
  year={2023},
  publisher={Elsevier}
}

@inproceedings{oeser2018effectiveness,
  title={Effectiveness of active balancing on high dispersion batteries},
  author={Oeser, David and Ziegler, Andreas and Arndt, Bernhard and Ackva, Ansgar},
  booktitle={2018 International IEEE Conference and Workshop in Obuda on Electrical and Power Engineering (CANDO-EPE)},
  pages={21--26},
  year={2018},
  organization={IEEE}
}

@article{pang2024effects,
  title={Effects of floating charge ageing on electrochemical impedance spectroscopy of lead-acid batteries},
  author={Pang, Zheyuan and Yang, Kun and Song, Zhengxiang and Chen, Guangyang and Niu, Pengcheng and Wang, Song and Meng, Jinhao},
  journal={Journal of Energy Storage},
  volume={87},
  pages={111322},
  year={2024},
  publisher={Elsevier}
}

@inproceedings{popp2024fast,
  title={Fast Screening and Sorting of Commercially Available Second-Life Batteries from Former Mobility Applications for Construction of Small Energy Storage Systems},
  author={Popp, Alexander and Spaeth, Utz and Schmuelling, Benedikt},
  booktitle={2024 IEEE Transportation Electrification Conference and Expo (ITEC)},
  pages={1--6},
  year={2024},
  organization={IEEE}
}

@article{quinard2019fast,
  title={Fast electrical characterizations of high-energy second life lithium-ion batteries for embedded and stationary applications},
  author={Quinard, Honorat and Redondo-Iglesias, Eduardo and Pelissier, Serge and Venet, Pascal},
  journal={Batteries},
  volume={5},
  number={1},
  pages={33},
  year={2019},
  publisher={MDPI}
}

@article{ramirez2023statistical,
  title={A statistical assessment of the state-of-health of LiFePO4 cells harvested from a hybrid-electric vehicle battery pack},
  author={Ramirez-Meyers, Katrina and Rawn, Barry and Whitacre, Jay F},
  journal={Journal of Energy Storage},
  volume={59},
  pages={106472},
  year={2023},
  publisher={Elsevier}
}

@article{ran2020data,
  title={Data-driven fast clustering of second-life lithium-ion battery: mechanism and algorithm},
  author={Ran, Aihua and Zhou, Zihao and Chen, Shuxiao and Nie, Pengbo and Qian, Kun and Li, Zhenlong and Li, Baohua and Sun, Hongbin and Kang, Feiyu and Zhang, Xuan and others},
  journal={Advanced Theory and Simulations},
  volume={3},
  number={8},
  pages={2000109},
  year={2020},
  publisher={Wiley Online Library}
}

@article{rasheed2023active,
  title={Active Reconditioning of Retired Lithium-ion Battery Packs from Electric Vehicles for Second Life Applications},
  author={Rasheed, Marium and Hassan, Rohail and Kamel, Mohamed and Wang, Hongjie and Zane, Regan and Tong, Shijie and Smith, Kandler},
  journal={IEEE Journal of Emerging and Selected Topics in Power Electronics},
  year={2023},
  publisher={IEEE}
}

@article{salinas2019second,
  title={A second life for li-ion cells rescued from notebook batteries},
  author={Salinas, Felipe and Kr{\"u}ger, Lars and Neupert, Steven and Kowal, Julia},
  journal={Journal of Energy Storage},
  volume={24},
  pages={100747},
  year={2019},
  publisher={Elsevier}
}

@article{schaltz2021incremental,
  title={Incremental capacity analysis applied on electric vehicles for battery state-of-health estimation},
  author={Schaltz, Erik and Stroe, Daniel-Ioan and N{\o}rregaard, Kjeld and Ingvardsen, Lasse Stenh{\o}j and Christensen, Andreas},
  journal={IEEE Transactions on Industry Applications},
  volume={57},
  number={2},
  pages={1810--1817},
  year={2021},
  publisher={IEEE}
}

@article{scholz2023experimental,
  title={Experimental Analysis of a Lithium-Ion Battery Pack after Long Service Life in a Conventional Electric Vehicle Considering Second-Life Applications},
  author={Scholz, Tobias and Pautzke, F and Schmuelling, B},
  journal={Proceedings of the ESREL, Southampton, UK},
  pages={3--8},
  year={2023}
}

@article{schuster2016correlation,
  title={Correlation between capacity and impedance of lithium-ion cells during calendar and cycle life},
  author={Schuster, Simon F and Brand, Martin J and Campestrini, Christian and Gleissenberger, Markus and Jossen, Andreas},
  journal={Journal of Power Sources},
  volume={305},
  pages={191--199},
  year={2016},
  publisher={Elsevier}
}

@article{schuster2015lithium,
  title={Lithium-ion cell-to-cell variation during battery electric vehicle operation},
  author={Schuster, Simon F and Brand, Martin J and Berg, Philipp and Gleissenberger, Markus and Jossen, Andreas},
  journal={Journal of Power Sources},
  volume={297},
  pages={242--251},
  year={2015},
  publisher={Elsevier}
}

@article{gasper2022predicting,
  title={Predicting battery capacity from impedance at varying temperature and state of charge using machine learning},
  author={Gasper, Paul and Schiek, Andrew and Smith, Kandler and Shimonishi, Yuta and Yoshida, Shuhei},
  journal={Cell Reports Physical Science},
  volume={3},
  number={12},
  year={2022},
  publisher={Elsevier}
}

@article{she2021battery,
  title={Battery state-of-health estimation based on incremental capacity analysis method: Synthesizing from cell-level test to real-world application},
  author={She, Chengqi and Zhang, Lei and Wang, Zhenpo and Sun, Fengchun and Liu, Peng and Song, Chunbao},
  journal={IEEE Journal of Emerging and Selected Topics in Power Electronics},
  volume={11},
  number={1},
  pages={214--223},
  year={2021},
  publisher={IEEE}
}

@article{sun2022quick,
  title={Quick Evaluation of the State-of-Health of Spent Lithium-Ion Battery Modules},
  author={Sun, Huiqin and Liu, Wei and Du, Zhichao and Li, Xinzhou and Fu, Zaiguo and Liao, Qiangqiang},
  journal={International Journal of Electrochemical Science},
  volume={17},
  number={3},
  pages={220318},
  year={2022},
  publisher={Elsevier}
}

@article{tong2013off,
  title={Off-grid photovoltaic vehicle charge using second life lithium batteries: An experimental and numerical investigation},
  author={Tong, Shi Jie and Same, Adam and Kootstra, Mark A and Park, Jae Wan},
  journal={Applied energy},
  volume={104},
  pages={740--750},
  year={2013},
  publisher={Elsevier}
}

@article{tong2017demonstration,
  title={Demonstration of reusing electric vehicle battery for solar energy storage and demand side management},
  author={Tong, Shijie and Fung, Tsz and Klein, Matthew P and Weisbach, David A and Park, Jae Wan},
  journal={Journal of Energy Storage},
  volume={11},
  pages={200--210},
  year={2017},
  publisher={Elsevier}
}

@inproceedings{tran2022used,
  title={Used lithium-ion batteries in second-life applications: Feasibility study},
  author={Tran, Minh and Messo, Tuomas and Luhtala, Roni and Sihvo, Jussi and Roinila, Tomi},
  booktitle={2022 IEEE Energy Conversion Congress and Exposition (ECCE)},
  pages={1--5},
  year={2022},
  organization={IEEE}
}

@article{vasquez2025,
title = {Comparative study of methodologies for SOH diagnosis and forecast of LFP and NMC lithium batteries used in electric vehicles},
journal = {Journal of Energy Storage},
volume = {105},
pages = {114725},
year = {2025},
author = {F.A. Vásquez and P. {Sara Gaitán} and Jorge A. Calderón},
}

@article{vignesh2024need,
  title={Need of voltage and capacity profile reconstruction in incremental capacity analysis for second life batteries from repurposer perspective},
  author={Vignesh, S and Che, Hang Seng and Selvaraj, Jeyraj and Tey, Kok Soon},
  journal={Journal of Energy Storage},
  volume={100},
  pages={113453},
  year={2024},
  publisher={Elsevier}
}

@inproceedings{wang2019study,
  title={Study on Characteristic Internal Resistance of Lithium Batteries Based on Double Pulse Test},
  author={Wang, Yiquan and Huang, Bixiong and Yan, Xiao and Lin, Wenheng and Wang, Gengjiong and Liu, Zhongcai},
  booktitle={Proceedings of the 2019 International Conference on Robotics, Intelligent Control and Artificial Intelligence},
  pages={258--262},
  year={2019}
}

@article{wang2023rapid,
  title={Rapid-regroup strategy for retired batteries based on short-time dynamic voltage and electrochemical impedance spectroscopy},
  author={Wang, Yuhang and Huang, Haihong and Wang, Haixin},
  journal={Journal of Energy Storage},
  volume={63},
  pages={107102},
  year={2023},
  publisher={Elsevier}
}

@article{wang2023state,
  title={State estimation and aging mechanism of 2nd life lithium-ion batteries: Non-destructive and postmortem combined analysis},
  author={Wang, Cun and Lin, Da and Zhao, Bo and Zhang, Xuesong and Wang, Yinfei and Wang, Yiwei and Cao, Wenjiong and Jiang, Fangming},
  journal={Electrochimica Acta},
  volume={443},
  pages={141996},
  year={2023},
  publisher={Elsevier}
}

@article{wanner2024diagnosis,
  title={Diagnosis and remaining useful life prediction of end-of-life battery systems for second-life applications},
  author={Wanner, Johannes and Singh, Soumya and Nagel, Tobias and Patel, Arpit and Baazouzi, Sabri and Grimm, Julian and Birke, Kai Peter},
  journal={Procedia CIRP},
  volume={126},
  pages={847--852},
  year={2024},
  publisher={Elsevier}
}

@article{wen2023fast,
  title={A Fast Estimation Algorithm for State of Health of Lithium-ion Battery Modules Based on Lorenz Plot},
  author={Wen, Xiankui and Zhong, Jingliang and Li, Xiang and Lin, Zhicheng and Wang, Luyan and Liao, Qiangqiang},
  journal={Electrochemistry},
  volume={91},
  number={6},
  pages={067004--067004},
  year={2023},
  publisher={The Electrochemical Society of Japan}
}

@article{white2020repurposed,
  title={Repurposed electric vehicle battery performance in second-life electricity grid frequency regulation service},
  author={White, Chris and Thompson, Ben and Swan, Lukas G},
  journal={Journal of Energy Storage},
  volume={28},
  pages={101278},
  year={2020},
  publisher={Elsevier}
}

@article{wu2023balancing,
  title={Balancing method of retired battery pack based on variable domain fuzzy control},
  author={Wu, Tiezhou and Chen, Liuliang and Xu, Yuhong and Zhang, Xiaoxing},
  journal={Journal of Electrochemical Energy Conversion and Storage},
  volume={20},
  number={4},
  pages={041005},
  year={2023},
  publisher={American Society of Mechanical Engineers}
}

@article{xiao2021discharge,
  title={Discharge curve-based formation of retired power batteries for secondary use},
  author={Xiao, Ziyu and Wu, Siqi},
  journal={International Journal of Low-Carbon Technologies},
  volume={16},
  number={3},
  pages={790--797},
  year={2021},
  publisher={Oxford University Press}
}

@article{xiong2020online,
  title={Online state-of-health estimation for second-use lithium-ion batteries based on weighted least squares support vector machine},
  author={Xiong, Wei and Mo, Yimin and Yan, Cong},
  journal={IEEE Access},
  volume={9},
  pages={1870--1881},
  year={2020},
  publisher={IEEE}
}

@article{xu2019study,
  title={Study on the performance evaluation and echelon utilization of retired LiFePO4 power battery for smart grid},
  author={Xu, Xiaolong and Mi, Jifu and Fan, Maosong and Yang, Kai and Wang, Hao and Liu, Jingbing and Yan, Hui},
  journal={Journal of Cleaner Production},
  volume={213},
  pages={1080--1086},
  year={2019},
  publisher={Elsevier}
}

@article{xu2020novel,
  title={A novel clustering algorithm for grouping and cascade utilization of retired Li-ion batteries},
  author={Xu, Zhicheng and Wang, Jun and Lund, Peter D and Fan, Qi and Dong, Ting and Liang, Yan and Hong, Jie},
  journal={Journal of Energy Storage},
  volume={29},
  pages={101303},
  year={2020},
  publisher={Elsevier}
}

@article{yang2020layered,
  title={A layered bidirectional active equalization method for retired power lithium-ion batteries for energy storage applications},
  author={Yang, Yang and Zhu, Wenchao and Xie, Changjun and Shi, Ying and Liu, Furong and Li, Weibo and Tang, Zebo},
  journal={Energies},
  volume={13},
  number={4},
  pages={832},
  year={2020},
  publisher={MDPI}
}

@article{yang2023fast,
  title={Fast screening of lithium-ion batteries for second use with pack-level testing and machine learning},
  author={Yang, Sijia and Zhang, Caiping and Jiang, Jiuchun and Zhang, Weige and Chen, Haoze and Jiang, Yan and Sauer, Dirk Uwe and Li, Weihan},
  journal={ETransportation},
  volume={17},
  pages={100255},
  year={2023},
  publisher={Elsevier}
}

@article{zhang2014novel,
  title={A novel data-driven fast capacity estimation of spent electric vehicle lithium-ion batteries},
  author={Zhang, Caiping and Jiang, Jiuchun and Zhang, Weige and Wang, Yukun and Sharkh, Suleiman M and Xiong, Rui},
  journal={Energies},
  volume={7},
  number={12},
  pages={8076--8094},
  year={2014},
  publisher={MDPI}
}

@article{zhang2022reliable,
  title={A reliable data-driven state-of-health estimation model for lithium-ion batteries in electric vehicles},
  author={Zhang, Chaolong and Zhao, Shaishai and Yang, Zhong and Chen, Yuan},
  journal={Frontiers in Energy Research},
  volume={10},
  pages={1013800},
  year={2022},
  publisher={Frontiers Media SA}
}

@article{zhou2020efficient,
  title={An efficient screening method for retired lithium-ion batteries based on support vector machine},
  author={Zhou, Zhongkai and Duan, Bin and Kang, Yongzhe and Shang, Yunlong and Cui, Naxin and Chang, Long and Zhang, Chenghui},
  journal={Journal of Cleaner Production},
  volume={267},
  pages={121882},
  year={2020},
  publisher={Elsevier}
}

@article{zhang2020performance,
  title={Performance assessment of retired EV battery modules for echelon use},
  author={Zhang, Youlang and Li, Yan and Tao, Yibin and Ye, Jilei and Pan, Aiqiang and Li, Xinzhou and Liao, Qiangqiang and Wang, Zhiqin},
  journal={Energy},
  volume={193},
  pages={116555},
  year={2020},
  publisher={Elsevier}
}

@article{zhang2020quick,
  title={A quick screening approach based on fuzzy C-means algorithm for the second usage of retired lithium-ion batteries},
  author={Zhang, Ying and Zhou, Zhongkai and Kang, Yongzhe and Zhang, Chenghui and Duan, Bin},
  journal={IEEE Transactions on Transportation Electrification},
  volume={7},
  number={2},
  pages={474--484},
  year={2020},
  publisher={IEEE}
}

@article{zhang2021state,
  title={State-of-health estimation of batteries in an energy storage system based on the actual operating parameters},
  author={Zhang, Qichao and Li, Xue and Zhou, Chun and Zou, Yang and Du, Zhichao and Sun, Mingguang and Ouyang, Yongsheng and Yang, Dong and Liao, Qiangqiang},
  journal={Journal of Power Sources},
  volume={506},
  pages={230162},
  year={2021},
  publisher={Elsevier}
}

@article{zhang2023state,
  title={State of health estimation of lithium-ion batteries based on the regional triangle},
  author={Zhang, Ya and Cai, Yongxiang and Liu, Wei and Dou, Zhenlan and Yao, Bin and Zhang, Bide and Liao, Qiangqiang and Fu, Zaiguo and Cheng, Zhiyuan},
  journal={Journal of Energy Storage},
  volume={69},
  pages={107918},
  year={2023},
  publisher={Elsevier}
}

@article{zhou2023health,
  title={Health feature extraction and efficient sorting of second-life lithium-ion batteries},
  author={Zhou, Zhongkai and Duan, Bin and Kang, Yongzhe and Shang, Yunlong and Zhang, Qi and Zhang, Chenghui},
  journal={IEEE Transactions on Energy Conversion},
  year={2023},
  publisher={IEEE}
}

@inproceedings{zou2023fast,
  title={Fast Estimation of State of Health for Retired Lithium-ion Batteries Based on Principal Component Regression},
  author={Zou, Zhanxin and Li, Yichao and Liu, Xuefeng and Zhang, Ze and Duan, Bin and Zhang, Chenghui},
  booktitle={2023 7th CAA International Conference on Vehicular Control and Intelligence (CVCI)},
  pages={1--6},
  year={2023},
  organization={IEEE}
}

@article{etxandi2024extending,
  title={Extending the electric vehicle battery first life: Performance beyond the current end of life threshold},
  author={Etxandi-Santolaya, Maite and Casals, Lluc Canals and Corchero, Cristina},
  journal={Heliyon},
  volume={10},
  number={4},
  year={2024},
  publisher={Elsevier}
}

@article{dubarry2022best,
  title={Best practices for incremental capacity analysis},
  author={Dubarry, Matthieu and Anse{\'a}n, David},
  journal={Frontiers in Energy Research},
  volume={10},
  pages={1023555},
  year={2022},
  publisher={Frontiers Media SA}
}

@article{shahjalal2022review,
  title={A review on second-life of Li-ion batteries: Prospects, challenges, and issues},
  author={Shahjalal, Mohammad and Roy, Probir Kumar and Shams, Tamanna and Fly, Ashley and Chowdhury, Jahedul Islam and Ahmed, Md Rishad and Liu, Kailong},
  journal={Energy},
  volume={241},
  pages={122881},
  year={2022},
  publisher={Elsevier}
}

@article{hu2022review,
  title={A review of second-life lithium-ion batteries for stationary energy storage applications},
  author={Hu, Xiaosong and Deng, Xinchen and Wang, Feng and Deng, Zhongwei and Lin, Xianke and Teodorescu, Remus and Pecht, Michael G},
  journal={Proceedings of the IEEE},
  volume={110},
  number={6},
  pages={735--753},
  year={2022},
  publisher={IEEE}
}

@article{azizighalehsari2024empowering,
  title={Empowering Electric Vehicles Batteries: A Comprehensive Look at the Application and Challenges of Second-Life Batteries},
  author={Azizighalehsari, Seyedreza and Venugopal, Prasanth and Pratap Singh, Deepak and Batista Soeiro, Thiago and Rietveld, Gert},
  journal={Batteries},
  volume={10},
  number={5},
  pages={161},
  year={2024},
  publisher={MDPI}
}

@misc{fitdist,
author={Mathworks},
year={2025},
url={https://www.mathworks.com/help/stats/fitdist.html}
}

@article{tran2022review,
  title={A review of lithium-ion battery thermal runaway modeling and diagnosis approaches},
  author={Tran, Manh-Kien and Mevawalla, Anosh and Aziz, Attar and Panchal, Satyam and Xie, Yi and Fowler, Michael},
  journal={Processes},
  volume={10},
  number={6},
  pages={1192},
  year={2022},
  publisher={MDPI}
}

@article{borner2017,
  title={Correlation of aging and thermal stability of commercial 18650-type lithium ion batteries},
  author={B{\"o}rner, M and Friesen, A and Gr{\"u}tzke, M and Stenzel, YP and Brunklaus, G and Haetge, J and Nowak, S and Schappacher, FM and Winter, M},
  journal={Journal of power sources},
  volume={342},
  pages={382--392},
  year={2017},
  publisher={Elsevier}
}

@article{carter2019,
  title={Mechanical collapse as primary degradation mode in mandrel-free 18650 Li-ion cells operated at 0 C},
  author={Carter, Rachel and Klein, Emily J and Atkinson III, Robert W and Love, Corey T},
  journal={Journal of Power Sources},
  volume={437},
  pages={226820},
  year={2019},
  publisher={Elsevier}
}

@article{friesen2016,
  title={Impact of cycling at low temperatures on the safety behavior of 18650-type lithium ion cells: Combined study of mechanical and thermal abuse testing accompanied by post-mortem analysis},
  author={Friesen, Alex and Horsthemke, Fabian and M{\"o}nnighoff, Xaver and Brunklaus, Gunther and Krafft, Roman and B{\"o}rner, Markus and Risthaus, Tim and Winter, Martin and Schappacher, Falko M},
  journal={Journal of power sources},
  volume={334},
  pages={1--11},
  year={2016},
  publisher={Elsevier}
}

@article{gerosa2025,
  title={Impact of Testing Method on Safety Assessment of Aged Li-ion Cells: Part I--Li Plating as Main Aging Mechanism},
  author={Gerosa, Gabriela G and Feinauer, Max and Hogrefe, Christin and H{\"a}fele, Samuel and Bischof, Katharina and W{\"o}rz, Michael and B{\"o}se, Olaf and Wohlfahrt-Mehrens, Margret and H{\"o}lzle, Markus and Waldmann, Thomas},
  journal={Journal of The Electrochemical Society},
  volume={172},
  number={3},
  pages={030502},
  year={2025},
  publisher={IOP Publishing}
}

@article{kabra2025,
  title={Quantifying the effect of degradation modes on Li-ion battery thermal instability and safety},
  author={Kabra, Venkatesh and Karmakar, Avijit and Vishnugopi, Bairav S and Mukherjee, Partha P},
  journal={Energy Storage Materials},
  volume={74},
  pages={103878},
  year={2025},
  publisher={Elsevier}
}

@article{zhang2024_2,
  title={Thermal runaway of Li-ion battery with different aging histories},
  author={Zhang, Liwen and Liu, Lu and Terekhov, Alexander and Warnberg, Douglas and Zhao, Peng},
  journal={Process Safety and Environmental Protection},
  volume={185},
  pages={910--917},
  year={2024},
  publisher={Elsevier}
}

@article{liu2023,
  title={Thermal stability of lithium-ion battery subjected to inhomogeneous aging},
  author={Liu, Jialong and Zhou, Longfei and Zhang, Yun and He, Tengfei and Wang, Zhirong},
  journal={Process Safety and Environmental Protection},
  volume={180},
  pages={992--1002},
  year={2023},
  publisher={Elsevier}
}

@article{xie2020,
  title={Influence of cycling aging and ambient pressure on the thermal safety features of lithium-ion battery},
  author={Xie, Song and Ren, Lixiang and Yang, Xiaoyong and Wang, Haibin and Sun, Qiang and Chen, Xiantao and He, Yuanhua},
  journal={Journal of Power Sources},
  volume={448},
  pages={227425},
  year={2020},
  publisher={Elsevier}
}

@article{zhang2016,
  title={The evolution of lithium-ion cell thermal safety with aging examined in a battery testing calorimeter},
  author={Zhang, Jianbo and Su, Laisuo and Li, Zhe and Sun, Ying and Wu, Ningning},
  journal={Batteries},
  volume={2},
  number={2},
  pages={12},
  year={2016},
  publisher={MDPI}
}

@article{abada2018,
  title={Combined experimental and modeling approaches of the thermal runaway of fresh and aged lithium-ion batteries},
  author={Abada, Sara and Petit, Martin and Lecocq, Amandine and Marlair, Guy and Sauvant-Moynot, Val{\'e}rie and Huet, Fran{\c{c}}ois},
  journal={Journal of Power Sources},
  volume={399},
  pages={264--273},
  year={2018},
  publisher={Elsevier}
}

@article{liu2020,
  title={Aging mechanisms and thermal stability of aged commercial 18650 lithium ion battery induced by slight overcharging cycling},
  author={Liu, Jialong and Duan, Qiangling and Ma, Mina and Zhao, Chunpeng and Sun, Jinhua and Wang, Qingsong},
  journal={Journal of power sources},
  volume={445},
  pages={227263},
  year={2020},
  publisher={Elsevier}
}

@article{wang2020,
  title={Thermal runaway behavior and features of LiFePO4/graphite aged batteries under overcharge},
  author={Wang, Cong-jie and Zhu, Yan-li and Gao, Fei and Qi, Chuang and Zhao, Peng-long and Meng, Qing-fen and Wang, Jian-yong and Wu, Qi-bing},
  journal={International Journal of Energy Research},
  volume={44},
  number={7},
  pages={5477--5487},
  year={2020},
  publisher={Wiley Online Library}
}

@article{waldmann2017,
  title={Electrochemical, post-mortem, and ARC analysis of Li-ion cell safety in second-life applications},
  author={Waldmann, Thomas and Quinn, Jason B and Richter, Karsten and Kasper, Michael and Tost, Alexander and Klein, Andreas and Wohlfahrt-Mehrens, Margret},
  journal={Journal of The Electrochemical Society},
  volume={164},
  number={13},
  pages={A3154},
  year={2017},
  publisher={IOP Publishing}
}

@article{kuntz2023,
  title={Evolution of safety behavior of high-power and high-energy commercial Li-Ion cells after electric vehicle aging},
  author={Kuntz, Pierre and Lonardoni, Lo{\"\i}c and Genies, Sylvie and Raccurt, Olivier and Aza{\"\i}s, Philippe},
  journal={Batteries},
  volume={9},
  number={8},
  pages={427},
  year={2023},
  publisher={MDPI}
}

@article{feng2020,
  title={Mitigating thermal runaway of lithium-ion batteries},
  author={Feng, Xuning and Ren, Dongsheng and He, Xiangming and Ouyang, Minggao},
  journal={Joule},
  volume={4},
  number={4},
  pages={743--770},
  year={2020},
  publisher={Elsevier}
}

@article{moy2024second,
  title={Second-life lithium-ion battery aging dataset based on grid storage cycling},
  author={Moy, Kevin and Khan, Muhammad Aadil and Fasolato, Simone and Pozzato, Gabriele and Allam, Anirudh and Onori, Simona},
  journal={Data in Brief},
  volume={57},
  pages={111046},
  year={2024},
  publisher={Elsevier}
}

@article{wang2025deep,
  title={Deep sorting of reused batteries for enabling long-term consistency grouping with unknown prior conditions},
  author={Wang, Shuquan and Gao, Feng and Tian, Hao},
  journal={Cell Reports Physical Science},
  year={2025},
  publisher={Elsevier}
}

@article{wheeler2025impact,
  title={On the impact of the negative electrode in Graphite/LFP cells during first and second life aging experiments},
  author={Wheeler, William and Bultel, Yann and Venet, Pascal and Riviere, Elie and Sari, Ali},
  journal={Journal of Energy Storage},
  volume={129},
  pages={117288},
  year={2025},
  publisher={Elsevier}
}

@article{mishra2023prisma,
  title={PRISMA for review of management literature--method, merits, and limitations--an academic review},
  author={Mishra, Vinaytosh and Mishra, Monu Pandey},
  journal={Advancing methodologies of conducting literature review in management domain},
  pages={125--136},
  year={2023},
  publisher={Emerald Publishing Limited}
}

@inproceedings{jesus2026impact,
  title={Impact of degradation pathways in the thermal stability of Li-ion cells},
  author={Cruz Rodriguez, Jesus Armando and Lecompte, Matthieu and Abada, Sara and Redondo-Iglesias, Eduardo and Pelissier, Serge},
  booktitle={2026 IEEE Vehicle Power and Propulsion Conference (VPPC)},
  pages={1--5},
  year={2026},
  organization={IEEE}
}

@article{cruz2025evaluating,
  title={Evaluating Safety Behavior of Li-Ion Batteries Regarding Negative Electrode Composition},
  author={Cruz Rodriguez, Jesus Armando and Lecompte, Matthieu and Redondo-Iglesias, Eduardo and Pelissier, Serge and Abada, Sara},
  journal={SAE International Journal of Advances and Current Practices in Mobility 369593},
  volume={8},
  number={1},
  pages={341--356},
  year={2025},
  publisher={SAE International 400 Commonwealth Drive, Warrendale, PA, United States}
}

@article{wang2025research,
  title={Research on aging evolution and safety characteristics of lithium-ion batteries cycling at low temperature},
  author={Wang, Ruiheng and Xue, Bing},
  journal={Batteries},
  volume={11},
  number={11},
  pages={396},
  year={2025},
  publisher={MDPI}
}

@article{zhang2024_1,
  title={Lithium-ion battery thermal safety evolution during high-temperature nonlinear aging},
  author={Zhang, Guangxu and Shen, Wei and Wei, Xuezhe},
  journal={Fuel},
  volume={362},
  pages={130845},
  year={2024},
  publisher={Elsevier}
}

@inproceedings{hassini2026ageing,
  title={Ageing of High-Capacity Prismatic Second-Life Batteries Repurposed in a Mobile Charging Station},
  author={Hassini, Marwan and Redondo-Iglesias, Eduardo and Venet, Pascal},
  booktitle={2026 IEEE Vehicle Power and Propulsion Conference (VPPC)},
  pages={1--6},
  year={2026},
  organization={IEEE}
}
\biboptions{sort&compress}





\end{document}